\documentstyle[amsfonts,myart,12pt]{article}
\normalbaselines
\font\tenbm=cmmib10
\font\sevenbm=cmmib7
\font\fivebm=cmmib5
\newfam\bmfam
\textfont\bmfam=\tenbm \scriptfont\bmfam=\sevenbm
\scriptscriptfont\bmfam=\fivebm
{\count0=\number\bmfam \multiply\count0 by "100
\def\defbgreek#1#2#3{{\count1=\count0 \advance\count1 by "#2#3
  \global\mathchardef#1=\count1 }}
\defbgreek\balpha  0B \defbgreek\brho       1A
\defbgreek\bbeta   0C \defbgreek\bsigma     1B
\defbgreek\bgamma  0D \defbgreek\btau       1C
\defbgreek\bdelta  0E \defbgreek\bupsilon   1D
\defbgreek\bepsilon0F \defbgreek\bphi       1E
\defbgreek\bzeta   10 \defbgreek\bchi       1F
\defbgreek\bmeta   11 \defbgreek\bpsi       20
\defbgreek\btheta  12 \defbgreek\bomega     21
\defbgreek\biota   13 \defbgreek\bvarepsilon22
\defbgreek\bkappa  14 \defbgreek\bvartheta  23
\defbgreek\blambda 15 \defbgreek\bvarpi     24
\defbgreek\bmu     16 \defbgreek\bvarrho    25
\defbgreek\bnu     17 \defbgreek\bvarsigma  26
        \defbgreek\bxi     18 \defbgreek\bvarphi    27
\defbgreek\bpi     19}
\newtheorem{opred}{Definition}[section]
\input{tcilatex}
\input tcilatex.tex

\begin{document}

\author{Yuri A. Rylov}
\title{The Euclidean geometry deformations and capacities of their application to
microcosm space-time geometry}
\date{Institute for Problems in Mechanics, Russian Academy of Sciences \\
101-1 ,Vernadskii Ave., Moscow, 119526, Russia \\
email: rylov@ipmnet.ru\\
Web site: {$http://rsfq1.physics.sunysb.edu/\symbol{126}rylov/yrylov.htm$}}
\maketitle

\begin{abstract}
Usually a Riemannian geometry is considered to be the most general geometry,
which could be used as a space-time geometry. In fact, any Riemannian
geometry is a result of some deformation of the Euclidean geometry. Class of
these Riemannian deformations is restricted by a series of unfounded
constraints. Eliminating these constraints, one obtains a more wide class of
possible space-time geometries (T-geometries). Any T-geometry is described
by the world function completely. T-geometry is a powerful tool for the
microcosm investigations due to three its characteristic features: (1) Any
geometric object is defined in all T-geometries at once, because its
definition does not depend on the form of world function. (2) Language of
T-geometry does not use external means of description such as coordinates
and curves; it uses only primordially geometrical concepts: subspaces and
world function. (3) There is no necessity to construct the complete
axiomatics of T-geometry, because it uses deformed Euclidean axiomatics, and
one can investigate only interesting geometric relations. Capacities of
T-geometries for the microcosm description are discussed in the paper. When
the world function is symmetric and T-geometry is nondegenerate, the
particle mass is geometrized, and nonrelativistic quantum effects are
described as geometric ones, i.e. without a reference to principles of
quantum theory. When world function is asymmetric, the future is not
geometrically equivalent to the past, and capacities of T-geometry increase
multiply. Antisymmetric component of the world function generates some
metric fields, whose influence on geometry properties is especially strong
in the microcosm.
\end{abstract}


\newpage

\section{Introduction}

In the case, when the existing theory cannot explain observed physical
phenomena, a choice of an appropriate space-time geometry is the most
effective and simplest method of solution of arising problems. The most
reliable and doubtless conceptions of contemporary theoretical physics: the
special relativity theory and the general relativity theory were created by
means of a change of the space-time geometry. It is a common practice to
consider the Riemannian conception of geometry to be the most general
conception of geometry. It is common practice to think that further
development of usual geometry is impossible. To increase the geometry
capacities, some authors tries to provide geometry by such unusual
properties as stochasticity and noncommutativity.

In reality the Riemannian conception of geometry is not the most general
conception of geometry. List of geometries, generated by the Riemannian
conception of geometry, is restricted by a series of unfounded constraints.
Imposition of these constraints was generated by a series of historical
reasons and cannot be justified. For instance, there is no necessity for
introducing stochasticity in geometry. It is sufficient to eliminate some
constraints, imposed on the Riemannian geometry. After elimination of these
constraints the motion of particles in such a space-time geometry becomes to
be stochastic automatically \cite{R91}, although the geometry in itself
remains to be deterministic.

To understand this, let us consider the problem, what is the geometry, in
general, and the Riemannian geometry, in particular. Well known
mathematician Felix Klein \cite{K37} supposed that only such a construction
on the point set is a geometry, where all points of the set have the same
properties. For instance, Felix Klein insisted that Euclidean geometry and
Lobachevsky geometry are geometries, whereas the Riemannian geometries are
not geometries at all. As a rule the Riemannian geometries are not uniform,
and their points have different properties. According to the Felix Klein
opinion, they should be called as ''Riemannian topographies'' or as
''Riemannian geographies''.

It is hardly relevant now to discuss the question what is the correct name
for the Riemannian geometry, but it is very important to understand, why
Felix Klein insisted on different names for the Euclidean geometry and for
the Riemannian one. The fact is that one can formulate axiomatics (system of
axioms), determining the Euclidean geometry as a self-sufficient
construction, which does not need auxiliary means for its construction.
There is no axiomatics for the Riemannian geometry. First, it is very
difficult (technically complicated) to construct axiomatics for each of
possible Riemannian geometries. Second, there is no necessity in such
axiomatics. In applications of the Riemannian geometry to the space-time
model only relations between the physical objects (world lines of particles,
vectors, etc.) are important. The geometry in itself is less interesting.
All relations of the Riemannian geometry are obtained as a result of
modification (deformation) of corresponding Euclidean relations.

Practically, this deformation is realized by a replacement of infinitesimal
Euclidean interval $ds_{{\rm E}}^{2}=\eta _{ik}dx^{i}dx^{k}$,\quad $\eta
_{ik}=$const by the infinitesimal interval $ds_{{\rm R}%
}^{2}=g_{ik}dx^{i}dx^{k}$, where $g_{ik}$ is a function of the point $x$.
Such a replacement is a change of distances between the points of the
space-time, what is a space-time deformation by definition. Thus, the
Riemannian geometry is not a self-sufficient construction (it has not its
own axiomatics). The Riemannian geometry is a deformed Euclidean geometry.
The Riemannian deformation of the space-time, converting Euclidean geometry
to the Riemannian one, form a class of deformations, restricted by a series
of constraints.

In general, any deformation is described by a change of distances $\rho $
between all pairs of space points. In the case of the space-time this
distance may be real (timelike), or imaginary (spacelike). It is more
reasonable to use the quantity $\Sigma (P,Q)=\frac{1}{2}\rho ^{2}\left(
P,Q\right) $, known as world function \cite{S60}. Here $\rho \left(
P,Q\right) $ is the distance between the points $P$ and $Q$. The world
function is real always, and it is very convenient at description of
geometry. The world function contains complete information on geometry. This
property is the most remarkable property of the world function. In
application to the Euclidean geometry, as a special case of Riemannian
geometry, this property is formulated in the form of a theorem \cite
{R00,R01,R02}, which states that, if and only if the world function
satisfies some Euclideaness conditions, formulated in terms of the world
function, the corresponding geometry is Euclidean. These conditions will be
written down, as soon as corresponding mathematical technique is developed.
Now it is important only that the Euclideaness conditions contain references
only to the world function and finite subspaces of the whole space. The
dimension of the space, and all other parameters of the Euclidean geometry
are determined by the form of the world function.

In the case of Euclidean geometry all information on geometry is contained
in the world function. This property remains to be valid also, if the world
function does not satisfy the Euclideaness conditions and the geometry is
not Euclidean. Then any choice of the world function $\Sigma $ corresponds
to some geometry ${\cal G}_{\Sigma }$. This circumstance can be interpreted
in the sense, that the world function $\Sigma $ describes deformation of the
Euclidean space, and any deformation $\Sigma $ corresponds to some geometry $%
{\cal G}_{\Sigma }$, which can be interpreted as a result of the Euclidean
geometry deformation.

From this viewpoint the Riemannian geometry is a result of the Euclidean
geometry deformation, when the world function between the points $x$ and $%
x^{\prime }$ is determined by the relation 
\begin{equation}
\Sigma \left( x,x^{\prime }\right) =\frac{1}{2}\left( \int\limits_{{\cal L}%
_{g}}\sqrt{g_{ik}dx^{i}dx^{k}}\right) ^{2}  \label{a1.2}
\end{equation}
where integration is produced along the shortest curve (geodesic) ${\cal L}%
_{g}$ between the points $x$ and $x^{\prime }$.

The relation (\ref{a1.2}) describes the Riemannian deformation of the
Euclidean space. This deformation is determined by the dimension $n$ of the
space and by the metric tensor $g_{ik}$, which is a set of $n(n+1)/2$
functions of one space point $x$. Information contained in these $n(n+1)/2$
functions is much less, than information included in one function $\Sigma $
of two space points $x$ and $x^{\prime }$. In other words, the Riemannian
deformation is a deformation of a very special form. This raises the
question. What are foundations for consideration of the Riemannian
deformation as the most general admissible deformation of the space-time?
Why the Riemannian geometry is the most general possible geometry of the
space-time?

There are no reasonable foundations for pretention of the Riemannian
geometry to the role of the most general space-time geometry. Consideration
of the Riemannian geometry as the unique possible space-time geometry is a
delusion which should be rejected. The question what is the reason of this
delusion is important and interesting. We shall not discuss it, restricting
ourselves by the remark that this delusion is an associative delusion \cite
{R001}. In other words, it is a delusion of the same sort, which stimulated
the scientific community to believe the Ptolemaic doctrine for a long time.
In other time (in the middle of XIX century) a delusion of the same sort
stimulated rejection of the idea of non-Euclidean geometry.

As soon as we assume that there are non-Riemannian deformations of the
Euclidean geometry, generating more general geometries, than Riemannian ones
(we shall refer to them as T-geometries), the simple and evident idea
arises, that many properties of particles in the microcosm can be explained
as specific properties of the microcosm geometry. In this connection it is
relevant to mention that the special relativity theory has solved problems
of motion with large velocities by means of simple change of the space-time
geometry. The general relativity has solved problems of relativistic
gravitation, changing the Minkowski geometry by the most general Riemannian
geometry, connecting the form of geometry with the matter distribution in
the space-time.

Description of nonrelativistic quantum effects can be obtained also by means
a simple change of the space-time geometry \cite{R91}. The arising
space-time geometry depends explicitly on the quantum constant. The particle
motion in such a geometry appears to be stochastic (quantum mechanics
principles are not mentioned at such a description of quantum effects). Such
an explanation of quantum effects differs from the conventional
quantum-mechanical explanation by absence of any additional suppositions.
One considers simply all possible geometries, generated by arbitrary
deformations (but not only Riemannian ones) of Euclidean geometry. One
chooses among these geometries the geometry which corresponds to the best
advantage to the experimental data.

As far as the world function as a function of two space points contains much
more information, than the metric tensor components, the capacities of
explanation of different effects, reserved in application of T-geometries as
space-time geometries, appear to be much more, than other capacities of
explanation, used now in the elementary particle theory and in the quantum
field theory. One does not need to make additional suppositions on
properties of physical phenomena in the microcosm. It is sufficient to
consider all possible T-geometries and to choose this one, which agrees with
experimental data. Of course, the problem of choosing the appropriate
geometry is not a simple problem, because for its solution one needs to
study very large massive of data. But it is important that there is no
necessity to invent anything. It is sufficient to investigate the existing
data.

From viewpoint of common sense and logic the strategy of the microcosm
investigation, based on the dominating role of geometry seems to be most
encouraging. Besides, this strategy appeared to be successful at
construction of special relativity, general relativity and explanation of
nonrelativistic quantum effects.

Idea of the geometry description in terms of only distance is a very old
idea. There were attempts to carry out this program, using so called
distance geometry \cite{M28,B53}, when some constraints, imposed on the
metric of metrical space were removed. Unfortunately, T-geometry has not
been constructed, because external means of description (in particular,
concept of a curve) were used. In previous papers \cite{R00,R01,R02} one
considered the symmetric T-geometry, i.e. T-geometry with symmetric world
function $\Sigma (P,Q)=\Sigma (Q,P)$. Such a restriction on the world
function $\Sigma (Q,P)$ seems to be usual and conventional. In the present
paper one considers non-symmetric T-geometry, when the world function is
asymmetric. Asymmetric world function is associated with the situation, when
the future and the past are not equivalent geometrically. One cannot test
experimentally, whether the future and the past are equivalent
geometrically, because one can measure the time only from the past to the
future. We do not insist that the future and the past are not equivalent
geometrically. But investigation of geometrical description reserves, hidden
in such an asymmetry, seems to be useful.

The main constraints on the Riemannian deformations are as follows: (1)
fixed dimension, (2) continuity, (3) impossibility of deformation of
one-dimensional curve into a surface, or into a point. These specific
constraints are conditioned by application of a coordinate system at the
description of Euclidean and Riemannian geometries. Indeed, any coordinate
system has fixed number of coordinates (dimension). Coordinates change
continuously, and this property is attributed to geometry, because
coordinates label the space points. Finally, transformations of coordinates
transform one-dimensional curve to one-dimensional curve, and this property
of coordinate system is attributed to geometry in itself.

There are methods of separation of geometric properties from the properties
of the coordinate system. One considers description of geometry in all
possible coordinate systems. The properties common for all these
descriptions are properties of the considered geometry. But there are no
coordinate transformations from $n$ coordinates to $m$ coordinates ($m\neq n$%
). There are no coordinate transformations, which transform one-dimensional
curve to $n$-dimensional surface ($n\neq 1$), and it is a common practice to
attribute these properties of the coordinate systems to geometry in itself.

Thus, constraints on the means of description are attributed to the geometry
in itself. To remove these constraints, generated by the means of
description, one should remove all external means of descriptions and use
the language, which uses only concepts which are attributes of the geometry
in itself. It means that the geometry is to be described in terms of
subspaces and world functions between points of these subspaces.

Practically a use of only finite subspaces of the whole space appears to be
sufficient. As a result the description of geometry is carried out in terms
of finite number of points and world function between pairs of them. Such a
description, which does not contain any external means of description, will
be referred to as $\sigma $-immanent description. The $\sigma $-immanent
description is convenient in the sense that it admits one to deal with
geometry directly. One does not need to consider coordinate systems and
group of their transformations. Sometimes we shall use the coordinate
description to connect $\sigma $-immanent description with conventional
description of geometry. But construction of T-geometry is produced in the $%
\sigma $-immanent form.

In the second section the main statements of T-geometry are formulated.
Concepts of a multivector, scalar $\Sigma $-product and collinearity are
introduced in the third sectioon. The fourth section is devoted to
investigation of the tube properties. In the fifth section one investigates
a connection between the particle motion stochasticity and the T-geometry
dondegeneracy. The particle dynamics in the nondegenerate space-time
geometry is investigated in the sixth section. Asymmetric T-geometry on a
manifold is investigated in the seventh section. The eighth section is
devoted to of the world function properties in vicinity of coincidence of
its arguments. Properties different sorts of curvature tensors are
investigated in the nineth section. The tenth section is devoted to
investigation of gradient lines. The nondegeneracy conditions of the neutral
first order tube are investigated in the eleventh section. Examples of the
first order tubes are considered in the twelfth section.

\section{T-geometry and $\Sigma $-space. Coordinate-free description.}

Let us yield necessary definitions.

\begin{opred}
T-geometry is the set of all statements about properties of all geometric
objects .
\end{opred}

The T-geometry is constructed on the point set $\Omega $ by giving the world
function $\Sigma $. The $\Sigma $-space $V=\{\Sigma ,\Omega \}$ is obtained
from the metric space after removal of the constraints, imposed on the
metric $\rho $, and introduction of the world function $\Sigma $ 
\begin{equation}
\Sigma \left( P,Q\right) =\frac{1}{2}\rho ^{2}\left( P,Q\right) ,\qquad
P,Q\in \Omega  \label{g2.1}
\end{equation}
instead of the metric $\rho $:

\begin{opred}
\label{dg2.1}$\Sigma $-space $V=\{\Sigma ,\Omega \}$ is nonempty set $\Omega 
$ of points $P$ with given on $\Omega \times \Omega $ real function $\Sigma $
\begin{equation}
\Sigma :\quad \Omega \times \Omega \rightarrow {\Bbb R},\qquad \Sigma
(P,P)=0,\qquad \forall P\in \Omega .  \label{g2.2}
\end{equation}
\end{opred}

The function $\Sigma $ is known as the world function \cite{S60}, or $\Sigma 
$-function. The metric $\rho $ may be introduced in $\Sigma $-space by means
of the relation (\ref{g2.1}). If $\Sigma $ is positive, the metric $\rho $
is also positive, but if $\Sigma $ is negative, the metric is imaginary.

\begin{opred}
\label{dg2.2}. Nonempty point set $\Omega ^{\prime }\subset \Omega $ of $%
\Sigma $-space $V=\{\Sigma ,\Omega \}$ with the world function $\Sigma
^{\prime }=\Sigma |_{\Omega ^{\prime }\times \Omega ^{\prime }}$, which is a
contraction $\Sigma $ on $\Omega ^{\prime }\times \Omega ^{\prime }$, is
called $\Sigma $-subspace $V^{\prime }=\{\Sigma ^{\prime },\Omega ^{\prime
}\}$ of $\Sigma $-space $V=\{\Sigma ,\Omega \}$.
\end{opred}

Further the world function $\Sigma ^{\prime }=\Sigma |_{\Omega ^{\prime
}\times \Omega ^{\prime }}$, which is a contraction of $\Sigma $ will be
denoted as $\Sigma $. Any $\Sigma $-subspace of $\Sigma $-space is a $\Sigma 
$-space. In T-geometry a geometric object ${\cal O}$ is described by means
of skeleton-envelope method. It means that any geometric object ${\cal O}$
is defined as follows.

\begin{opred}
\label{d1.7} Geometric object ${\cal O}$ is some $\Sigma $-subspace of $%
\Sigma $-space, which can be represented as a set of intersections and joins
of elementary geometric objects (EGO).
\end{opred}

\begin{opred}
\label{dd3.1}Elementary geometric object ${\cal E}\subset \Omega $ is a set
of zeros of the envelope function 
\begin{equation}
f_{{\cal P}^{n}}:\qquad \Omega \rightarrow {\Bbb R},\qquad {\cal P}%
^{n}\equiv \left\{ P_{0},P_{1},...P_{n}\right\} \in \Omega ^{n+1}
\label{b1.4}
\end{equation}
i.e. 
\begin{equation}
{\cal E}={\cal E}_{f}\left( {\cal P}^{n}\right) =\left\{ R|f_{{\cal P}%
^{n}}\left( R\right) =0\right\}  \label{b1.4a}
\end{equation}
The finite set ${\cal P}^{n}\subset \Omega $ of parameters of the envelope
function $f_{{\cal P}^{n}}$ is the skeleton of elementary geometric object
(EGO). The set ${\cal E}\subset \Omega $ of points forming EGO is called the
envelope of its skeleton ${\cal P}^{n}$. The envelope function $f_{{\cal P}%
^{n}}$ is an algebraic function of $s$ arguments $w=\left\{
w_{1},w_{2},...w_{s}\right\} $, $s=(n+2)(n+1)$. Each of arguments $%
w_{k}=\Sigma \left( Q_{k},L_{k}\right) $ is a $\Sigma $-function of two
arguments $Q_{k},L_{k}\in \left\{ R,{\cal P}^{n}\right\} $.
\end{opred}

For continuous T-geometry the envelope ${\cal E}$ is usually a continual set
of points. The envelope function $f_{{\cal P}^{n}}$, determining EGO is a
function of the running point $R\in \Omega $ and of parameters ${\cal P}%
^{n}\in \Omega ^{n+1}$. Thus, any elementary geometric object is determined
by its skeleton ${\cal P}^{n}$ and by the form of the envelope function $f_{%
{\cal P}^{n}}$.

Let us investigate T-geometry on the $\Sigma $-space $V=\left\{ \Sigma
,\Omega \right\} $. For some special choice $\Sigma _{{\rm E}}$ of $\Sigma $%
-function, the $\Sigma $-space $V$ turns to a $\Sigma $-subspace $V_{{\rm E}%
}^{\prime }=\left\{ \Sigma _{{\rm E}},\Omega \right\} $ of a $n$-dimensional
proper Euclidean space $V_{{\rm E}}=\left\{ \Sigma _{{\rm E}},\Omega _{{\rm E%
}}\right\} $, \ $\Omega \subset \Omega _{{\rm E}}$. (It will be shown). Then
all relations between geometric objects in $V_{{\rm E}}^{\prime }$ are
relations of proper Euclidean geometry. Replacement of $\Sigma _{{\rm E}}$
by $\Sigma $ means a deformation of $V_{{\rm E}}^{\prime }$, because world
function $\Sigma $ describes distances between two points, and change of
these distances is a deformation of the space. We shall use concept of
deformation in a wide meaning, including in this term any increase and any
reduction of number of points in the set $\Omega $. Then any transition from 
$\left\{ \Sigma _{{\rm E}},\Omega _{{\rm E}}\right\} $ to $\left\{ \Sigma
,\Omega \right\} $ is a deformation of $\left\{ \Sigma _{{\rm E}},\Omega _{%
{\rm E}}\right\} $.

Let us write Euclidean relations between geometric objects in $V_{{\rm E}%
}^{\prime }$ in the $\sigma $-immanent form (i.e. in the form, which
contains references only to geometrical objects and $\Sigma $-function).
Replacing the world function $\Sigma _{{\rm E}}$ by $\Sigma $ in these
relations, one obtains the relations between the geometric objects in the $%
\Sigma $-space $V=\left\{ \Sigma ,\Omega \right\} $.

Geometry in the proper Euclidean space is known very well, and one uses
deformation, described by world function, to establish T-geometry in
arbitrary $\Sigma $-space. Considering deformations of Euclidean space, one
goes around the problem of axiomatics in the $\Sigma $-space $V=\left\{
\Sigma ,\Omega \right\} $. One uses only Euclidean axiomatics. T-geometry of
arbitrary $\Sigma $-space is obtained as a result of ''deformation of proper
Euclidean geometry''. This point is very important, because axiomatics of
arbitrary T-geometry is very complicated. It is relatively simple only for
highly symmetric spaces. Investigation of arbitrary deformations is much
simpler, than investigations of arbitrary axiomatics. Formally, a work with
deformations of $\Sigma $-spaces is manipulations with the world function.
These manipulations may be carried out without mention of space deformations.

Description of EGOs by means (\ref{b1.4}) is carried out in the
deform-invariant form (invariant with respect to $\Sigma $-space
deformations). The envelope function $f_{{\cal P}^{n}}$ as a function of
arguments $w_{k}=\Sigma \left( Q_{k},L_{k}\right) ,$ $Q_{k},L_{k}\in \left\{
R,{\cal P}^{n}\right\} $ does not depend on the form of the world function $%
\Sigma $. Thus, definition of the envelope function is invariant with
respect to deformations (deform-invariant), and the envelope function
determines any EGO in all $\Sigma $-spaces at once.

Let ${\cal E}_{{\rm E}}$ be EGO in the Euclidean geometry ${\cal G}_{{\rm E}%
} $. Let ${\cal E}_{{\rm E}}$ be described by the skeleton ${\cal P}^{n}$
and the envelope function $f_{{\cal P}^{n}}$ in the Euclidean $\Sigma $%
-space $V_{{\rm E}}=\left\{ \Sigma _{{\rm E}},\Omega \right\} $. Then the
EGO ${\cal E}$ in the T-geometry ${\cal G}$, described by the same skeleton $%
{\cal P}^{n}$ and the same envelope function $f_{{\cal P}^{n}}$ in the $%
\Sigma $-space $V=\left\{ \Sigma ,\Omega \right\} $, is an analog in ${\cal G%
}$ of the Euclidean EGO ${\cal E}_{{\rm E}}$. T-geometry ${\cal G}$ may be
considered to be a result of deformation of the Euclidean geometry ${\cal G}%
_{{\rm E}}$, when distances $\sqrt{\Sigma \left( P,Q\right) +\Sigma \left(
Q,P\right) }$ between the pairs of points $P$ and $Q$ are changed. At such a
deformation the Euclidean EGO ${\cal E}_{{\rm E}}$ transforms to its analog $%
{\cal E}$.

The Euclidean space has the most powerful group of motion, and the same
envelope ${\cal E}_{{\rm E}}$ may be generated by the envelope function $f_{%
{\cal P}^{n}}$ with different values ${\cal P}_{(1)}^{n},$ ${\cal P}%
_{(2)}^{n},...$ of the skeleton ${\cal P}^{n}$, or even by another envelope
function $f_{\left( 1\right) {\cal Q}^{m}}$. It means that the Euclidean EGO 
${\cal E}_{{\rm E}}$ may have several analogs ${\cal E}_{(1)},{\cal E}%
_{(2)},...$ in the geometry ${\cal G}$. In other words, deformation of the
Euclidean space may split EGOs, (but not only deform them). Note that the
splitting may be interpreted as a kind of deformation.

Concept of curve is defined as the continuous mapping 
\begin{equation}
{\cal L}:\;\;\left[ 0,1\right] \rightarrow \Omega ,\qquad \left[ 0,1\right]
\subset {\Bbb R},  \label{g1.4}
\end{equation}
It is a common practice to consider the curve ${\cal L}\left( \left[ 0,1%
\right] \right) \subset \Omega $ to be an important geometrical object of
geometry. From point of view of T-geometry the set of points ${\cal L}\left( %
\left[ 0,1\right] \right) \subset \Omega $ cannot be considered to be EGO,
because the mapping (\ref{g1.4}) is not deform-invariant. Indeed, let us
consider a sphere ${\cal S}_{P_{0}P_{1}}$, passing through the point $P_{1}$
and having its center at the point $P_{0}$. It is described by the envelope
function 
\begin{equation}
f_{P_{0}P_{1}}\left( R\right) =\sqrt{\Sigma \left( P_{0},R\right) +\Sigma
\left( R,P_{0}\right) }-\sqrt{\Sigma \left( P_{1},R\right) +\Sigma \left(
R,P_{1}\right) }  \label{g1.4a}
\end{equation}
In the two-dimensional proper Euclidean space the envelope function (\ref
{g1.4a}) describes a one-dimensional circumference ${\cal L}_{1}$, whereas
in the three-dimensional proper Euclidean space the envelope function (\ref
{g1.4a}) describes a two-dimensional sphere ${\cal S}_{2}$. The point set $%
{\cal L}_{1}$ can be represented as the continuous mapping (\ref{g1.4}),
whereas the surface ${\cal S}_{2}$ cannot. Transition from two-dimensional
Euclidean space to three-dimensional Euclidean space is a space deformation.
Thus, deformation of the $\Sigma $-space may destroy the property of EGO of
being a curve (\ref{g1.4}).

Application of objects, defined by the property (\ref{g1.4}) for
investigation of T-geometries is inconvenient, because the T-geometry
investigation is founded on deform-invariant methods. Formally, one cannot
choose appropriate envelope function for description of the set (\ref{g1.4}%
), because the envelope function is deform-invariant, whereas the set (\ref
{g1.4}) is not. Hence, (\ref{g1.4}) is incompatible with the definition \ref
{dd3.1} of EGO.

The nonsymmetric T-geometry, considered in this paper can be investigated by
the same methods, as the symmetric one. The world function $\Sigma $ in the
nonsymmetric T-geometry is presented in the form 
\begin{eqnarray}
\Sigma \left( P,Q\right) &=&G\left( P,Q\right) +A\left( P,Q\right) ,\qquad
P,Q\in \Omega  \label{b2.2} \\
G\left( P,Q\right) &=&G\left( Q,P\right) ,\qquad A\left( P,Q\right)
=-A\left( Q,P\right)  \label{g1.2} \\
G\left( P,Q\right) &=&\frac{1}{2}\left( \Sigma \left( P,Q\right) +\Sigma
\left( Q,P\right) \right) ,  \label{c2.3a} \\
A\left( P,Q\right) &=&\frac{1}{2}\left( \Sigma \left( P,Q\right) -\Sigma
\left( Q,P\right) \right)  \label{g1.3}
\end{eqnarray}
where $G$ denotes the symmetric part of the world function $\Sigma $,
whereas $A$ denotes its antisymmetric part.

Motives for consideration of nonsymmetric T-geometry are as follows. In the
symmetric T-geometry the distance from the point $P$ to the point $Q$ is the
same as the distance from the point $Q$ to the point $P$. In the asymmetric
T-geometry it is not so. Apparently, it is not important for spacelike
distances in the space-time, because it can be tested experimentally for
spacelike distances. In the case, when interval between points $P$ and $Q$
is timelike, one uses watch to measure this interval. But the watch can
measure the time interval only in one direction, and one cannot be sure that
the time interval is the same in opposite direction.

If the antisymmetric part $A$ of the world function does not vanish, it
means that the future and the past are not equivalent geometrically. We do
not insist that this fact takes place, but we admit this. It is useful to
construct a nonsymmetric T-geometry, to apply it to the space-time and to
obtain the corollaries of asymmetry which could be tested experimentally.
The symmetrical part of the world function generates the field of the metric
tensor $g_{ik}$. In a like way the antisymmetric part generates some vector
force filed $a_{i}$. Maybe, existence of this field can be tested
experimentally. For construction of nonsymmetric T-geometry one does not
need to make any additional supposition. It is sufficient to remove the
constraint $\Sigma \left( P,Q\right) =\Sigma \left( Q,P\right) $ and to
apply mathematical technique developed for the symmetric T-geometry with
necessary modifications.

Besides, there is a hope that nonsymmetric T-geometry will be useful in the
elementary particle theory, where the main object is a superstring. The
first order tubes (main objects of T-geometry) are associated with world
tubes of strings and branes. In the nonsymmetric T-geometry antisymmetric
variables appear. They are absent in the conventional symmetric T-geometry,
but antisymmetric variables are characteristic for the superstring theory.

Two important general remarks.

1. Nonsymmetric T-geometry, as well as the symmetric one, is considered on
an arbitrary set $\Omega $ of points $P$. It is formulated in the scope of
the purely metric conception of geometry \cite{R02}, which is very simple,
because it uses only very simple tools for the geometry description. The
T-geometry formulated in terms of the world function $\Sigma $ and finite
subsets ${\cal P}^{n}\equiv \left\{ P_{0},P_{1},...,P_{n}\right\} $ of the
set $\Omega $. Mathematically it means, that the purely metric conception of
geometry uses only mappings

\begin{equation}
m_{n}:\;\;I_{n}\rightarrow \Omega ,\qquad I_{n}\equiv \left\{
0,1,...n\right\} \subset \left\{ 0\right\} \cup {\Bbb N}  \label{c2.8c}
\end{equation}
whereas the topology-metric conception of geometry \cite{T59,ABN86,BGP92}
uses much more complicated mappings (\ref{g1.4}), known as curves ${\cal L}$%
. Both mappings (\ref{c2.8c}) and (\ref{g1.4}) are methods of the geometry
description (and construction). But the method (\ref{c2.8c}) is much
simpler. It can be studied exhaustively, whereas the set of mappings (\ref
{g1.4}) cannot.

2. The nonsymmetric T-geometry will be mainly interpreted as a symmetric
T-geometry determined by the two-point scalar $G\left( P,Q\right) $ with
some additional metric structures, introduced to the symmetric geometry by
means of the additional two-point scalar $A\left( P,Q\right) $. For
instance, in the symmetric space-time T-geometry the world line of a free
particle is described by a geodesic. In the nonsymmetric space-time
T-geometry there are, in general, several different types of geodesics. This
fact may be interpreted in the sense, that a free particle has some internal
degrees of freedom, and it may be found in different states. In these
different states the free particle interacts differently with the force
fields, generated by the two-point scalar $A\left( P,Q\right) $. Several
different types of geodesics are results of this interaction.

\begin{opred}
\label{dg2.3}. $\Sigma $-space $V=\{\Sigma ,\Omega \}$ is called
isometrically embeddable in $\Sigma $-space $V^{\prime }=\{\Sigma ^{\prime
},\Omega ^{\prime }\}$, if there exists such a monomorphism $f:\Omega
\rightarrow \Omega ^{\prime }$, that $\Sigma (P,Q)=\Sigma ^{\prime
}(f(P),f(Q))$,\quad $\forall P,\forall Q\in \Omega ,\quad f(P),f(Q)\in
\Omega ^{\prime }$,
\end{opred}

Any $\Sigma $-subspace $V^{\prime }$ of $\Sigma $-space $V=\{\Sigma ,\Omega
\}$ is isometrically embeddable in it.

\begin{opred}
\label{dd1.1}. Two $\Sigma $-spaces $V=\{\Sigma ,\Omega \}$ and $V^{\prime
}=\{\Sigma ^{\prime },\Omega ^{\prime }\}$ are called to be isometric
(equivalent), if $V$ is isometrically embeddable in $V^{\prime }$, and $%
V^{\prime }$ is isometrically embeddable in $V$.
\end{opred}

\begin{opred}
\label{dd1.2}The $\Sigma $-space $M=\{\Sigma ,\Omega \}$ is called a finite $%
\Sigma $-space, if the set $\Omega $ contains a finite number of points.
\end{opred}

\begin{opred}
\label{dd1.3}. The $\Sigma $-subspace $M_{n}({\cal P}^{n})=\{\Sigma ,{\cal P}%
^{n}\}$ of the $\Sigma $-space $V=\{\Sigma ,\Omega \}$, consisting of $n+1$
points ${\cal P}^{n}=\left\{ P_{0},P_{1},...,P_{n}\right\} $ is called the $%
n $th order $\Sigma $-subspace .
\end{opred}

The T-geometry is a set of all propositions on properties of $\Sigma $%
-subspaces of $\Sigma $-space $V=\{\Sigma ,\Omega \}$. Presentation of
T-geometry is produced in the language, containing only references to $%
\Sigma $-function and constituents of $\Sigma $-space, i.e. to its $\Sigma $%
-subspaces.

\begin{opred}
\label{d3} A description is called $\sigma $-immanent, if it does not
contain any references to objects or concepts other, than finite subspaces
of the $\Sigma $-space and its world function (metric).
\end{opred}

$\sigma $-immanence of description provides independence of the description
on the method of description. In this sense the $\sigma $-immanence of a
description in T-geometry reminds the concept of covariance in Riemannian
geometry. Covariance of some geometric relation in Riemannian geometry means
that the considered relation is valid in all coordinate systems and, hence,
describes only the properties of the Riemannian geometry in itself.
Covariant description provides cutting-off from the coordinate system
properties, considering the relation in all coordinate systems at once. The $%
\sigma $-immanence provides truncation from the methods of description by
absence of a reference to objects, which do not relate to geometry in itself
(coordinate system, concept of curve, dimension).

The idea of constructing the T-geometry is very simple. Relations of proper
Euclidean geometry are written in the $\sigma $-immanent form and declared
to be valid for any $\Sigma $-function. This results that any relation of
proper Euclidean geometry corresponds to some relation of T-geometry.

\section{Multivectors as basic objects of T-geometry. Scalar $\Sigma $%
-product and concept of collinearity}

The basic elements of T-geometry are finite $\Sigma $-subspaces $M_{n}({\cal %
P}^{n})$, i.e. finite sets 
\begin{equation}
{\cal P}^{n}=\{P_{0},P_{1},\ldots ,P_{n}\}\subset \Omega  \label{a1.9}
\end{equation}

The simplest finite subset is a nonzero vector $\overrightarrow{{\cal P}^{1}}%
={\bf P}_{0}{\bf P}_{1}\equiv \overrightarrow{P_{0}P_{1}}$. The vector $%
\overrightarrow{P_{0}P_{1}}$ is an ordered set of two points $\left\{
P_{0},P_{1}\right\} $. The scalar product $\left( {\bf P}_{0}{\bf P}_{1}.%
{\bf Q}_{0}{\bf Q}_{1}\right) $ of two vectors ${\bf P}_{0}{\bf P}_{1}$ and $%
{\bf Q}_{0}{\bf Q}_{1}$

\begin{equation}
\left( {\bf P}_{0}{\bf P}_{1}.{\bf Q}_{0}{\bf Q}_{1}\right) =\Sigma \left(
P_{0},Q_{1}\right) -\Sigma \left( P_{1},Q_{1}\right) -\Sigma \left(
P_{0},Q_{0}\right) +\Sigma \left( P_{1},Q_{0}\right)  \label{a2.1a}
\end{equation}
is the main construction of T-geometry, and we substantiate this definition.

$\sigma $-immanent expression for scalar product $\left( {\bf P}_{0}{\bf P}%
_{1}.{\bf Q}_{0}{\bf Q}_{1}\right) $ of two vectors ${\bf P}_{0}{\bf P}_{1}$
and ${\bf Q}_{0}{\bf Q}_{1}$ in the proper Euclidean space has the form (\ref
{a2.1a}). This relation can be easily proved as follows.

In the proper Euclidean space three vectors ${\bf P}_{0}{\bf P}_{1}$, ${\bf P%
}_{0}{\bf Q}_{1}$, and ${\bf P}_{1}{\bf Q}_{1}$ are coupled by the relation 
\begin{equation}
|{\bf P}_{1}{\bf Q}_{1}|^{2}=|{\bf P}_{0}{\bf Q}_{1}-{\bf P}_{0}{\bf P}%
_{1}|^{2}=|{\bf P}_{0}{\bf P}_{1}|^{2}+|{\bf P}_{0}{\bf Q}_{1}|^{2}-2({\bf P}%
_{0}{\bf P}_{1}.{\bf P}_{0}{\bf Q}_{1})  \label{g2.3}
\end{equation}
where $({\bf P}_{0}{\bf P}_{1}.{\bf P}_{0}{\bf Q}_{1})$ denotes the scalar
product of two vectors ${\bf P}_{0}{\bf P}_{1}$ and ${\bf P}_{0}{\bf Q}_{1}$
in the proper Euclidean space, and $|{\bf P}_{0}{\bf P}_{1}|^{2}\equiv ({\bf %
P}_{0}{\bf P}_{1}.{\bf P}_{0}{\bf P}_{1})$. It follows from (\ref{g2.3}) 
\begin{equation}
({\bf P}_{0}{\bf P}_{1}.{\bf P}_{0}{\bf Q}_{1})={\frac{1}{2}}\left( |{\bf P}%
_{0}{\bf Q}_{1}|^{2}+|{\bf P}_{0}{\bf P}_{1}|^{2}-|{\bf P}_{1}{\bf Q}%
_{1}|^{2}\right)  \label{g2.4}
\end{equation}
Substituting the point $Q_{1}$ by $Q_{0}$ in (\ref{g2.4}), one obtains 
\begin{equation}
({\bf P}_{0}{\bf P}_{1}.{\bf P}_{0}{\bf Q}_{0})={\frac{1}{2}}\{|{\bf P}_{0}%
{\bf Q}_{0}|^{2}+|{\bf P}_{0}{\bf P}_{1}|^{2}-|{\bf P}_{1}{\bf Q}_{0}|^{2}\}
\label{g2.5}
\end{equation}
Subtracting (\ref{g2.5}) from (\ref{g2.4}) and using the properties of the
scalar product in the proper Euclidean space, one obtains 
\begin{equation}
({\bf P}_{0}{\bf P}_{1}.{\bf Q}_{0}{\bf Q}_{1})={\frac{1}{2}}\{|{\bf P}_{0}%
{\bf Q}_{1}|^{2}+|{\bf P}_{1}{\bf Q}_{0}|^{2}-|{\bf P}_{0}{\bf Q}_{0}|^{2}-|%
{\bf P}_{1}{\bf Q}_{1}|^{2}\}  \label{g2.6}
\end{equation}
Taking into account that in the proper Euclidean geometry $|{\bf P}_{0}{\bf Q%
}_{1}|^{2}=2\Sigma \left( P_{0},Q_{1}\right) =2G\left( P_{0},Q_{1}\right) $,
one obtains the relation\ (\ref{a2.1a}) from the relation (\ref{g2.6}).

In the Euclidean geometry the world function is symmetric, and the order of
arguments in the rhs of (\ref{a2.1a}) is not essential. In the asymmetric
T-geometry the order of arguments in the rhs of (\ref{a2.1a}) is essential.
The order has been chosen in such a way that 
\begin{eqnarray}
\left( {\bf P}_{0}{\bf P}_{1}.{\bf Q}_{0}{\bf Q}_{1}\right) _{{\rm s}}
&\equiv &\frac{1}{2}\left( \left( {\bf P}_{0}{\bf P}_{1}.{\bf Q}_{0}{\bf Q}%
_{1}\right) +\left( {\bf Q}_{0}{\bf Q}_{1}.{\bf P}_{0}{\bf P}_{1}\right)
\right)  \nonumber \\
&=&G\left( P_{0},Q_{1}\right) -G\left( P_{1},Q_{1}\right) -G\left(
P_{0},Q_{0}\right) +G\left( P_{1},Q_{0}\right)  \label{g2.7}
\end{eqnarray}
\begin{eqnarray}
\left( {\bf P}_{0}{\bf P}_{1}.{\bf Q}_{0}{\bf Q}_{1}\right) _{{\rm a}}
&\equiv &\frac{1}{2}\left( \left( {\bf P}_{0}{\bf P}_{1}.{\bf Q}_{0}{\bf Q}%
_{1}\right) -\left( {\bf Q}_{0}{\bf Q}_{1}.{\bf P}_{0}{\bf P}_{1}\right)
\right)  \nonumber \\
&=&A\left( P_{0},Q_{1}\right) -A\left( P_{1},Q_{1}\right) -A\left(
P_{0},Q_{0}\right) +A\left( P_{1},Q_{0}\right)  \label{g2.8}
\end{eqnarray}
It follows from (\ref{a2.1a}) that 
\begin{equation}
\left( {\bf P}_{0}{\bf P}_{1}.{\bf Q}_{0}{\bf Q}_{1}\right) =-\left( {\bf P}%
_{1}{\bf P}_{0}.{\bf Q}_{0}{\bf Q}_{1}\right) ,\qquad \left( {\bf P}_{0}{\bf %
P}_{1}.{\bf Q}_{0}{\bf Q}_{1}\right) =-\left( {\bf P}_{0}{\bf P}_{1}.{\bf Q}%
_{1}{\bf Q}_{0}\right)  \label{g2.9}
\end{equation}
Thus, the scalar product $\left( {\bf P}_{0}{\bf P}_{1}.{\bf Q}_{0}{\bf Q}%
_{1}\right) $ of two vectors ${\bf P}_{0}{\bf P}_{1}$ and ${\bf Q}_{0}{\bf Q}%
_{1}$ is antisymmetric with respect to permutation $P_{0}\leftrightarrow
P_{1}$ of points determining the vector ${\bf P}_{0}{\bf P}_{1}$, as well as
with respect to permutation $Q_{0}\leftrightarrow Q_{1}$.

\begin{opred}
\label{d.2.6} The finite $\Sigma $-space $M_{n}({\cal P}^{n})=\{\Sigma ,%
{\cal P}^{n}\}$ is called oriented $\overrightarrow{M_{n}({\cal P}^{n})}$,
if the order of its points ${\cal P}^{n}=\{P_{0},P_{1},\ldots P_{n}\}$ is
fixed.
\end{opred}

\begin{opred}
\label{d3.1.2b}. The $n$th order multivector $m_{n}$ is the mapping 
\begin{equation}
m_{n}:\qquad I_{n}\rightarrow \Omega ,\qquad I_{n}\equiv \left\{
0,1,...,n\right\}  \label{g2.10}
\end{equation}
\end{opred}

The set $I_{n}$ has a natural ordering, which generates an ordering of
images $m_{n}(k)\in \Omega $ of points $k\in I_{n}$. The ordered list of
images of points in $I_{n}$ has one-to-one connection with the multivector
and may be used as the multivector descriptor. Different versions of the
point list will be used for writing the $n$th order multivector descriptor: 
\[
\overrightarrow{P_{0}P_{1}...P_{n}}\equiv {\bf P}_{0}{\bf P}_{1}...{\bf P}%
_{n}\equiv \overrightarrow{{\cal P}^{n}} 
\]
Originals of points $P_{k}$ in $I_{n}$ are determined by the order of the
point $P_{k}$ in the list of descriptor. Index of the point $P_{k}$ has
nothing to do with the original of $P_{k}$. Further we shall use descriptor $%
\overrightarrow{P_{0}P_{1}...P_{n}}$ of the multivector instead of the
multivector. In this sense the $n$th order multivector $\overrightarrow{%
P_{0}P_{1}...P_{n}}$ in the $\Sigma $-space $V=\{\Sigma ,\Omega \}$ may be
defined as the ordered set $\{P_{l}\},\quad l=0,1,\ldots n$ of $n+1$ points $%
P_{0},P_{1},...,P_{n}$, belonging to the $\Sigma $-space $V$. Some points
may be identical. The point $P_{0}$ is the origin of the multivector $%
\overrightarrow{P_{0}P_{1}...P_{n}}$. Image $m_{n}\left( I_{n}\right) $ of
the set $I_{n}$ contains $k$ points ($k\leq n+1).$ The set of all $n$th
order multivectors $m_{n}$ constitutes the set $\Omega
^{n+1}=\bigotimes\limits_{k=1}^{n+1}\Omega $, and any multivector $%
\overrightarrow{{\cal P}^{n}}\in \Omega ^{n+1}$.

\begin{opred}
\label{d3.1.6b}. The scalar $\Sigma $-product $(\overrightarrow{{\cal P}^{n}}%
.\overrightarrow{{\cal Q}^{n}})$ of two $n$th order multivectors $%
\overrightarrow{{\cal P}^{n}}$ and $\overrightarrow{{\cal Q}^{n}}$ is the
real number 
\begin{equation}
(\overrightarrow{{\cal P}^{n}}.\overrightarrow{{\cal Q}^{n}})=\det \Vert (%
{\bf P}_{0}{\bf P}_{i}.{\bf Q}_{0}{\bf Q}_{k})\Vert
,\,\,\,\,\,\,\,\,\,\,\,i,k=1,2,...n  \label{g2.11}
\end{equation}
\begin{eqnarray}
({\bf P}_{0}{\bf P}_{i}.{\bf Q}_{0}{\bf Q}_{k}) &\equiv &\Sigma
(P_{0},Q_{k})+\Sigma (P_{i},Q_{0})-\Sigma (P_{0},Q_{0})-\Sigma (P_{i},Q_{k}),
\label{g2.12} \\
P_{0},P_{i},Q_{0},Q_{k} &\in &\Omega ,\qquad \overrightarrow{{\cal P}^{n}},%
\overrightarrow{{\cal Q}^{n}}\in \Omega ^{n+1}  \nonumber
\end{eqnarray}
\end{opred}

Operation of permutation of the multivector points can be effectively
defined in the $\Sigma $-space. Let us consider two $n$th order multivectors 
$\overrightarrow{{\cal P}^{n}}=\overrightarrow{P_{0}P_{1}P_{2}...P_{n}}$ and 
$\overrightarrow{{\cal P}_{(k\leftrightarrow l)}^{n}}=\overrightarrow{%
P_{0}P_{1}...P_{k-1}P_{l}P_{k+1}...P_{l-1}P_{k}P_{l+1}...P_{n}},\;\;(n\geq 1)
$, which is a result of permutation of points $P_{k},$ $P_{l},\;\;(k<l)$.
The scalar $\Sigma $-product $(\overrightarrow{{\cal P}^{n}}.\overrightarrow{%
{\cal Q}^{n}})$ is defined by the relation (\ref{g2.11}). One can show that 
\begin{equation}
(\overrightarrow{{\cal P}^{n}}.\overrightarrow{{\cal Q}^{n}})=-(%
\overrightarrow{{\cal P}_{(k\leftrightarrow l)}^{n}}.\overrightarrow{{\cal Q}%
^{n}})\qquad k\neq l,\qquad l,k=0,1,2,...n,\qquad \forall \overrightarrow{%
{\cal Q}^{n}}\in \Omega ^{n+1},  \label{g2.14}
\end{equation}
As far as the relation (\ref{g2.14}) is valid for permutation of any two
points of the multivector $\overrightarrow{{\cal P}^{n}}$ and for any
multivector $\overrightarrow{{\cal Q}^{n}}\in \Omega ^{n+1},$ one may write 
\begin{equation}
\overrightarrow{{\cal P}_{(i\leftrightarrow k)}^{n}}=-\overrightarrow{{\cal P%
}^{n}},\qquad i,k=0,1,...n,\qquad i\neq k,\qquad n\geq 1.  \label{g2.15}
\end{equation}
Thus, a change of the $n$th order multivector sign $(n\geq 1)$
(multiplication by the number $a=-1$) may be always defined as an odd
permutation of points.

Let us consider the relation 
\begin{equation}
\overrightarrow{{\cal P}^{n}}T\overrightarrow{{\cal R}^{n}}:\qquad (%
\overrightarrow{{\cal P}^{n}}.\overrightarrow{{\cal Q}^{n}})=(%
\overrightarrow{{\cal R}^{n}}.\overrightarrow{{\cal Q}^{n}})\wedge (%
\overrightarrow{{\cal Q}^{n}}.\overrightarrow{{\cal P}^{n}})=(%
\overrightarrow{{\cal Q}^{n}}.\overrightarrow{{\cal R}^{n}}),\qquad \forall 
\overrightarrow{{\cal Q}^{n}}\in \Omega ^{n+1},  \label{g2.16}
\end{equation}
between two $n$th order multivectors $\overrightarrow{{\cal P}^{n}}\in
\Omega ^{n+1}$ and $\overrightarrow{{\cal R}^{n}}\in \Omega ^{n+1}.$ The
relation (\ref{g2.16}) is reflexive, symmetric and transitive, and it may be
considered as an equivalence relation.

\begin{opred}
\label{d3.1.6a}. Two $n$th order multivectors $\overrightarrow{{\cal P}^n}%
\in \Omega ^{n+1}$ and $\overrightarrow{{\cal R}^n}\in \Omega ^{n+1}$ are
equivalent $\overrightarrow{{\cal P}^n}=\overrightarrow{{\cal R}^n}$, if the
relations (\ref{g2.16}) takes place.
\end{opred}

\begin{opred}
\label{d3.1.6b0}. If the $n$th order multivector $\overrightarrow{{\cal N}%
^{n}}$ satisfies the relations 
\begin{equation}
(\overrightarrow{{\cal N}^{n}}.\overrightarrow{{\cal Q}^{n}})=0\wedge (%
\overrightarrow{{\cal Q}^{n}}.\overrightarrow{{\cal N}})=0,\qquad \forall 
\overrightarrow{{\cal Q}^{n}}\in \Omega ^{n+1},  \label{g2.17}
\end{equation}
$\overrightarrow{{\cal N}^{n}}$ is the null $n$th order multivector.
\end{opred}

\begin{opred}
\label{d3.1.6c}. The length $|\overrightarrow{{\cal P}^{n}}|$ of the
multivector $\overrightarrow{{\cal P}^{n}}$ is the number 
\begin{equation}
|\overrightarrow{{\cal P}^{n}}|=\left\{ 
\begin{array}{c}
\mid \sqrt{(\overrightarrow{{\cal P}^{n}}.\overrightarrow{{\cal P}^{n}})}%
\mid =|\sqrt{F_{n}({\cal P}^{n})}|,\quad (\overrightarrow{{\cal P}^{n}}.%
\overrightarrow{{\cal P}^{n}})\geq 0 \\ 
i\mid \sqrt{(\overrightarrow{{\cal P}^{n}}.\overrightarrow{{\cal P}^{n}})}%
\mid =i|\sqrt{F_{n}({\cal P}^{n})}|,\quad (\overrightarrow{{\cal P}^{n}}.%
\overrightarrow{{\cal P}^{n}})<0
\end{array}
\right. \qquad \overrightarrow{{\cal P}^{n}}\in \Omega ^{n+1}  \label{g2.19}
\end{equation}
where the quantity $F_{n}({\cal P}^{n})$ is defined by the relations 
\begin{equation}
F_{n}:\quad \Omega ^{n+1}\rightarrow {\Bbb R},\qquad \Omega
^{n+1}=\bigotimes\limits_{k=1}^{n+1}\Omega ,\qquad n=1,2,\ldots
\label{g2.20}
\end{equation}
\begin{equation}
F_{n}\left( \overrightarrow{{\cal P}^{n}}\right) =\det ||\left( {\bf P}_{0}%
{\bf P}_{i}.{\bf P}_{0}{\bf P}_{k}\right) ||,\qquad P_{0},P_{i},P_{k}\in
\Omega ,\qquad i,k=1,2,...n  \label{g2.21}
\end{equation}
\begin{equation}
\left( {\bf P}_{0}{\bf P}_{i}.{\bf P}_{0}{\bf P}_{k}\right) \equiv \Sigma
\left( P_{i},P_{0}\right) +\Sigma \left( P_{0},P_{k}\right) -\Sigma \left(
P_{i},P_{k}\right) ,\qquad i,k=1,2,...n,  \label{a1.7}
\end{equation}
\end{opred}

The function (\ref{g2.20}) is a symmetric function of all its arguments $%
{\cal P}^{n}=\{ P_{0},P_{1},...,$ $P_{n}\} $, i.e. it is invariant with
respect to permutation of any points $P_{i},$ $P_{k}$, \ $i,k=0,1,...n$. It
follows from representation 
\[
F_{n}\left( \overrightarrow{{\cal P}^{n}}\right) =F_{n}\left( {\cal P}%
^{n}\right) =\left( \overrightarrow{{\cal P}^{n}}.\overrightarrow{{\cal P}%
^{n}}\right) 
\]
and the relation (\ref{g2.15}). It means that the squared length $|%
\overrightarrow{{\cal P}^{n}}|^{2}=\left| M\left( {\cal P}^{n}\right)
\right| ^{2}$ of any multivector $\overrightarrow{{\cal P}^{n}}$ does not
depend on the order of points. The squared length of any finite subset $%
{\cal P}^{n}$ is unique.

In the case, when multivector $\overrightarrow{{\cal P}^{n}}$ does not
contain similar points, it coincides with the oriented finite $\Sigma $%
-subspace $\overrightarrow{M_{n}({\cal P}^{n})}$, and it is a constituent of 
$\Sigma $-space. In the case, when at least two points of multivector
coincide, the multivector length vanishes, and the multivector is considered
to be a null multivector. The null multivector $\overrightarrow{{\cal P}^{n}}
$ is not a finite $\Sigma $-subspace $M_{n}({\cal P}^{n})$, or an oriented
finite $\Sigma $-subspace $\overrightarrow{M_{n}({\cal P}^{n})}$, but a use
of null multivectors assists in creation of a more simple technique, because
the null multivectors $\overrightarrow{{\cal P}^{n}}$ play a role of zeros.
Essentially, the multivectors are basic objects of T-geometry. As to
continual geometric objects, which are analogs of planes, sphere, ellipsoid,
etc., they are constructed by means of skeleton-envelope method (see \cite
{R02}) with multivectors, or finite $\Sigma $-subspaces used as skeletons.
As a consequence the T-geometry is presented $\sigma $-immanently, i.e.
without references to objects, external with respect to $\Sigma $-space.

The usual vector ${\bf P}_{0}{\bf P}_{1}\equiv \overrightarrow{P_{0}P_{1}}%
\equiv \overrightarrow{{\cal P}^{1}}=\left\{ P_{0},P_{1}\right\}
,\;\;P_{0},P_{1}\in \Omega $ is a special case of multivector. The squared
length $|{\bf P}_{0}{\bf P}_{1}|^{2}$ of the vector ${\bf P}_{0}{\bf P}_{1}$
is defined by the relation (\ref{a2.1a}). This gives 
\begin{equation}
|{\bf P}_{0}{\bf P}_{1}|^{2}\equiv \left( {\bf P}_{0}{\bf P}_{1}.{\bf P}_{0}%
{\bf P}_{1}\right) =\Sigma \left( P_{0},P_{1}\right) +\Sigma \left(
P_{1},P_{0}\right) =2G\left( P_{0},P_{1}\right)  \label{g2.22}
\end{equation}
The following quantities are also associated with the vector ${\bf P}_{0}%
{\bf P}_{1}$%
\begin{eqnarray}
|{\bf P}_{1}{\bf P}_{0}|^{2} &\equiv &\left( {\bf P}_{1}{\bf P}_{0}.{\bf P}%
_{1}{\bf P}_{0}\right) =2G\left( P_{1},P_{0}\right) ,  \label{g2.23} \\
\left( {\bf P}_{0}{\bf P}_{1}.{\bf P}_{1}{\bf P}_{0}\right) &=&-\Sigma
\left( P_{0},P_{1}\right) -\Sigma \left( P_{1},P_{0}\right) =-2G\left(
P_{0},P_{1}\right) ,  \label{g2.24}
\end{eqnarray}
It is rather unexpected that $|{\bf P}_{0}{\bf P}_{1}|^{2}=2G\left(
P_{0},P_{1}\right) $, but it is well that the vector ${\bf P}_{0}{\bf P}_{1}$
has only one length, but not two $\sqrt{2\Sigma \left( P_{0},P_{1}\right) }$
and $\sqrt{2\Sigma \left( P_{1},P_{0}\right) }$, as one could expect.

\begin{opred}
\label{d1.13} The squared length $\left| M\left( {\cal P}^{n}\right) \right|
^{2}$ of the $n$th order $\Sigma $-subspace $M\left( {\cal P}^{n}\right)
\subset \Omega $ of the $\Sigma $-space $V=\left\{ \Sigma ,\Omega \right\} $
is the real number. 
\[
\left| M\left( {\cal P}^{n}\right) \right| ^{2}=F_{n}\left( {\cal P}%
^{n}\right) , 
\]
where $M\left( {\cal P}^{n}\right) =\left\{ P_{0},P_{1},...,P_{n},\right\}
\subset \Omega $ with all different $P_{i}\in \Omega $, \ $i=0,1,...n$,\ \ $%
\overrightarrow{{\cal P}^{n}}\in \Omega ^{n+1},$ and the quantity $F_{n}(%
{\cal P}^{n})$ is defined by the relations (\ref{g2.21}) -- (\ref{a1.7}).
\end{opred}

The meaning of the written relations is as follows. In the special case,
when the $\Sigma $-space is Euclidean space, its $\Sigma $-function is
symmetric. It coincides with $\Sigma $-function of Euclidean space. Any two
points $P_{0},P_{i}$ determine the vector ${\bf P}_{0}{\bf P}_{i}$. The
relation (\ref{a1.7}) is a $\sigma $-immanent expression for the scalar $%
\Sigma $-product $\left( {\bf P}_{0}{\bf P}_{i}.{\bf P}_{0}{\bf P}%
_{k}\right) $ of two vectors. Then the relation (\ref{g2.21}) is the Gram's
determinant for $n$ vectors ${\bf P}_{0}{\bf P}_{i},\quad i=1,2,\ldots n$,
and $\sqrt{F_{n}({\cal P}^{n})}/n!$ is the Euclidean volume of the $(n+1)$%
-hedron with vertices at the points ${\cal P}^{n}$.

Now we enable to formulate in terms of the world function the necessary and
sufficient condition of that the $\Sigma $-space is the $n$-dimensional
Euclidean space

\noindent I. 
\begin{equation}
\Sigma \left( P,Q\right) =\Sigma \left( Q,P\right) ,\qquad P,Q\in \Omega
\label{a3.4a}
\end{equation}

\noindent II. 
\begin{equation}
\exists {\cal P}^{n}\subset \Omega ,\qquad F_{n}({\cal P}^{n})\neq 0,\qquad
F_{n+1}(\Omega ^{n+2})=0,  \label{a3.4}
\end{equation}

\noindent III. 
\begin{equation}
\Sigma (P,Q)={\frac{1}{2}}\sum_{i,k=1}^{n}g^{ik}({\cal P}^{n})[x_{i}\left(
P\right) -x_{i}\left( Q\right) ][x_{k}\left( P\right) -x_{k}\left( Q\right)
],\qquad \forall P,Q\in \Omega ,  \label{a3.5}
\end{equation}
where the quantities $x_{i}\left( P\right) $, $x_{i}\left( Q\right) $ are
defined by the relations 
\begin{equation}
x_{i}\left( P\right) =\left( {\bf P}_{0}{\bf P}_{i}.{\bf P}_{0}{\bf P}%
\right) ,\qquad x_{i}\left( Q\right) =\left( {\bf P}_{0}{\bf P}_{i}.{\bf P}%
_{0}{\bf Q}\right) ,\qquad i=1,2,...n  \label{a3.5a}
\end{equation}
The contravariant components $g^{ik}({\cal P}^{n}),$ $(i,k=1,2,\ldots n)$ of
metric tensor are defined by its covariant components $g_{ik}({\cal P}^{n}),$
$(i,k=1,2,\ldots n)$ by means of relations 
\begin{equation}
\sum_{k=1}^{n}g_{ik}({\cal P}^{n})g^{kl}({\cal P}^{n})=\delta
_{i}^{l},\qquad i,l=1,2,\ldots n,  \label{a3.11}
\end{equation}
where covariant components $g_{ik}({\cal P}^{n})$ are defined by relations 
\begin{equation}
g_{ik}({\cal P}^{n})=\left( {\bf P}_{0}{\bf P}_{i}.{\bf P}_{0}{\bf P}%
_{k}\right) ,\qquad i,k=1,2,\ldots n  \label{a3.9}
\end{equation}

\noindent IV.\quad The relations 
\begin{equation}
\left( {\bf P}_{0}{\bf P}_{i}.{\bf P}_{0}{\bf P}\right) =x_{i},\qquad
x_{i}\in {\Bbb R},\qquad i=1,2,\ldots n,  \label{a3.12}
\end{equation}
considered to be equations for determination of $P\in \Omega $, have always
one and only one solution.

\begin{remark}
\label{r3} The condition (\ref{a3.4}) is a corollary of the condition
(\ref{a3.5}). It is formulated in the form of a special condition, in order
that a determination of dimension were separated from determination of a
coordinate system.
\end{remark}

The condition II determines the space dimension. The condition III describes 
$\sigma $-immanently the scalar $\Sigma $-product properties of the proper
Euclidean space. Setting $n+1$ points ${\cal P}^{n}$, satisfying the
condition II, one determines $n$-dimensional basis of vectors in Euclidean
space. Relations (\ref{a3.9}), (\ref{a3.11}) determine covariant and
contravariant components of the metric tensor, and the relations (\ref{a3.5a}%
) determine covariant coordinates of points $P$ and $Q$ at this basis. The
relation (\ref{a3.5}) determines the expression for $\Sigma $-function for
two arbitrary points in terms of coordinates of these points. Finally, the
condition IV describes continuity of the set $\Omega $ and a possibility of
the manifold construction on it. Necessity of conditions I -- IV for
Euclideaness of $\Sigma $-space is evident. One can prove their sufficiency 
\cite{R01}. The connection of conditions I -- IV with the Euclideaness of
the $\Sigma $-space can be formulated in the form of a theorem. 
\begin{theorem}
\label{c2}The $\Sigma  $-space $V=\{\Sigma  ,\Omega \}$ is the
$n$-dimensional Euclidean  space, if and only if $\sigma $-immanent
conditions I -- IV are fulfilled.
\end{theorem}
\begin{remark}
\label{r2} For the $\Sigma$-space were proper Euclidean, the eigenvalues
of the matrix $g_{ik}({\cal P}^n),\quad i,k=1,2,\ldots n$ must have the same
sign, otherwise it is pseudo-Euclidean.
\end{remark}
The theorem states that it is sufficient to know metric (world function) to
construct the Euclidean geometry. Concepts of topological space and curve,
which are used usually in metric geometry for increasing its informativity,
appear to be excess in the sense that they are not needed for construction
of geometry. Proof of this theorem can be found in \cite{R01}.

\begin{opred}
\label{d3.1.5c} Two $n$th order multivectors $\overrightarrow{{\cal P}^{n}}$%
, $\overrightarrow{{\cal Q}^{n}}$ are neutrally collinear ($n$-collinear) $%
\overrightarrow{{\cal P}^{n}}\parallel _{\left( {\rm n}\right) }%
\overrightarrow{{\cal Q}^{n}}$, if 
\begin{equation}
(\overrightarrow{{\cal P}^{n}}.\overrightarrow{{\cal Q}^{n}})(%
\overrightarrow{{\cal Q}^{n}}.\overrightarrow{{\cal P}^{n}})=|%
\overrightarrow{{\cal P}^{n}}|^{2}\cdot |\overrightarrow{{\cal Q}^{n}}|^{2}
\label{a2.10}
\end{equation}
\end{opred}

\begin{opred}
\label{d3.6} The $n$th order multivector $\overrightarrow{{\cal P}^{n}}$ is $%
f$-collinear to $n$th order multivector $\overrightarrow{{\cal Q}^{n}}$, $%
\;\;\left( \overrightarrow{{\cal P}^{n}}\parallel _{\left( {\rm f}\right) }%
\overrightarrow{{\cal Q}^{n}}\right) $, if 
\begin{equation}
(\overrightarrow{{\cal P}^{n}}.\overrightarrow{{\cal Q}^{n}})^{2}=|%
\overrightarrow{{\cal P}^{n}}|^{2}\cdot |\overrightarrow{{\cal Q}^{n}}|^{2}
\label{a2.10a}
\end{equation}
\end{opred}

\begin{opred}
\label{d3.7} The $n$th order multivector $\overrightarrow{{\cal P}^{n}}$ is $%
p$-collinear to $n$th order multivector $\overrightarrow{{\cal Q}^{n}}$,\ \ $%
\left( \overrightarrow{{\cal P}^{n}}\parallel _{\left( {\rm p}\right) }%
\overrightarrow{{\cal Q}^{n}}\right) $, if 
\begin{equation}
(\overrightarrow{{\cal Q}^{n}}.\overrightarrow{{\cal P}^{n}})^{2}=|%
\overrightarrow{{\cal P}^{n}}|^{2}\cdot |\overrightarrow{{\cal Q}^{n}}|^{2}
\label{a2.10b}
\end{equation}
\end{opred}

Here indices ''f'' and ''p'' are associated with the terms ''future'' and
''past'' respectively.

In the symmetric T-geometry there is only one type of collinearity, because
the three mentioned types of collinearity coincide in the symmetric
T-geometry. The property of the neutral collinearity is commutative, i.e. if 
$\overrightarrow{{\cal P}^{n}}\parallel _{\left( {\rm n}\right) }%
\overrightarrow{{\cal Q}^{n}}$, then $\overrightarrow{{\cal Q}^{n}}\parallel
_{\left( {\rm n}\right) }\overrightarrow{{\cal P}^{n}}$. The property of $p$%
-collinearity and $f$-collinearity are not commutative, in general. Instead,
one has according to (\ref{a2.10a}) and (\ref{a2.10b}) that, if $%
\overrightarrow{{\cal P}^{n}}\parallel _{\left( {\rm p}\right) }%
\overrightarrow{{\cal Q}^{n}}$, then $\overrightarrow{{\cal Q}^{n}}\parallel
_{\left( {\rm f}\right) }\overrightarrow{{\cal P}^{n}}$.

\begin{opred}
\label{d3.15e}. The $n$th order multivector $\overrightarrow{{\cal P}^{n}} $
is $f$-parallel to the $n$th order multivector $\overrightarrow{{\cal Q}^{n}}
$ $\left( \overrightarrow{{\cal P}^{n}}\uparrow \uparrow _{({\rm f})}%
\overrightarrow{{\cal Q}^{n}}\right) $, if 
\begin{equation}
(\overrightarrow{{\cal P}^{n}}.\overrightarrow{{\cal Q}^{n}})=|%
\overrightarrow{{\cal P}^{n}}|\cdot |\overrightarrow{{\cal Q}^{n}}|
\label{a2.11}
\end{equation}
The $n$th order multivector $\overrightarrow{{\cal P}^{n}}$ is $f$%
-antiparallel to the $n$th order multivector $\overrightarrow{{\cal Q}^{n}}$ 
$\left( \overrightarrow{{\cal P}^{n}}\uparrow \downarrow _{({\rm f})}%
\overrightarrow{{\cal Q}^{n}}\right) $, if 
\begin{equation}
(\overrightarrow{{\cal P}^{n}}.\overrightarrow{{\cal Q}^{n}})=-|%
\overrightarrow{{\cal P}^{n}}|\cdot |\overrightarrow{{\cal Q}^{n}}|
\label{a2.12}
\end{equation}
\end{opred}

\begin{opred}
\label{d3.15ee}The $n$th order multivector $\overrightarrow{{\cal P}^{n}}$
is $p$-parallel to the $n$th order multivector $\overrightarrow{{\cal Q}^{n}}
$ $\left( \overrightarrow{{\cal P}^{n}}\uparrow \uparrow _{({\rm p})}%
\overrightarrow{{\cal Q}^{n}}\right) $, if 
\begin{equation}
(\overrightarrow{{\cal Q}^{n}}.\overrightarrow{{\cal P}^{n}})=|%
\overrightarrow{{\cal P}^{n}}|\cdot |\overrightarrow{{\cal Q}^{n}}|
\label{a2.12a}
\end{equation}
The $n$th order multivector $\overrightarrow{{\cal P}^{n}}$ is $p$%
-antiparallel to the $n$th order multivector $\overrightarrow{{\cal Q}^{n}}$ 
$\left( \overrightarrow{{\cal P}^{n}}\uparrow \uparrow _{({\rm p})}%
\overrightarrow{{\cal Q}^{n}}\right) $, if 
\begin{equation}
(\overrightarrow{{\cal Q}^{n}}.\overrightarrow{{\cal P}^{n}})=-|%
\overrightarrow{{\cal P}^{n}}|\cdot |\overrightarrow{{\cal Q}^{n}}|
\label{a2.12b}
\end{equation}
\end{opred}

The $f$-parallelism and the $p$-parallelism are connected as follows. If $%
\overrightarrow{{\cal P}^{n}}\uparrow \uparrow _{({\rm p})}\overrightarrow{%
{\cal Q}^{n}}$, then $\overrightarrow{{\cal Q}^{n}}\uparrow \uparrow _{({\rm %
f})}\overrightarrow{{\cal P}^{n}}$ and vice versa.

Vector ${\bf P}_{0}{\bf P}_{1}=\overrightarrow{{\cal P}^{1}}$ as well as the
vector ${\bf Q}_{0}{\bf Q}_{1}=\overrightarrow{{\cal Q}^{1}}$ are the first
order multivectors. If ${\bf P}_{0}{\bf P}_{1}\uparrow \uparrow _{({\rm f})}%
{\bf Q}_{0}{\bf Q}_{1}$, then ${\bf P}_{1}{\bf P}_{0}\uparrow \downarrow _{(%
{\rm f})}{\bf Q}_{0}{\bf Q}_{1}$ and ${\bf P}_{0}{\bf P}_{1}\uparrow
\downarrow _{({\rm f})}{\bf Q}_{1}{\bf Q}_{0}$

\section{Tubes in $\Sigma $-space and their properties.}

The simplest geometrical object in T-geometry is the $n$th order tube ${\cal %
T}\left( {\cal P}^{n}\right) $, which is determined by its skeleton ${\cal P}%
^{n}$. The tube is an analog of Euclidean $n$-dimensional plane, which is
also determined by $n+1$ points ${\cal P}^{n},$ not belonging to a $\left(
n-1\right) $-dimensional plane.

\begin{opred}
\label{d1.14} $n$th order $\Sigma $-subspace $M\left( {\cal P}^{n}\right) =%
{\cal P}^{n}$\ of nonzero length $\;\left| M\left( {\cal P}^{n}\right)
\right| ^{2}=\left| {\cal P}^{n}\right| ^{2}=F_{n}\left( {\cal P}^{n}\right)
\neq 0$ determines the $n$th order tube ${\cal T}$ $\left( {\cal P}%
^{n}\right) $ by means of relation 
\begin{equation}
{\cal T}\left( {\cal P}^{n}\right) \equiv {\cal T}_{{\cal P}^{n}}=\left\{
P_{n+1}|F_{n+1}\left( {\cal P}^{n+1}\right) =0\right\} ,\qquad P_{i}\in
\Omega ,\qquad i=0,1\ldots n+1,  \label{b1.3}
\end{equation}
where the function $F_{n}$ is defined by the relations (\ref{g2.20}) -- (\ref
{a1.7})
\end{opred}

The shape of the tube ${\cal T}\left( {\cal P}^{n}\right) $ does not depend
on the order of points of the multivector $\overrightarrow{{\cal P}^{n}}$.
The basic point ${\cal P}^{n}$, determining the tube ${\cal T}_{{\cal P}^{n}}
$, belong to ${\cal T}_{{\cal P}^{n}}$.

The first order tube ${\cal T}_{P_{0}P_{1}}$ can be defined by means of
concept of $n$-collinearity (\ref{a2.10}) 
\begin{eqnarray}
{\cal T}\left( {\cal P}^{1}\right) &\equiv &{\cal T}_{({\rm n}%
)P_{0}P_{1}}=\left\{ R|F_{2}\left( P_{0},P_{1},R\right) =0\right\} \equiv
\left\{ R\left| \overrightarrow{P_{0}P_{1}}||_{({\rm n)}}\overrightarrow{%
P_{0}R}\right. \right\}  \nonumber \\
&\equiv &\left\{ R\left| \;|\overrightarrow{P_{0}P_{1}}|^{2}|\overrightarrow{%
P_{0}R}|^{2}-\left( \overrightarrow{P_{0}P_{1}}.\overrightarrow{P_{0}R}%
\right) \left( \overrightarrow{P_{0}R}.\overrightarrow{P_{0}P_{1}}\right)
=0\right. \right\}  \label{f2.3}
\end{eqnarray}
As far as there are concepts of $f$-collinearity and of $p$-collinearity,
one can define also the first order $f$-tube and $p$-tube on the basis of
these collinearities. The first order $f$-tube is defined by the relation 
\begin{equation}
{\cal T}_{({\rm f})P_{0}P_{1}}=\left\{ R\left| \overrightarrow{P_{0}P_{1}}%
||_{({\rm f})}\overrightarrow{P_{0}R}\right. \right\} =\left\{ R\left| \;|%
\overrightarrow{P_{0}P_{1}}|^{2}|\overrightarrow{P_{0}R}|^{2}-\left( 
\overrightarrow{P_{0}P_{1}}.\overrightarrow{P_{0}R}\right) ^{2}=0\right.
\right\}  \label{f2.4}
\end{equation}
The first order $p$-tube is defined as follows 
\begin{equation}
{\cal T}_{({\rm p})P_{0}P_{1}}=\left\{ R\left| \overrightarrow{P_{0}P_{1}}%
||_{({\rm p})}\overrightarrow{P_{0}R}\right. \right\} =\left\{ R\left| \;|%
\overrightarrow{P_{0}P_{1}}|^{2}|\overrightarrow{P_{0}R}|^{2}-\left( 
\overrightarrow{P_{0}R}.\overrightarrow{P_{0}P_{1}}\right) ^{2}=0\right.
\right\}  \label{f2.5}
\end{equation}

In the symmetric T-geometry all three tubes (\ref{f2.3}) -- (\ref{f2.5})
coincide. In the nonsymmetric T-geometry they are different, in general. The
tubes (\ref{f2.3}), (\ref{f2.4}), (\ref{f2.5}) can be divided into segments,
each of them is determined by one of factors of expressions (\ref{f2.3}) -- (%
\ref{f2.5}).

In all cases the factorization of the expressions 
\[
F_{({\rm f})}\left( P_{0},P_{1},R\right) =\left| \overrightarrow{P_{0}P_{1}}%
\right| ^{2}\left| \overrightarrow{P_{0}R}\right| ^{2}-\left( 
\overrightarrow{P_{0}P_{1}}.\overrightarrow{P_{0}R}\right) ^{2} 
\]
\[
F_{({\rm p})}\left( P_{0},P_{1},R\right) =\left| \overrightarrow{P_{0}P_{1}}%
\right| ^{2}\left| \overrightarrow{P_{0}R}\right| ^{2}-\left( 
\overrightarrow{P_{0}R}.\overrightarrow{P_{0}P_{1}}\right) ^{2}, 
\]
\begin{equation}
F_{({\rm n})}\left( P_{0},P_{1},R\right) =\left| \overrightarrow{P_{0}P_{1}}%
\right| ^{2}\left| \overrightarrow{P_{0}R}\right| ^{2}-\left( 
\overrightarrow{P_{0}P_{1}}.\overrightarrow{P_{0}R}\right) \left( 
\overrightarrow{P_{0}R}.\overrightarrow{P_{0}P_{1}}\right)  \label{f2.2b}
\end{equation}
has similar form 
\begin{equation}
F_{({\rm q})}\left( P_{0},P_{1},R\right) =-F_{({\rm q}0)}F_{({\rm q}1)}F_{(%
{\rm q}2)}F_{({\rm q}3)}  \label{f2.0}
\end{equation}
Here index $q$ runs values $f,p,n$, and factorization of expressions $F_{(%
{\rm q})}\left( P_{0},P_{1},R\right) ,$ $q=f,p,n$ has a similar form 
\begin{equation}
F_{({\rm q}0)}=F_{({\rm q}0)}\left( P_{0},P_{1},R\right) =\sqrt{G_{0R}}+%
\sqrt{G_{01}}+\sqrt{G_{1R}-\eta _{{\rm q}}}  \label{f2.0a}
\end{equation}
\begin{eqnarray}
F_{({\rm q}1)} &=&F_{({\rm q}1)}\left( P_{0},P_{1},R\right) =\sqrt{G_{0R}}-%
\sqrt{G_{01}}+\sqrt{G_{1R}-\alpha _{{\rm q}}\eta _{{\rm q}}}  \nonumber \\
F_{({\rm q}2)} &=&F_{({\rm q}2)}\left( P_{0},P_{1},R\right) =\sqrt{G_{0R}}+%
\sqrt{G_{01}}-\sqrt{G_{1R}-\eta _{{\rm q}}}  \label{f2.1} \\
F_{({\rm q}3)} &=&F_{({\rm q}3)}\left( P_{0},P_{1},R\right) =\sqrt{G_{0R}}-%
\sqrt{G_{01}}-\sqrt{G_{1R}-\alpha _{{\rm q}}\eta _{{\rm q}}}  \nonumber
\end{eqnarray}
where for brevity one uses designations 
\begin{eqnarray*}
G_{ik} &=&G\left( P_{i},P_{k}\right) ,\qquad A_{ik}=A\left(
P_{i},P_{k}\right) ,\qquad i,k=0,1 \\
G_{iR} &=&G\left( P_{i},R\right) ,\qquad A_{iR}=A\left( P_{i},R\right)
,\qquad i=0,1
\end{eqnarray*}
\begin{eqnarray}
\eta _{{\rm f}} &=&-\eta _{{\rm p}}=A_{10}+A_{0R}+A_{R1},\qquad \eta _{{\rm n%
}}=\frac{\eta _{{\rm f}}^{2}}{\sqrt{4G_{01}G_{0R}+\eta _{{\rm f}}^{2}}+2%
\sqrt{G_{01}G_{0R}}}  \label{f2.2ee} \\
\alpha _{{\rm p}} &=&\alpha _{{\rm f}}=1,\qquad \alpha _{{\rm n}}=-1 
\nonumber
\end{eqnarray}
In the symmetric T-geometry, when $A\left( P,Q\right) =0$,\ \ $\forall
P,Q\in \Omega $, and $\eta =0$, all expressions (\ref{f2.1}), for $F_{\left( 
{\rm n}i\right) },F_{\left( {\rm f}i\right) },F_{\left( {\rm p}i\right)
},\;\;i=0,1,2,3$ coincide.

Factorizations (\ref{f2.0}) -- (\ref{f2.1}) determine division of the tubes
into segments. As it follows from (\ref{f2.2ee}) $\eta _{{\rm p}}=\eta _{%
{\rm f}}=\eta _{{\rm n}}=0$, provided $R=P_{0}$, or $R=P_{1}$. Then one can
see that 
\begin{eqnarray}
P_{0},P_{1} &\in &{\cal T}_{({\rm q})[P_{0}P_{1}]}=\left\{ R|F_{({\rm q}%
1)}\left( P_{0},P_{1},R\right) =0\right\}  \nonumber \\
&=&\left\{ R|\sqrt{G_{0R}}-\sqrt{G_{01}}+\sqrt{G_{1R}-\alpha _{{\rm q}}\eta
_{{\rm q}}}=0\right\}  \label{g2.3aa}
\end{eqnarray}
\begin{eqnarray}
P_{0} &\in &{\cal T}_{({\rm q})P_{0}]P_{1}}=\left\{ R|F_{({\rm q}2)}\left(
P_{0},P_{1},R\right) =0\right\}  \nonumber \\
&=&\left\{ R|\sqrt{G_{0R}}+\sqrt{G_{01}}-\sqrt{G_{1R}-\eta _{{\rm q}}}%
=0\right\}  \label{g2.3ab}
\end{eqnarray}
\begin{eqnarray}
P_{1} &\in &{\cal T}_{({\rm q})P_{0}[P_{1}}=\left\{ R|F_{({\rm q}3)}\left(
P_{0},P_{1},R\right) =0\right\}  \nonumber \\
&=&\left\{ R|\sqrt{G_{0R}}-\sqrt{G_{01}}-\sqrt{G_{1R}-\alpha _{{\rm q}}\eta
_{{\rm q}}}=0\right\}  \label{g2.3ac}
\end{eqnarray}

The tube segment 
\begin{equation}
{\cal T}_{({\rm q}0)P_{0}P_{1}}=\left\{ R|F_{({\rm q}0)}\left(
P_{0},P_{1},R\right) =0\right\}  \label{g2.3ad}
\end{equation}
determined (\ref{f2.0a}), does not contain basic points $P_{0}$, $P_{1}$, in
general. ${\cal T}_{({\rm q}0)P_{0}P_{1}}=\emptyset $ for any timelike tube $%
{\cal T}_{P_{0}P_{1}}$.

In the relations (\ref{g2.3aa}) -- (\ref{g2.3ad}) index $q$ runs values $%
f,p,n$. Values of $\eta _{{\rm q}}$, $\alpha _{{\rm q}}$ are determined by
the relations (\ref{f2.2ee}).

The tube segments may be classified by the number of basic points $%
P_{0},P_{1}$, belonging to the segment. The segment ${\cal T}_{({\rm q}%
)[P_{0}P_{1}]}$, containing two basic points will be referred to as the
internal tube segment. The segments ${\cal T}_{({\rm q})P_{0}[P_{1}}$ and $%
{\cal T}_{({\rm q})P_{0}]P_{1}}$ contain one basic point. They will be
referred to as external tube segments, or as tube rays, directed along the
vectors ${\bf P}_{0}{\bf P}_{1}$ and ${\bf P}_{1}{\bf P}_{0}$ respectively.
The segment ${\cal T}_{({\rm q}0)P_{0}P_{1}}$, which does not contain basic
points, will be referred to as null segment. As a rule it is empty.

In the geometry of Minkowski the timelike tube ${\cal T}_{P_{0}P_{1}}$,
determined by the timelike vector ${\bf P}_{0}{\bf P}_{1}$, is the straight,
passing through the points $P_{0}$ and $P_{1}$. ${\cal T}_{[P_{0}P_{1}]}$ is
the segment $[P_{0},P_{1}]$ of the straight between the points $P_{0}$ and $%
P_{1}$. The tube rays ${\cal T}_{P_{0}[P_{1}}$, ${\cal T}_{P_{0}]P_{1}}$ are
rays of the straight $[P_{1},\infty )$ and $(-\infty ,P_{0}]$. The null
segment ${\cal T}_{(0)P_{0}P_{1}}$ is empty.

\begin{opred}
\label{d3.1.9}. Section ${\cal S}_{n;P}$ of the tube ${\cal T}({\cal P}^{n})$
at the point $P\in {\cal T}({\cal P}^{n})$ is the set ${\cal S}_{n;P}({\cal T%
}({\cal P}^{n}))$ of points, belonging to the tube ${\cal T}({\cal P}^{n})$ 
\begin{equation}
{\cal S}_{n;P}({\cal T}({\cal P}^{n}))=\{P^{\prime }\mid
\bigwedge_{l=0}^{l=n}\Sigma (P_{l},P^{\prime })=\Sigma (P_{l},P)\},\qquad
P\in {\cal T}({\cal P}^{n}),\qquad P^{\prime }\in \Omega .  \label{a2.38}
\end{equation}
\end{opred}

Let us note that ${\cal S}_{n;P}({\cal T}({\cal P}^{n}))\subset {\cal T}(%
{\cal P}^{n})$, because $P\in {\cal T}({\cal P}^{n})$. Indeed, whether the
point $P$ belongs to ${\cal T}({\cal P}^{n})$ depends only on values of $n+1$
quantities $\Sigma (P_{l},P),\;\;l=0,1,...n$. In accordance with (\ref{a2.38}%
) these quantities are the same for both points $P$ and $P^{\prime }$.
Hence, the running point $P^{\prime }\in {\cal T}({\cal P}^{n})$, if $P\in 
{\cal T}({\cal P}^{n})$.

In the proper Euclidean space the $n$th order tube is $n$-dimensional plane,
containing points ${\cal P}^{n}$, and its section ${\cal S}_{n;P}({\cal T}(%
{\cal P}^{n}))$ at the point $P$ consists of one point $P$.

\begin{opred}
\label{d3.10}Section ${\cal S}_{n;P}$ of the tube ${\cal T}({\cal P}^{n})$
at the point $P\in {\cal T}({\cal P}^{n})$ is minimal, if ${\cal S}%
_{n;P}\left( {\cal T}({\cal P}^{n})\right) =\left\{ P\right\} $.
\end{opred}

\begin{opred}
\label{d3.11}The first order tube ${\cal T}_{P_{0}P_{1}}$ is degenerate, if
its section at any point $P\in {\cal T}_{P_{0}P_{1}}$ is minimal.
\end{opred}

Minimality of the first tube section means that the first order tube
degenerates to a curve, and any section of the tube consists of one point.
It means that there is only one vector $\overrightarrow{P_{0}R}$, $R\in 
{\cal T}_{P_{0}P_{1}}$ of fixed length, which is parallel, or antiparallel
to the vector $\overrightarrow{P_{0}P_{1}}$. As far as in the nonsymmetric
T-geometry there is several types of parallelism, there is several types of
degeneracy, in general. In the symmetric T-geometry there is only one type
of the degeneracy.

\begin{opred}
\label{d3.12}The $\Sigma $-space $V=\left\{ \Sigma ,\Omega \right\} $ is
degenerate on the set ${\cal T}$ of the first order tubes, if the set ${\cal %
T}$ contains only degenerate tubes ${\cal T}_{P_{0}P_{1}}$.
\end{opred}

\begin{opred}
\label{d3.12a}The $\Sigma $-space $V=\left\{ \Sigma ,\Omega \right\} $ is
locally $f$-degenerate, if all first order tubes ${\cal T}_{({\rm f}%
)P_{0}P_{1}}$ are degenerate.
\end{opred}

\begin{opred}
\label{d3.13}The $\Sigma $-space $V=\left\{ \Sigma ,\Omega \right\} $ is
locally $p$-degenerate, if all first order tubes ${\cal T}_{({\rm p}%
)P_{0}P_{1}}$ are degenerate.
\end{opred}

\begin{opred}
\label{d3.14}The $\Sigma $-space $V=\left\{ \Sigma ,\Omega \right\} $ is
locally $n$-degenerate, if all first order tubes ${\cal T}_{({\rm n}%
)P_{0}P_{1}}\equiv {\cal T}_{P_{0}P_{1}}$ are degenerate.
\end{opred}

Note that the Riemannian space considered to be a $\Sigma $-space is locally
degenerate.

\section{Nondegenerate tubes in the space-time and their interpretation}

Nondegeneracy of the first order tube ${\cal T}_{P_{0}P_{1}}$ means that the
tube is not a one-dimensional curve, although it is an analog of the
Euclidean straight line. In the Minkowski space-time geometry ${\cal G}_{%
{\rm M}}$ the timelike straight line describes the world line of a free
particle. One should expect that the nondegenerate first order tube ${\cal T}%
_{P_{0}P_{1}}$ describes also the free particle in the nondegenerate
space-time geometry ${\cal G}_{{\rm D}}$.

Let us describe $\sigma $-immanently a particle of the mass $m$ in ${\cal G}%
_{{\rm M}}$. World line of the particle is a broken line ${\cal T}_{{\rm br}%
} $ consisting of rectilinear internal segments ${\cal T}_{\left[
P_{i}P_{i+1}\right] }$, $i=0,\pm 1,\pm 2$... 
\begin{equation}
{\cal T}_{{\rm br}}=\bigcup_{i}{\cal T}_{\left[ P_{i}P_{i+1}\right] }
\label{g5.1}
\end{equation}
It is supposed that all segments ${\cal T}_{\left[ P_{i}P_{i+1}\right] }$
has the same length $\mu $, and the quantity $\mu $ is proportional to the
particle mass $m$. 
\begin{equation}
m=b\mu ,\qquad b=\text{const,}  \label{g5.2}
\end{equation}
where $b$ is some universal constant transforming the length of a segment to
its mass. The particle momentum ${\bf p}_{i}$ on the segment ${\cal T}_{%
\left[ P_{i}P_{i+1}\right] }$ is defined by the relation 
\begin{equation}
\left( {\bf p}_{i}.{\bf Q}_{0}{\bf Q}_{1}\right) =bc\left( {\bf P}_{i}{\bf P}%
_{i+1}.{\bf Q}_{0}{\bf Q}_{1}\right) ,\qquad \forall Q_{0},Q_{1}\in {\Bbb R}%
^{4}  \label{g5.2a}
\end{equation}
where $c$ is the speed of the light. It means that the momentum ${\bf p}_{i}$
is proportional to the vector ${\bf P}_{i}{\bf P}_{i+1}$, determining the
segment ${\cal T}_{\left[ P_{i}P_{i+1}\right] }$. This can be written
symbolically in the form 
\begin{equation}
{\bf p}_{i}=bc{\bf P}_{i}{\bf P}_{i+1},\qquad \left| {\bf p}_{i}\right|
^{2}=b^{2}c^{2}\mu ^{2}=m^{2}c^{2},\text{\qquad }i=0,\pm 1,\pm 2...
\label{g5.2b}
\end{equation}
Segment ${\cal T}_{\left[ P_{i}P_{i+1}\right] }$ is defined by the equation (%
\ref{g2.3aa}). In the case of the Minkowski geometry ${\cal G}_{{\rm M}}$,
as well as in the case of any symmetric T-geometry one obtains 
\begin{equation}
{\cal T}_{\left[ P_{i}P_{i+1}\right] }=\left\{ R|\sqrt{G\left(
P_{i},P_{i+1}\right) }-\sqrt{G\left( P_{i},R\right) }-\sqrt{G\left(
R,P_{i+1}\right) }\right\}  \label{g5.3}
\end{equation}

Formulae (\ref{g5.1}) -- (\ref{g5.3}) carry out $\sigma $-immanent
description of the world line of a particle. It means that these relations (%
\ref{g5.1}) -- (\ref{g5.3}) describe the particle world tube in any
symmetric T-geometry.

If the particle is free, one should add the parallelism condition ${\bf P}%
_{i}{\bf P}_{i+1}\uparrow \uparrow {\bf P}_{i+1}{\bf P}_{i+2}$: 
\begin{equation}
\left( {\bf P}_{i}{\bf P}_{i+1}.{\bf P}_{i+1}{\bf P}_{i+2}\right) =\left| 
{\bf P}_{i}{\bf P}_{i+1}\right| \cdot \left| {\bf P}_{i+1}{\bf P}%
_{i+2}\right| ,\qquad i=0,\pm 1,\pm 2...  \label{g5.4}
\end{equation}
In ${\cal G}_{{\rm M}}$ relations (\ref{g5.1}) -- (\ref{g5.4}) describe $%
\sigma $-immanently the world line of a free particle of mass $m.$ It is a
timelike straight line, because in the Minkowski geometry there is only one
timelike vector ${\bf P}_{i+1}{\bf P}_{i+2}$ of length $\mu $, which is
parallel to the vector ${\bf P}_{i}{\bf P}_{i+1}$. Then, if the vector ${\bf %
P}_{i}{\bf P}_{i+1}$ is fixed, the point ${\bf P}_{i+2}$ is determined
uniquely. It means, that if one segment, for instance ${\cal T}_{\left[
P_{0}P_{1}\right] }$, is fixed, positions of all other segments ${\cal T}_{%
\left[ P_{i}P_{i+1}\right] }$, \ $i=0,\pm 1,\pm 2...$ and the whole broken
tube ${\cal T}_{{\rm br}}$ are determined uniquely. In other words, motion
of the free particle in the Minkowski geometry ${\cal G}_{{\rm M}}$ is
deterministic.

Equations (\ref{g5.1}) -- (\ref{g5.4}), written in the $\sigma $-immanent
form, determine the free particle world tube in the case of any symmetric
T-geometry ${\cal G}_{{\rm D}}$. Let us consider the case, when the
space-time geometry is symmetric and nondegenerate. Then there are many
timelike vectors ${\bf P}_{i+1}{\bf P}_{i+2}$ of length $\mu $, which are
parallel to the vector ${\bf P}_{i}{\bf P}_{i+1}$. At fixed vector ${\bf P}%
_{i}{\bf P}_{i+1}$, the point $P_{i+2}$ is not determined uniquely. Then at
fixed segment ${\cal T}_{\left[ P_{0}P_{1}\right] }$ the positions of all
other segments ${\cal T}_{\left[ P_{i}P_{i+1}\right] }$, $i=0,\pm 1,\pm 2...$
and the whole broken tube ${\cal T}_{{\rm br}}$ are not determined uniquely.
It means that the world tube of a free particle is stochastic.

Let us consider the case of geometry ${\cal G}_{{\rm D}}$ with the symmetric
world function 
\begin{equation}
{\cal G}_{{\rm D}}:\qquad G=\Sigma =\Sigma _{{\rm M}}+D\left( \Sigma _{{\rm M%
}}\right)  \label{g5.5}
\end{equation}
\begin{equation}
D\left( \Sigma _{{\rm M}}\right) =\left\{ 
\begin{array}{cc}
d & \text{if \ }\sigma _{0}<\Sigma _{{\rm M}} \\ 
\frac{d}{\sigma _{0}}\Sigma _{{\rm M}} & \text{if \ }0<\Sigma _{{\rm M}%
}<\sigma _{0} \\ 
0 & \text{if \ }\Sigma _{{\rm M}}<0
\end{array}
\right. ,\qquad d=\frac{\hbar }{2bc}=\text{const}  \label{g5.6}
\end{equation}
where $\Sigma _{{\rm M}}$ is the world function in ${\cal G}_{{\rm M}}$, $c$
is the speed of the light, $\hbar $ is the quantum constant, and the
constant $b$ is defined by the relation (\ref{g5.2}).The quantity $\sigma
_{0}\approx d\approx 10^{-20}$cm$^{2}$. The function $D$ is the distortion
function, and the constant $d$ is an integral distortion. The distortion
function is the quantity, responsible for nondegeneracy of the space-time
geometry ${\cal G}_{{\rm D}}$. Geometry ${\cal G}_{{\rm D}}$, described by
the world function (\ref{g5.5}), (\ref{g5.6}), is uniform and isotropic. The
tube segment ${\cal T}_{\left[ P_{0}P_{1}\right] }$ has the shape of a
hallow tube. Radius $R$ of the tube is approximately $R\approx \sqrt{3d/2}$.
More exactly the shape of the tube is described by the relation \cite{R92} 
\begin{equation}
R^{2}\left( \tau \right) =\frac{3}{2}d+\frac{2\mu ^{2}d}{\mu ^{2}-2d}\left(
\tau -\frac{1}{2}\right) ^{2},\qquad \frac{\sqrt{2d}}{\mu }<\tau <1-\frac{%
\sqrt{2d}}{\mu }  \label{g5.7}
\end{equation}
where $R\left( \tau \right) $ is radius of the tube segment ${\cal T}_{\left[
P_{0}P_{1}\right] }$ as a function of the parameter $\tau $ along the axis
of this segment ($\tau =0$ at the point $P_{0}$, and $\tau =1$ at the point $%
P_{1}$). $\mu $ is the length of the segment ${\cal T}_{\left[ P_{0}P_{1}%
\right] }$.

Vectors ${\bf P}_{i}{\bf P}_{i+1}$ and ${\bf P}_{i+1}{\bf P}_{i+2}$ are
parallel in ${\cal G}_{{\rm D}},$ But they are not parallel in ${\cal G}_{%
{\rm M}}$. The angle $\vartheta _{{\rm D}}$ between ${\bf P}_{i}{\bf P}%
_{i+1} $ and ${\bf P}_{i+1}{\bf P}_{i+2}$ is equal to $0$, because 
\begin{equation}
\cosh \vartheta _{{\rm D}}=\frac{\left( {\bf P}_{i}{\bf P}_{i+1}.{\bf P}%
_{i+1}{\bf P}_{i+2}\right) }{\mu ^{2}}=1  \label{g5.8}
\end{equation}
In ${\cal G}_{{\rm M}}$ the angle $\vartheta _{{\rm M}}$ between ${\bf P}_{i}%
{\bf P}_{i+1}$ and ${\bf P}_{i+1}{\bf P}_{i+2}$ is determined by the
relation 
\begin{equation}
\cosh \vartheta _{{\rm M}}=\frac{\left( {\bf P}_{i}{\bf P}_{i+1}.{\bf P}%
_{i+1}{\bf P}_{i+2}\right) _{{\rm M}}}{\left| {\bf P}_{i}{\bf P}%
_{i+1}\right| _{{\rm M}}\cdot \left| {\bf P}_{i+1}{\bf P}_{i+2}\right| _{%
{\rm M}}}=\frac{\left( {\bf P}_{i}{\bf P}_{i+1}.{\bf P}_{i+1}{\bf P}%
_{i+2}\right) +d}{\mu ^{2}-2d}  \label{g5.9}
\end{equation}
where index ''M'' means that the corresponding quantity is calculated in $%
{\cal G}_{{\rm M}}$. Taking into account (\ref{g5.8}) and supposing that $%
\sqrt{d}\ll \mu $, one obtains from (\ref{g5.9}) 
\begin{equation}
\cosh \vartheta _{{\rm M}}=1+\frac{3d}{\mu ^{2}},\qquad \vartheta _{{\rm M}%
}^{2}=\frac{6d}{\mu ^{2}}  \label{g5.10}
\end{equation}
The angle $\vartheta _{{\rm M}}$ describes intensity of stochasticity of the
particle motion. The diffusion displacement $\lambda $ of a particle
determined this stochasticity is described by the quantity 
\[
\lambda =\mu \left\langle \vartheta _{{\rm M}}^{2}\right\rangle \approx 
\frac{3\hbar }{b\mu c}=\frac{3\hbar }{mc} 
\]
This is rather close to the particle Compton wave length.

One can see from (\ref{g5.10}), that this stochasticity is large for the
particle of small mass $m=b\mu $. It is rather unexpected, because, dealing
with general relativity, one thinks that influence of space-time geometry on
the particle motion does not depend on its mass. This dependence (\ref{g5.10}%
) on the particle mass is a corollary of geometrization of the mass in the
nondegenerate T-geometry. Indeed, the geometrical mass $\mu $ of the
particle can be determined from the shape of the world tube (\ref{g5.1}).
The geometrical mass $\mu $ is the distance between the adjacent points $%
P_{i}$ and $P_{i+1}$, where the tube radius vanishes. In ${\cal G}_{{\rm M}}$
this radius vanishes everywhere, and the mass cannot be determined from the
shape of the world tube (line).

Geometrization of the particle mass is very important phenomenon, which is
essential for effective description of physical phenomena of microcosm.

\section{Particle dynamics in the nondegenerate space-time geometry}

In the Minkowski space-time geometry ${\cal G}_{{\rm M}}$ the particle
motion is deterministic, and one can describe a single particle, writing
dynamic equations for its world line. In ${\cal G}_{{\rm D}}$ it is
impossible due to the world line stochasticity. In ${\cal G}_{{\rm D}}$ one
uses statistical method of the particle motion description, when one
describes motion of many identical particles. This method is described in
details in papers \cite{R99,R002}. We consider here only characteristic
features of this method, which are essential for understanding of geometric
origin of nonrelativistic quantum phenomena.

Let us consider in ${\cal G}_{{\rm M}}$ a deterministic dynamic system $%
{\cal S}_{{\rm d}}$, consisting of deterministic particle. The dynamic
system ${\cal S}_{{\rm d}}$ is described by the Lagrangian function $L\left(
t,{\bf x},{\bf \dot{x}}\right) $, where ${\bf x=}\left\{
x^{1},x^{2},x^{3}\right\} $ are coordinates of the particle in some inertial
coordinate system, and ${\bf \dot{x}}$ is its velocity. By definition a pure
statistical ensemble ${\cal E}_{{\rm p}}\left[ {\cal S}_{{\rm d}}\right] $
of dynamic systems ${\cal S}_{{\rm d}}$ is such an ensemble, whose
distribution function $F_{{\rm p}}\left( t,{\bf x},{\bf p}\right) $ may be
represented in the form 
\begin{equation}
{\cal E}_{{\rm p}}\left[ {\cal S}_{{\rm d}}\right] :\qquad F_{{\rm p}}\left(
t,{\bf x},{\bf p}\right) =\rho \left( t,{\bf x}\right) \delta \left( {\bf p}-%
{\bf P}\left( t,{\bf x}\right) \right)  \label{g6.1}
\end{equation}
where $\rho \left( t,{\bf x}\right) $ and ${\bf P}\left( t,{\bf x}\right)
=\left\{ P_{\alpha }\left( t,{\bf x}\right) \right\} $, $\alpha =1,2,3$ are
functions of only time $t$ and ${\bf x}$. In other words, the pure ensemble $%
{\cal E}_{{\rm p}}\left[ {\cal S}_{{\rm d}}\right] $ is a dynamic system,
considered in the configuration space of coordinates ${\bf x}$. It is a
fluidlike continuous dynamic system, which can be described by the action 
\cite{R002} 
\begin{equation}
{\cal E}_{{\rm p}}\left[ {\cal S}_{{\rm d}}\right] :\quad {\cal A}[j,\varphi
,{\bf \xi }]=\int \{L\left( x^{0},{\bf x,j/}j^{0}\right)
j^{0}-b_{0}j^{i}[\partial _{i}\varphi +g^{\alpha }({\bf \xi })\partial
_{i}\xi _{\alpha }]\}{\rm d}^{4}x,  \label{g6.2}
\end{equation}
where $j^{i}=\left\{ j^{0},{\bf j}\right\} ,\varphi ,{\bf \xi }$ are
dependent variables, which are considered to be functions of $x=\left\{
x^{0},{\bf x}\right\} =\left\{ t,{\bf x}\right\} $. $L\left( x^{0},{\bf x,j/}%
j^{0}\right) =L\left( x^{0},{\bf x,}d{\bf x/}dx^{0}\right) {\cal \ }$is the
Lagrangian function of ${\cal S}_{{\rm d}}$. $b_{0}$ is an arbitrary
constant and $g^{\alpha }({\bf \xi }),\;\;\alpha =1,2,3$ are arbitrary
functions of the argument ${\bf \xi }=\left\{ \xi _{1},\xi _{2},\xi
_{3}\right\} $. These functions describe initial state of the statistical
ensemble ${\cal E}_{{\rm p}}\left[ {\cal S}_{{\rm d}}\right] $. The
4-current $j^{i}$, describes the fluid flow. The action (\ref{g6.2}) as well
as dynamic equations, generated by this action, contain derivatives $%
j^{k}\partial _{k}$ only in the direction of the vector $j^{i}$. It means
that the system of dynamic equations, which are partial differential
equations, form essentially a system of ordinary differential equations,
describing a single dynamic system ${\cal S}_{{\rm d}}$. Thus, Lagrangian
function $L$ describes both dynamic systems ${\cal S}_{{\rm d}}$ and ${\cal E%
}_{{\rm p}}\left[ {\cal S}_{{\rm d}}\right] $.

If the dynamic system ${\cal S}_{{\rm d}}$ is subjected influence of some
stochastic agent, it turns to stochastic system ${\cal S}_{{\rm st}}$, and
parameters of the statistical ensemble ${\cal E}_{{\rm p}}\left[ {\cal S}_{%
{\rm st}}\right] $ stops to be constant. They become to be functions of the
ensemble state $j^{i}$ and of derivatives $\partial _{k}j^{i}$. The action
for ${\cal E}_{{\rm p}}\left[ {\cal S}_{{\rm st}}\right] $ stops to depend
on only $j^{k}\partial _{k}$. In this case dynamic equations contain
derivatives in all directions, and the system of dynamic equations cannot be
reduced to a system of ordinary equations. Physically it means that there is
no dynamic equation for the single system ${\cal S}_{{\rm st}}$, although
they exist for the statistical ensemble ${\cal E}_{{\rm p}}\left[ {\cal S}_{%
{\rm st}}\right] $.

Thus, if we want to describe a deterministic dynamic system and a stochastic
system as different partial cases of a dynamic system and to describe their
dynamics by similar method, we should describe dynamics of a pure
statistical ensemble, but not dynamics of a single dynamic system. In this
sense the concept of dynamics of the pure statistical ensemble is more
general and fundamental, than concept of the single system dymanics. Such an
idea is not new \cite{V66}.

If the dynamic system ${\cal S}_{{\rm d}}$ is a free particle, its
Lagrangian function has the form 
\begin{equation}
L\left( {\bf \dot{x}}\right) =-mc^{2}\sqrt{1-\frac{{\bf \dot{x}}^{2}}{c^{2}}}%
,\qquad {\bf \dot{x}\equiv }\frac{d{\bf x}}{dt}  \label{g6.4}
\end{equation}
The Lagrangian function depends on the only parameter $m$. In the case of
space-time geometry ${\cal G}_{{\rm D}}$ the mass $m$ depends on the
ensemble state $\rho =j^{0}.$ The mass $m$ is modified as follows \cite{R91} 
\begin{equation}
m^{2}\rightarrow m_{{\rm q}}^{2}=m^{2}+\frac{\hbar ^{2}}{4c^{2}}\left( {\bf %
\nabla }\ln \rho \right) ^{2},\qquad \rho =j^{0}  \label{g6.5}
\end{equation}
Let us substitute (\ref{g6.4}), (\ref{g6.5}) in the action (\ref{g6.2}). One
obtains in the nonrelativistic approximation

\begin{equation}
{\cal A}[j,\varphi ,{\bf \xi }]=\int \{-\rho mc^{2}+\frac{m{\bf j}^{2}}{%
2\rho }-\frac{\hbar ^{2}\rho }{8m}\left( {\bf \nabla }\ln \rho \right)
^{2}-b_{0}j^{i}[\partial _{i}\varphi +g^{\alpha }({\bf \xi })\partial
_{i}\xi _{\alpha }]\}{\rm d}^{4}x,  \label{g6.6}
\end{equation}
Any ideal fluid can be described in terms of wave function \cite{R99}.
Describing the action (\ref{g6.6}) in terms of wave function and considering
the special case, when the fluid flow is irrotational and the wave function
has only one component, one obtains instead of (\ref{g6.6}) 
\begin{eqnarray}
{\cal A}[\psi ,\psi ^{\ast }] &=&\int \left\{ \frac{ib_{0}}{2}(\psi ^{\ast
}\partial _{0}\psi -\partial _{0}\psi ^{\ast }\cdot \psi )-\frac{b_{0}^{2}}{%
2m}{\bf \nabla }\psi ^{\ast }\cdot {\bf \nabla }\psi -mc^{2}\psi ^{\ast
}\psi \right.  \nonumber \\
&&\left. +\frac{b_{0}^{2}}{8m\psi ^{\ast }\psi }(\nabla \left( \psi ^{\ast
}\psi \right) )^{2}-\frac{\hbar ^{2}}{8m\psi ^{\ast }\psi }(\nabla \left(
\psi ^{\ast }\psi \right) )^{2}\right\} {\rm d}^{4}x,  \label{g6.7}
\end{eqnarray}
The last term in the action (\ref{g6.7}) describes influence of geometrical
stochasticity. This term contains the quantum constant $\hbar $. If one sets 
$\hbar =0$ in (\ref{g6.7}), the action becomes to describe statistical
ensemble of free deterministic particles. The action (\ref{g6.7}) generates
nonlinear dynamic equation for the wave function $\psi $. The dynamic
equation becomes to be a linear differential equation, provided one sets $%
b_{0}=\hbar $, because in this case two last terms in (\ref{g6.7})
compensate each other. Note that the constant $b_{0}$ is an arbitrary
integration constant, which can take any value, in particular $b_{0}=\hbar $%
. After this substitution the action (\ref{g6.7}) takes the form 
\begin{equation}
{\cal A}[\psi ,\psi ^{\ast }]=\int \left\{ \frac{i\hbar }{2}(\psi ^{\ast
}\partial _{0}\psi -\partial _{0}\psi ^{\ast }\cdot \psi )-\frac{\hbar ^{2}}{%
2m}{\bf \nabla }\psi ^{\ast }\cdot {\bf \nabla }\psi -mc^{2}\psi ^{\ast
}\psi \right\} {\rm d}^{4}x  \label{g6.8}
\end{equation}
Now all terms in the action becomes to be quantum, because they contain the
quantum constant. One cannot obtain the action for the statistical ensemble
of deterministic (classical) particles, setting $\hbar =0$, as it is
possible to make in the action (\ref{g6.7}). One cannot suppress the
geometrical (quantum) stochasticity in (\ref{g6.8}), setting $\hbar =0.$ In
the action (\ref{g6.8}) quantum properties (the constant $\hbar $) are
attributed to those terms, which are purely dynamic in (\ref{g6.7}), and one
cannot indicate the term, responsible for quantum effects. This is the price
which is paid for linearity of the dynamic equation.

Possibility of compensation of two last terms (dynamic and quantum) in the
action (\ref{g6.7}) is connected with the nonrelativistic character of the
Hamiltonian function for a classical particle, which has the form $H_{{\rm nr%
}}=mc^{2}+{\bf p}^{2}/\left( 2m\right) .$ For instance, if the Hamiltonian
function $H_{{\rm nr}}$ is replaced by the relativistic one $H_{{\rm r}}=%
\sqrt{m^{2}c^{4}+{\bf p}^{2}c^{2}}$ identification $b_{0}=\hbar $ does not
lead to the linear dynamic equation for the wave function.

From geometric viewpoint, the linearity of the dynamic equation for the wave
function is valid only for nonrelativistic case. From this viewpoint it
seems to be doubtful, that the linearity should be used as a principle of
the relativistic quantum theory, although from practical viewpoint the
linearity is a very useful property of dynamic equation.

Thus, removing unfounded constraints on geometry and using space-time
geometry ${\cal G}_{{\rm D}}$, one can freely explain nonrelativistic
quantum effects. There is no necessity to invent and to use quantum
principles. Quantum effects appear to be corollary of quantum (or geometric)
stochasticity, generated by nondegenerate character of the space-time
geometry.

Properties of the world function (\ref{g5.6}) at the coinciding points ($%
0<\Sigma _{{\rm M}}<\sigma _{0}$) are of no importance for stochastic
behavior of particles, because it depends only on integral distortion $d$.
From this viewpoint application of methods of differential geometry to
investigation of the world function properties is useless, because
differential geometry studies properties of the world function $\Sigma
\left( x,x^{\prime }\right) $ in the limit, when $x\rightarrow x^{\prime }$.
But methods of differential geometry are useful for investigation of
antisymmetric component $A\left( x,x^{\prime }\right) $ of the nonsymmetric
world function. We shall see that the antisymmetric component $A\left(
x,x^{\prime }\right) $ generates fields in the space-time and generates as a
rule additional nondegeneracy of geometry. In other words, $A\left(
x,x^{\prime }\right) $ generates not only interaction between particles, but
also their additional stochasticity. In this sense the fields, generated by $%
A\left( x,x^{\prime }\right) $, are quantum fields.

\section{ Asymmetric T-geometry on manifold}

We have considered T-geometry in the coordinate-free form. But to discover a
connection between the T-geometry and usual differential geometry, one needs
to introduce coordinates and to consider the T-geometry on a manifold. It is
important also from the viewpoint of the asymmetric T-geometry application
as a possible space-time geometry. The asymmetric T-geometry on the manifold
may be considered to be a conventional symmetric geometry (for instance,
Riemannian) with additional force fields $a_{i}\left( x\right) $, $%
a_{ikl}\left( x\right) $, generated on the manifold by the antisymmetric
component $A$ of the world function. Testing experimentally existence of
these force fields, one can conclude whether the antisymmetric component $A$
exists and how large it is.

Let it be possible to attribute $n+1$ real numbers $x=\left\{ x^{i}\right\}
, $ $\;i=0,1,...n$ to any point $P$ in such a way, that there be one-to-one
correspondence between the point $P$ and the set $x$ of $n+1$ coordinates $%
\left\{ x^{i}\right\} ,$ $\;i=0,1,...n$. All points $x$ form a set ${\cal M}%
_{n+1}$. Then the world function $\Sigma \left( P,P^{\prime }\right) $ is a
function $\Sigma \left( x,x^{\prime }\right) $%
\begin{equation}
\Sigma :\;\;\;{\cal M}_{n+1}\times {\cal M}_{n+1}\rightarrow {\Bbb R},\qquad
\Sigma \left( x,x\right) =0,\qquad \forall x\in {\cal M}_{n+1}  \label{c3.1}
\end{equation}
of coordinates $x,x^{\prime }\in {\cal M}_{n+1}\subset {\Bbb R}^{n+1}$ of
points $P,P^{\prime }\in \Omega $. Two-point quantities ($\Sigma $-function
and their derivatives) are designed as a rule by capital characters.
One-point quantities are designed by small characters.

Let the function $\Sigma \left( x,x^{\prime }\right) $ be multiply
differentiable. Then the set ${\cal M}_{n+1}\subset {\Bbb R}^{n+1}$ may be
called the $\left( n+1\right) $-dimensional manifold. One can differentiate $%
\Sigma \left( x,x^{\prime }\right) $ with respect to $x^{i}$ and with
respect to $x^{\prime i}$, forming two-point tensors. For instance, 
\begin{eqnarray*}
\Sigma _{,k}\left( x,x^{\prime }\right) &\equiv &\frac{\partial }{\partial
x^{k}}\Sigma \left( x,x^{\prime }\right) ,\qquad \Sigma _{,k^{\prime
}}\left( x,x^{\prime }\right) \equiv \frac{\partial }{\partial x^{\prime k}}%
\Sigma \left( x,x^{\prime }\right) \\
\Sigma _{,kl^{\prime }}\left( x,x^{\prime }\right) &\equiv &\Sigma
_{,l^{\prime }k}\left( x,x^{\prime }\right) \equiv \frac{\partial ^{2}}{%
\partial x^{k}\partial x^{\prime l}}\Sigma \left( x,x^{\prime }\right) ,
\end{eqnarray*}
are two-point tensors. Here indices after comma mean differentiation with
respect to $x^{k}$, if the index $k$ has not a prime, and differentiation
with respect to $x^{\prime k}$, if the index $k$ has a prime. The first
argument of the two-point quantity is denoted by unprimed variable, whereas
the second one is denoted by primed one. Primed indices relate to the second
argument of the two-point quantity, whereas the unprimed ones relate to the
first argument.

$\Sigma _{k}\equiv \Sigma _{,k}=\Sigma _{,k}\left( x,x^{\prime }\right) $ is
a vector at the point $x$ and a scalar at the point $x^{\prime }.$ Vice
versa $\Sigma _{k^{\prime }}\equiv \Sigma _{,k^{\prime }}=\Sigma
_{,k^{\prime }}\left( x,x^{\prime }\right) $ is a vector at the point $%
x^{\prime }$ and a scalar at the point $x$. The quantity $\Sigma
_{,kl^{\prime }}=\Sigma _{,kl^{\prime }}\left( x,x^{\prime }\right) $ is a
vector at the point $x$ and a vector at the point $x^{\prime }$. Other
derivatives are not tensors. For instance, $\Sigma _{,kl}\left( x,x^{\prime
}\right) \equiv \Sigma _{,lk}\left( x,x^{\prime }\right) \equiv \frac{%
\partial ^{2}}{\partial x^{k}\partial x^{l}}\Sigma \left( x,x^{\prime
}\right) $ is a scalar at the point $x^{\prime }$, but it is not a tensor at
the point $x.$

To construct tensors of higher rank by means of differentiation, let us
introduce covariant derivatives. Let $\Sigma _{,kl^{\prime }}\equiv \Sigma
_{kl^{\prime }}\equiv \Sigma _{l^{\prime }k}$ and $\det ||\Sigma
_{kl^{\prime }}||\neq 0$. The quantity $\Sigma _{kl^{\prime }}$ will be
referred to as covariant fundamental metric tensor. One can introduce also
contravariant fundamental metric tensor $\Sigma ^{ik^{\prime }}\equiv \Sigma
^{k^{\prime }i}$, defining it by the relation 
\begin{equation}
\Sigma ^{ik^{\prime }}\Sigma _{lk^{\prime }}=\delta _{l}^{i}\qquad \Sigma
^{i^{\prime }k}\Sigma _{l^{\prime }k}=\delta _{l^{\prime }}^{i^{\prime }}
\label{c3.4}
\end{equation}
Let us note that the quantity 
\begin{equation}
\tilde{\Gamma}_{kl}^{i}\left( x,x^{\prime }\right) \equiv \Sigma
^{is^{\prime }}\Sigma _{,kls^{\prime }},\qquad \Sigma _{,kls^{\prime
}}\equiv \frac{\partial ^{3}\Sigma }{\partial x^{k}\partial x^{l}\partial
x^{\prime s}}  \label{c3.6}
\end{equation}
is a scalar at the point $x^{\prime }$ and a Christoffel symbol at the point 
$x$. Vice versa, the quantity 
\begin{equation}
\tilde{\Gamma}_{k^{\prime }l^{\prime }}^{i^{\prime }}\left( x,x^{\prime
}\right) \equiv \Sigma ^{si^{\prime }}\Sigma _{,k^{\prime }l^{\prime
}s},\qquad \Sigma _{,k^{\prime }l^{\prime }s}\equiv \frac{\partial
^{3}\Sigma }{\partial x^{\prime k}\partial x^{\prime l}\partial x^{s}}
\label{c3.7}
\end{equation}
is a scalar at the point $x$ and a Christoffel symbol at the point $%
x^{\prime }$.

In the same way one can introduce two other Christoffel symbols on the basis
of the function $G$%
\begin{eqnarray}
\Gamma _{kl}^{i}\left( x,x^{\prime }\right) &\equiv &G^{is^{\prime
}}G_{,kls^{\prime }},\qquad G_{,kls^{\prime }}\equiv \frac{\partial ^{3}G}{%
\partial x^{k}\partial x^{l}\partial x^{\prime s}},\qquad G^{is^{\prime
}}G_{,ks^{\prime }}=\delta _{k}^{i}  \label{c3.7a} \\
\Gamma _{k^{\prime }l^{\prime }}^{i^{\prime }}\left( x,x^{\prime }\right)
&\equiv &G^{si^{\prime }}G_{,k^{\prime }l^{\prime }s},\qquad G_{,k^{\prime
}l^{\prime }s}\equiv \frac{\partial ^{3}G}{\partial x^{\prime k}\partial
x^{\prime l}\partial x^{s}}  \label{c3.7b}
\end{eqnarray}

Using Christoffel symbols (\ref{c3.6}) - (\ref{c3.7b}), one can introduce
two covariant derivatives $\tilde{\nabla}_{i}^{x^{\prime }}$, $\nabla
_{i}^{x^{\prime }}$ with respect to $x^{i}$ and two covariant derivatives $%
\tilde{\nabla}_{i^{\prime }}^{x}$, $\nabla _{i^{\prime }}^{x}$ with respect
to $x^{\prime i}.$ For instance, the quantities 
\begin{eqnarray}
\Sigma _{ik} &\equiv &\tilde{\nabla}_{k}^{x^{\prime }}\tilde{\nabla}%
_{i}^{x^{\prime }}\Sigma \equiv \Sigma _{,i||k}=\Sigma _{,ik}-\tilde{\Gamma}%
_{ik}^{s}\left( x,x^{\prime }\right) \Sigma _{,s}=\Sigma _{,ik}-\Sigma
^{ls^{\prime }}\Sigma _{,iks^{\prime }}\Sigma _{,l}  \label{c3.8} \\
G_{ik} &\equiv &\nabla _{k}^{x^{\prime }}\nabla _{i}^{x^{\prime }}G\equiv
G_{,i|k}=G_{,ik}-\Gamma _{ik}^{s}\left( x,x^{\prime }\right)
G_{,s}=G_{,ik}-G^{ls^{\prime }}G_{,iks^{\prime }}G_{,l}  \label{c3.8a}
\end{eqnarray}
are scalars at the point $x^{\prime }$ and second rank tensors at the point $%
x$. Here symbol ''$||$'' before the index denotes covariant derivative with
the Christoffel symbol $\tilde{\Gamma}_{ik}^{s}$, and the symbol ''$|$''
before the index denotes covariant derivative with the Christoffel symbol $%
\Gamma _{ik}^{s}$. In the same way one obtains 
\begin{eqnarray}
\Sigma _{i^{\prime }k^{\prime }} &\equiv &\tilde{\nabla}_{k^{\prime }}^{x}%
\tilde{\nabla}_{i^{\prime }}^{x}\Sigma \equiv \Sigma _{,i^{\prime
}||k^{\prime }}=\Sigma _{,i^{\prime }k^{\prime }}-\tilde{\Gamma}_{i^{\prime
}k^{\prime }}^{s^{\prime }}\Sigma _{,s^{\prime }}=\Sigma _{,i^{\prime
}k^{\prime }}-\Sigma ^{l^{\prime }s}\Sigma _{,i^{\prime }k^{\prime }s}\Sigma
_{,l^{\prime }}  \label{c3.9} \\
G_{i^{\prime }k^{\prime }} &\equiv &\nabla _{k^{\prime }}^{x}\nabla
_{i^{\prime }}^{x}G\equiv G_{,i^{\prime }|k^{\prime }}=G_{,i^{\prime
}k^{\prime }}-\Gamma _{i^{\prime }k^{\prime }}^{s^{\prime }}G_{,s^{\prime
}}=G_{,i^{\prime }k^{\prime }}-G^{l^{\prime }s}G_{,i^{\prime }k^{\prime
}s}G_{,l^{\prime }}  \label{c3.9a}
\end{eqnarray}

Covariant derivatives $\tilde{\nabla}_{k}^{x^{\prime }}$, $\tilde{\nabla}%
_{i}^{x^{\prime }}$ with respect to $x$ commute, as well as $\nabla
_{k}^{x^{\prime }}$, $\nabla _{i}^{x^{\prime }}$, i.e. 
\begin{equation}
\left( \tilde{\nabla}_{k}^{x^{\prime }}\tilde{\nabla}_{i}^{x^{\prime }}-%
\tilde{\nabla}_{i}^{x^{\prime }}\tilde{\nabla}_{k}^{x^{\prime }}\right)
T_{ml^{\prime }}^{sp^{\prime }}\equiv 0,\qquad \left( \nabla _{k}^{x^{\prime
}}\nabla _{i}^{x^{\prime }}-\nabla _{i}^{x^{\prime }}\nabla _{k}^{x^{\prime
}}\right) T_{ml^{\prime }}^{sp^{\prime }}\equiv 0  \label{c3.10}
\end{equation}
where $T_{ml^{\prime }}^{sp^{\prime }}$ is an arbitrary tensor at points $x$
and $x^{\prime }$. Unprimed indices are associated with the point $x$, and
primed ones with the point $x^{\prime }$. The covariant derivatives commute,
because the Riemann-Christoffel curvature tensors $\tilde{R}_{i.kl}^{.j}$, $%
R_{i.kl}^{.j}$ constructed respectively of Christoffel symbols $\tilde{\Gamma%
}_{il}^{s}$ and $\Gamma _{il}^{s}$ vanish identically 
\begin{eqnarray}
\tilde{R}_{i.lm}^{.s} &\equiv &\tilde{\Gamma}_{il,m}^{s}-\tilde{\Gamma}%
_{im,l}^{s}+\tilde{\Gamma}_{il}^{j}\tilde{\Gamma}_{jm}^{s}-\tilde{\Gamma}%
_{im}^{j}\tilde{\Gamma}_{jl}^{s}\equiv 0  \label{c3.11} \\
R_{i.lm}^{.s} &\equiv &\Gamma _{il,m}^{s}-\Gamma _{im,l}^{s}+\Gamma
_{il}^{j}\Gamma _{jm}^{s}-\Gamma _{im}^{j}\Gamma _{jl}^{s}\equiv 0
\label{c3.11a}
\end{eqnarray}
One can test the identity (\ref{c3.11}), substituting (\ref{c3.6}) into (\ref
{c3.11}).

Covariant derivatives $\tilde{\nabla}_{k^{\prime }}^{x}$, $\tilde{\nabla}%
_{i^{\prime }}^{x}$ with respect to $x^{\prime }$ commute as well as $\nabla
_{k^{\prime }}^{x}$, $\nabla _{i^{\prime }}^{x}$. Commutativity of covariant
derivatives $\tilde{\nabla}_{i}^{x^{\prime }}$, $\tilde{\nabla}%
_{k}^{x^{\prime }}$ with respect to $x$ for all values of $x^{\prime }$
means that the covariant derivative $\tilde{\nabla}_{i}^{x^{\prime }}$, $%
\tilde{\nabla}_{k}^{x^{\prime }}$ are covariant derivatives in some flat
spaces $\tilde{E}_{x^{\prime }}$. The same is valid for covariant
derivatives $\nabla _{i}^{x^{\prime }}$, $\nabla _{k}^{x^{\prime }}$ which
are covariant derivatives in the flat spaces $E_{x^{\prime }}$. The spaces $%
\tilde{E}_{x^{\prime }},$ $E_{x^{\prime }}$ are associated with the $\Sigma $%
-spaces $V=\left\{ \Sigma ,{\cal M}_{n+1}\right\} $ and $V_{{\rm s}}=\left\{
G,{\cal M}_{n+1}\right\} $ respectively, given on ${\cal M}_{n+1}$ by means
of the world function $\Sigma $ and its symmetric part $G$. Any of two-point
invariant quantities $\Sigma $ and $G$ with nonvanishing determinants $\det
||\Sigma _{ik^{\prime }}||\neq 0,$ and $\det ||G_{ik^{\prime }}||\neq 0$
realize two sets of mappings. For instance, the quantity $\Sigma $ generates
mappings $V=\left\{ \Sigma ,{\cal M}_{n+1}\right\} \rightarrow \tilde{E}%
_{x^{\prime }}$, $\;V\rightarrow \tilde{E}_{x}$ The two-point quantity $G$
generates also two sets of mappings $V_{{\rm s}}=\left\{ G,{\cal M}%
_{n+1}\right\} \rightarrow E_{x^{\prime }}$, $\;V_{{\rm s}}\rightarrow E_{x}$%
. Mappings of any set are labelled by points $x$ or $x^{\prime }$ of the
manifold ${\cal M}_{n+1}$. In the case of $G$ both sets of mappings $%
V\rightarrow E_{x^{\prime }}$ and $\;V\rightarrow E_{x}$ coincide, but in
the case of $\Sigma $ the sets $V\rightarrow \tilde{E}_{x^{\prime }}$ and $%
\;V\rightarrow \tilde{E}_{x}$ are different, in general.

It is easy to see that 
\[
\tilde{\nabla}_{s}^{x^{\prime }}\Sigma _{ik^{\prime }}=\Sigma _{ik^{\prime
}||s}=\Sigma _{,ik^{\prime }s}-\tilde{\Gamma}_{is}^{p}\Sigma _{pk^{\prime
}}=\Sigma _{,ik^{\prime }s}-\Sigma ^{pl^{\prime }}\Sigma _{isl^{\prime
}}\Sigma _{pk^{\prime }}\equiv 0 
\]

The covariant derivatives have the following properties 
\begin{equation}
\tilde{\nabla}_{k}^{x^{\prime }}t_{l^{\prime }}^{i^{\prime }}\left(
x^{\prime }\right) =0,\qquad \nabla _{k}^{x^{\prime }}t_{l^{\prime
}}^{i^{\prime }}\left( x^{\prime }\right) =0,\qquad \tilde{\nabla}%
_{k^{\prime }}^{x}t_{l}^{i}\left( x\right) =0\qquad \nabla _{k^{\prime
}}^{x}t_{l}^{i}\left( x\right) =0  \label{c3.12}
\end{equation}
\begin{equation}
\tilde{\nabla}_{s}^{x^{\prime }}\Sigma _{ik^{\prime }}=\Sigma _{ik^{\prime
}||s}=0\qquad \Sigma _{ik^{\prime }||s^{\prime }}=0,\qquad G_{ik^{\prime
}|s}=0\qquad G_{ik^{\prime }|s^{\prime }}=0,  \label{c3.13}
\end{equation}
where $t_{l^{\prime }}^{i^{\prime }}\left( x^{\prime }\right) $ is an
arbitrary tensor at the point $x^{\prime }$, and $t_{l}^{i}\left( x\right) $
is an arbitrary tensor at the point $x$.

The considered mappings onto Euclidean spaces can serve as a powerful tool
for description of the $\Sigma $-space properties. Let us note in this
connection, that the Riemannian space may be considered to be a set of
infinitesimal pieces of Euclidean spaces glued in some way between
themselves. The way of gluing determines the character of the Riemannian
space in the sense, that different ways of gluing generate different
Riemannian spaces. The way of gluing is determined by the difference of the
metric tensor at the point $x$ and at the narrow point $x+dx$, where it has
the forms $g_{ik}\left( x\right) $ and $g_{ik}\left( x+dx\right) $
respectively. The metric tensor depends on a point and on a choice of the
coordinate system. It is rather difficult one to separate dependence on the
way of gluing from that on the choice of the coordinate system. Nevertheless
the procedure of separation has been well developed. It leads to the
curvature tensor, which is an indicator of the way of gluing.

In the case of the $\Sigma $-space one considers a set of {\it finite
Euclidean spaces} \ $E_{x^{\prime }}\;$(ins\-tead of its infinitesimal
pieces) and a set of mappings $\Sigma \rightarrow E_{x^{\prime }}$. Here the
''way of gluing'' is determined by the dependence of mapping on the
parameter $x^{\prime }$. It does not depend on a choice of the coordinate
system. This circumstance simplifies investigation. Differentiating the
mappings with respect to parameters $x^{\prime }$, one derives local
characteristics of the ''way of gluing'', which are modifications of the
curvature tensor. For instance, considering commutators of derivatives $%
\tilde{\nabla}_{i}^{x^{\prime }}$ and $\tilde{\nabla}_{i^{\prime }}^{x}$,
one can introduce two-point curvature tensor for the $\Sigma $-space, as it
have been made for the Riemannian space \cite{R62,R64}. We shall see this
further.

Let $\tilde{G}_{(x^{\prime })ik}$ be the metric tensor in the Euclidean
space $\tilde{E}_{x^{\prime }}$ at the point $x$. Then the Christoffel
symbol $\tilde{\Gamma}_{kl}^{i}=\Sigma ^{is^{\prime }}\Sigma _{,kls^{\prime
}}$ in the space $\tilde{E}_{x^{\prime }}$ can be written in the form

\begin{equation}
\tilde{\Gamma}_{kl}^{i}=\Sigma ^{is^{\prime }}\Sigma _{,kls^{\prime }}=\frac{%
1}{2}\tilde{G}_{(x^{\prime })}^{im}\left( \tilde{G}_{(x^{\prime })km,l}+%
\tilde{G}_{(x^{\prime })lm,k}-\tilde{G}_{(x^{\prime })kl,m}\right)
\label{c3.14}
\end{equation}
where $\tilde{G}_{(x^{\prime })}^{im}$ are contravariant components of the
metric tensor $\tilde{G}_{(x^{\prime })ik}$.

Let us consider the set of equations (\ref{c3.14}) as a system of linear
differential equations for determination of the metric tensor components $%
\tilde{G}_{(x^{\prime })ik}$, which is supposed to be symmetric. Solution of
this system has the form 
\begin{equation}
\tilde{G}_{(x^{\prime })ik}=\Sigma _{ip^{\prime }}\tilde{g}_{\left(
x^{\prime }\right) }^{p^{\prime }q^{\prime }}\Sigma _{kq^{\prime }}
\label{c3.15}
\end{equation}
where $\tilde{g}_{\left( x^{\prime }\right) }^{p^{\prime }q^{\prime }}=%
\tilde{g}_{\left( x^{\prime }\right) }^{p^{\prime }q^{\prime }}\left(
x^{\prime }\right) $ is some symmetric tensor at the point $x^{\prime }$.
This fact can be tested by a direct substitution of (\ref{c3.15}) in (\ref
{c3.14}). Taking the relation (\ref{c3.15}) at the coinciding points $%
x=x^{\prime }$ and denoting coincidence of points $x$ and $x^{\prime }$ by
means of square brackets, one obtains from (\ref{c3.15}) 
\begin{equation}
\tilde{g}_{\left( x^{\prime }\right) }^{p^{\prime }q^{\prime }}\left(
x^{\prime }\right) =\left[ \Sigma ^{lp^{\prime }}\right] _{x^{\prime }}\left[
\tilde{G}_{(x^{\prime })lm}\right] _{x^{\prime }}\left[ \Sigma ^{mq^{\prime
}}\right] _{x^{\prime }}  \label{c3.16}
\end{equation}
or 
\begin{equation}
\tilde{G}_{(x^{\prime })ik}=\Sigma _{ip^{\prime }}\left[ \Sigma ^{lp^{\prime
}}\right] _{x^{\prime }}\left[ \tilde{G}_{(x^{\prime })lm}\right]
_{x^{\prime }}\left[ \Sigma ^{mq^{\prime }}\right] _{x^{\prime }}\Sigma
_{kq^{\prime }}  \label{c3.17}
\end{equation}
The equation (\ref{c3.17}) can be written in the form 
\begin{eqnarray}
\tilde{G}_{(x^{\prime })ik}\left( x,x^{\prime }\right) &=&\tilde{P}%
_{(x^{\prime })}{}_{i.}^{.l^{\prime }}\tilde{P}_{(x^{\prime
})}{}_{k.}^{.m^{\prime }}\tilde{G}_{(x^{\prime })lm}\left( x^{\prime
},x^{\prime }\right) ,  \label{c3.18} \\
\tilde{P}_{(x^{\prime })}{}_{k.}^{.m^{\prime }} &\equiv &\tilde{P}%
_{(x^{\prime })}{}_{k.}^{.m^{\prime }}\left( x,x^{\prime }\right) \equiv
\Sigma _{kq^{\prime }}\left( x,x^{\prime }\right) \Sigma ^{mq^{\prime
}}\left( x^{\prime },x^{\prime }\right) \equiv \Sigma _{kq^{\prime }}\left[
\Sigma ^{mq^{\prime }}\right] _{x^{\prime }}  \label{c3.19}
\end{eqnarray}

The relation (\ref{c3.18}) means that the metric tensor$\ \tilde{G}%
_{(x^{\prime })ik}$ of the Euclidean space $\tilde{E}_{x^{\prime }}$ at the
point $x$ can be obtained as a result of the parallel transport of the
metric tensor from the point $x^{\prime }$ in $\tilde{E}_{x^{\prime }}$ by
means of the parallel transport tensor $\tilde{P}_{(x^{\prime
})}{}_{k.}^{.m^{\prime }}$. The parallel transport of the vector $%
b_{k^{\prime }}$ from the point $x^{\prime }$ to the point $x$ is defined by
the relation 
\[
b_{k}=\tilde{P}_{(x^{\prime })}{}_{k.}^{.m^{\prime }}b_{m^{\prime }}. 
\]
The parallel transport tensor has evident properties 
\begin{eqnarray}
\tilde{\nabla}_{i}^{x^{\prime }}\tilde{P}_{(x^{\prime })}{}_{k.}^{.m^{\prime
}} &\equiv &\tilde{P}_{(x^{\prime })}{}_{k.||i}^{.m^{\prime }}=0,
\label{c3.20} \\
\left[ \tilde{P}_{(x^{\prime })}{}_{k.}^{.m^{\prime }}\right] _{x^{\prime }}
&\equiv &\tilde{P}_{(x^{\prime })}{}_{k.}^{.m^{\prime }}\left( x^{\prime
},x^{\prime }\right) =\delta _{k^{\prime }}^{m^{\prime }}.  \label{c3.21}
\end{eqnarray}

In the same way one can obtain the parallel transport tensor $\tilde{P}%
_{(x)}{}_{k^{\prime }.}^{.m}$ in the Euclidean space $\tilde{E}_{x}$%
\begin{equation}
\tilde{P}_{(x)}{}_{k^{\prime }.}^{.m}\equiv \Sigma _{k^{\prime }q}\left[
\Sigma ^{mq^{\prime }}\right] _{x}\equiv \Sigma _{k^{\prime }q}\left(
x,x^{\prime }\right) \Sigma ^{mq^{\prime }}\left( x,x\right)  \label{c3.22}
\end{equation}
describing a parallel transport from the point $x$ to the point $x^{\prime }$
in $\tilde{E}_{x}$.

In the same way one can obtain the parallel transport tensors $P_{(x^{\prime
})}{}_{k.}^{.m^{\prime }}$ and $P_{(x)}{}_{k^{\prime }.}^{.m}$ respectively
in Euclidean spaces $E_{x^{\prime }}$ and $E_{x}$%
\begin{equation}
P_{(x^{\prime })}{}_{k.}^{.m^{\prime }}\equiv G_{kq^{\prime }}\left[
G^{mq^{\prime }}\right] _{x^{\prime }},\qquad P_{(x)}{}_{k^{\prime
}.}^{.m}\equiv G_{k^{\prime }q}\left[ G^{mq^{\prime }}\right] _{x}
\label{c3.23}
\end{equation}

Thus, the world function $\Sigma $ of the $\Sigma $-space $V=\left\{ \Sigma ,%
{\cal M}_{n+1}\right\} $ and its symmetric component $G$ determine Euclidean
spaces $\tilde{E}_{x^{\prime }}$, $\tilde{E}_{x}$, $E_{x^{\prime }}$, $E_{x}$%
, mappings of $V$ on them and the parallel transport of vectors and tensors
in these Euclidean spaces independently of that, whether or not the $\Sigma $%
-space $V=\left\{ \Sigma ,{\cal M}_{n+1}\right\} $ is degenerate in the
sense of definitions \ref{d3.12a} -- \ref{d3.14}.

\section{Derivatives of the world function at coincidence of points $x$ and $%
x^{\prime }$.}

Let the world function be represented in the form (\ref{b2.2}), (\ref{g1.2}%
). Let us expand the function $G$ and $A$ with respect to powers of $\xi
^{i}=x^{i}-x^{\prime i}$. Taking into account the symmetry relations (\ref
{g1.2}), one obtains 
\begin{eqnarray}
G\left( x,x^{\prime }\right) &=&\frac{1}{2}g_{ik}\left( x^{\prime }\right)
\xi ^{i}\xi ^{k}+\frac{1}{6}g_{ikl}\left( x^{\prime }\right) \xi ^{i}\xi
^{k}\xi ^{l}+\frac{1}{24}g_{iklm}\left( x^{\prime }\right) \xi ^{i}\xi
^{k}\xi ^{l}\xi ^{m}+...  \label{c5.3} \\
&=&\frac{1}{2}g_{ik}\left( x\right) \xi ^{i}\xi ^{k}-\frac{1}{6}%
g_{ikl}\left( x\right) \xi ^{i}\xi ^{k}\xi ^{l}+\frac{1}{24}g_{iklm}\left(
x\right) \xi ^{i}\xi ^{k}\xi ^{l}\xi ^{m}+...  \label{c5.4}
\end{eqnarray}

\begin{eqnarray}
A\left( x,x^{\prime }\right) &=&a_{i}\left( x^{\prime }\right) \xi ^{i}+%
\frac{1}{2}a_{ik}\left( x^{\prime }\right) \xi ^{i}\xi ^{k}+\frac{1}{6}%
a_{ikl}\left( x^{\prime }\right) \xi ^{i}\xi ^{k}\xi ^{l}+...  \label{c5.5}
\\
&=&a_{i}\left( x\right) \xi ^{i}-\frac{1}{2}a_{ik}\left( x\right) \xi
^{i}\xi ^{k}+\frac{1}{6}a_{ikl}\left( x\right) \xi ^{i}\xi ^{k}\xi ^{l}-...
\label{c5.6}
\end{eqnarray}
In relations (\ref{c5.3}) and (\ref{c5.5}) the functions $G\left(
x,x^{\prime }\right) $ and $A\left( x,x^{\prime }\right) $ are expanded at
the point $x^{\prime }$. In the relations (\ref{c5.4}) and (\ref{c5.6}) one
has the same expansions after transposition $x\leftrightarrow x^{\prime }$.

Differentiating relations (\ref{c5.3}) - (\ref{c5.6}) with respect to $x$
and $x^{\prime }$ and setting $x=x^{\prime }$ thereafter, one obtains
relations between the expansion coefficients and expressions for derivatives
of functions $\Sigma $, $G$, $A$ at the limit of coincidence $x=x^{\prime }$.

After calculations one obtains 
\begin{eqnarray}
g_{ikl} &=&\frac{1}{2}\left( g_{ik,l}+g_{li,k}+g_{kl,i}\right) ,\qquad
g_{ik,l}\equiv g_{ik,l}\left( x\right) \equiv \frac{\partial }{\partial x^{l}%
}g_{ik}\left( x\right)  \label{c5.7} \\
a_{ik} &=&\frac{1}{2}\left( a_{i,k}+a_{k,i}\right) ,\qquad a_{i,k}\equiv
a_{i,k}\left( x\right) \equiv \frac{\partial }{\partial x^{k}}a_{i}\left(
x\right)  \label{c5.8}
\end{eqnarray}
\[
a_{iklm}=\frac{1}{2}\left(
a_{ikl,m}+a_{klm,i}+a_{lmi,k}+a_{mik,l}-a_{ik,lm}-a_{lm,ik}\right) 
\]
Coefficients $g_{ik},$ $a_{i}$, $a_{ikl}$ are arbitrary and symmetric with
respect to transposition of indices. Using square brackets for designation
of coincidence $x=x^{\prime }$ and relations (\ref{c5.7}), (\ref{c5.8}), one
obtains 
\begin{eqnarray}
\left[ G_{,i}\left( x,x^{\prime }\right) \right] &\equiv &\left[ G_{,i}%
\right] =\left[ G_{,i^{\prime }}\right] =0,  \nonumber \\
\left[ G_{,ik}\left( x,x^{\prime }\right) \right] &\equiv &\left[ G_{,ik}%
\right] =\left[ G_{,i^{\prime }k^{\prime }}\right] =g_{ik},  \label{c5.9} \\
\left[ G_{,i^{\prime }k}\left( x,x^{\prime }\right) \right] &\equiv &\left[
G_{,i^{\prime }k}\right] =-g_{ik}  \nonumber
\end{eqnarray}

\begin{eqnarray}
\left[ G_{,ikl}\left( x,x^{\prime }\right) \right] &\equiv &\left[ G_{,ikl}%
\right] =\left[ G_{,i^{\prime }k^{\prime }l^{\prime }}\right] =\frac{1}{2}%
\left( g_{ik,l}+g_{li,k}+g_{kl,i}\right)  \nonumber \\
\left[ G_{,ikl^{\prime }}\left( x,x^{\prime }\right) \right] &\equiv &\left[
G_{,ikl^{\prime }}\right] =\frac{1}{2}\left(
g_{ik,l}-g_{li,k}-g_{kl,i}\right)  \label{c5.10} \\
\left[ G_{,ik^{\prime }l^{\prime }}\left( x,x^{\prime }\right) \right]
&\equiv &\left[ G_{,ik^{\prime }l^{\prime }}\right] =\frac{1}{2}\left(
g_{kl,i}-g_{li,k}-g_{ik,l}\right)  \nonumber
\end{eqnarray}
\begin{eqnarray}
\left[ G_{,iklm}\left( x,x^{\prime }\right) \right] &\equiv &\left[ G_{,iklm}%
\right] =\left[ G_{,i^{\prime }k^{\prime }l^{\prime }m^{\prime }}\right]
=g_{iklm}  \nonumber \\
\left[ G_{,iklm^{\prime }}\left( x,x^{\prime }\right) \right] &\equiv &\left[
G_{,iklm^{\prime }}\right] =\left[ G_{,i^{\prime }k^{\prime }l^{\prime }m}%
\right] =-g_{iklm}+g_{ikl,m}  \label{c5.10a} \\
\left[ G_{,ikl^{\prime }m^{\prime }}\left( x,x^{\prime }\right) \right]
&\equiv &\left[ G_{,ikl^{\prime }m^{\prime }}\right]
=g_{iklm}-g_{ikl,m}-g_{ikm,l}+g_{ik,ml}  \nonumber
\end{eqnarray}
\begin{equation}
\left[ A_{,i}\left( x,x^{\prime }\right) \right] \equiv \left[ A_{,i}\right]
=a_{i},\qquad \left[ A_{,i^{\prime }}\left( x,x^{\prime }\right) \right]
\equiv \left[ A_{,i^{\prime }}\right] =-a_{i}  \label{c5.11}
\end{equation}
\begin{eqnarray}
\left[ A_{,ik}\left( x,x^{\prime }\right) \right] &\equiv &\left[ A_{,ik}%
\right] =a_{ik}=\frac{1}{2}\left( a_{i,k}+a_{k,i}\right)  \nonumber \\
\left[ A_{,ik^{\prime }}\left( x,x^{\prime }\right) \right] &\equiv &\left[
A_{,ik^{\prime }}\right] =\frac{1}{2}\left( a_{i,k}-a_{k,i}\right)
\label{c5.12} \\
\left[ A_{,i^{\prime }k^{\prime }}\left( x,x^{\prime }\right) \right]
&\equiv &\left[ A_{,i^{\prime }k^{\prime }}\right] =-a_{ik}=-\frac{1}{2}%
\left( a_{i,k}+a_{k,i}\right)  \nonumber
\end{eqnarray}
\begin{eqnarray}
\left[ A_{,ikl}\left( x,x^{\prime }\right) \right] &\equiv &\left[ A_{,ikl}%
\right] =a_{ikl}  \nonumber \\
\left[ A_{,ikl^{\prime }}\left( x,x^{\prime }\right) \right] &\equiv &\left[
A_{,ikl^{\prime }}\right] =\frac{1}{2}a_{i,kl}+\frac{1}{2}a_{k,il}-a_{ikl}
\label{c5.12a} \\
\left[ A_{,ik^{\prime }l^{\prime }}\left( x,x^{\prime }\right) \right]
&\equiv &\left[ A_{,ik^{\prime }l^{\prime }}\right] =-\frac{1}{2}a_{l,ik}-%
\frac{1}{2}a_{k,il}+a_{ikl}  \nonumber
\end{eqnarray}
\begin{eqnarray}
\left[ A_{,iklm}\left( x,x^{\prime }\right) \right] &\equiv &\left[ A_{,iklm}%
\right] =a_{iklm}  \nonumber \\
\left[ A_{,iklm^{\prime }}\left( x,x^{\prime }\right) \right] &\equiv &\left[
A_{,iklm^{\prime }}\right] =-a_{iklm}+a_{ikl,m}  \nonumber \\
\left[ A_{,ikl^{\prime }m^{\prime }}\left( x,x^{\prime }\right) \right]
&\equiv &\left[ A_{,ikl^{\prime }m^{\prime }}\right]
=a_{iklm}-a_{ikl,m}-a_{ikm,l}+a_{ik,lm}  \label{c5.13a} \\
\left[ A_{,ik^{\prime }l^{\prime }m^{\prime }}\left( x,x^{\prime }\right) %
\right] &\equiv &\left[ A_{,ik^{\prime }l^{\prime }m^{\prime }}\right]
=-a_{iklm}-a_{klm,i}  \nonumber \\
\left[ A_{,i^{\prime }k^{\prime }l^{\prime }m^{\prime }}\left( x,x^{\prime
}\right) \right] &\equiv &\left[ A_{,i^{\prime }k^{\prime }l^{\prime
}m^{\prime }}\right] =-a_{iklm}  \nonumber
\end{eqnarray}

The first order coefficient $a_{k}\left( x\right) $ is a covariant vector at
the point $x$. The second order coefficients $g_{ik}\left( x\right) $ is the
second rank covariant tensors at the point $x.$ The second order coefficient 
$a_{ik}\left( x\right) $ and the third order coefficients $g_{ikl}\left(
x\right) $, $a_{ikl}\left( x\right) $ are not tensors, in general. The law
of their transformation at the coordinate transformation is more complicated.

According to (\ref{b2.2}), (\ref{c5.9}) and (\ref{c5.11}) 
\begin{equation}
\left[ \Sigma _{,i}\right] =a_{i}\left( x\right) ,\qquad \left[ \Sigma
_{,i^{\prime }}\right] =-a_{i}\left( x\right)  \label{c5.14}
\end{equation}
According to (\ref{b2.2}), (\ref{c5.9}) and (\ref{c5.12}) 
\begin{eqnarray}
\left[ \Sigma _{,ik}\right] &=&\sigma _{\left( {\rm f}\right) ik}=g_{ik}+%
\frac{1}{2}\left( a_{i,k}+a_{k,i}\right)  \nonumber \\
\left[ \Sigma _{,i^{\prime }k^{\prime }}\right] &=&\sigma _{\left( {\rm p}%
\right) ik}=g_{ik}-\frac{1}{2}\left( a_{i,k}+a_{k,i}\right)  \label{c5.15} \\
\left[ \Sigma _{ik^{\prime }}\right] &\equiv &\left[ \Sigma _{,ik^{\prime }}%
\right] =-\tilde{g}_{ik}=-g_{ik}+\frac{1}{2}\left( a_{i,k}-a_{k,i}\right) 
\nonumber
\end{eqnarray}
One obtains for the value $\left[ \Sigma ^{ik^{\prime }}\right] $ of the
quantity $\Sigma ^{ik^{\prime }}$ 
\begin{equation}
\left[ \Sigma ^{ik^{\prime }}\right] =-\tilde{g}^{ik},  \label{c5.16}
\end{equation}
where $\tilde{g}^{ik}$ is determined by the relation 
\begin{equation}
\tilde{g}^{il}\tilde{g}_{ik}=\tilde{g}^{il}\left( g_{ik}-\frac{1}{2}\left(
a_{i,k}-a_{k,i}\right) \right) =\delta _{k}^{l}  \label{c5.17}
\end{equation}
The quantity $a_{i}$ is a one-point vector, and $g_{ik}$ is a one-point
tensor. Then it follows from (\ref{c5.15}), (\ref{c5.17}), that $\tilde{g}%
_{ik}$ and $\tilde{g}^{ik}$ are also a one-point tensors, whereas $\sigma
_{\left( {\rm f}\right) ik}$ and $\sigma _{\left( {\rm p}\right) ik}$ are
not tensors, in general.

One obtains for the quantities $\left[ \Gamma _{kl}^{i}\right] ,$ $\left[
\Gamma _{k^{\prime }l^{\prime }}^{i^{\prime }}\right] $ and $\left[ \tilde{%
\Gamma}_{kl}^{i}\right] ,$ $\left[ \tilde{\Gamma}_{k^{\prime }l^{\prime
}}^{i^{\prime }}\right] $ the following expressions 
\begin{eqnarray}
\left[ \Gamma _{kl}^{i}\right] &=&\left[ \Gamma _{k^{\prime }l^{\prime
}}^{i^{\prime }}\right] =\gamma _{kl}^{i}\left( x\right) =\frac{1}{2}%
g^{si}\left( g_{ks,l}+g_{sl,k}-g_{lk,s}\right) ,\qquad g^{ik}g_{lk}=\delta
_{l}^{i}  \label{c5.18a} \\
\left[ \tilde{\Gamma}_{kl}^{i}\right] &=&\tilde{\gamma}_{\left( {\rm f}%
\right) }{}_{kl}^{i}=\tilde{g}^{is}g_{ps}\left( \gamma _{kl}^{p}+\beta
_{kl}^{p}\right) ,  \label{c5.18} \\
\left[ \tilde{\Gamma}_{k^{\prime }l^{\prime }}^{i^{\prime }}\right] &=&%
\tilde{\gamma}_{\left( {\rm p}\right) }{}_{kl}^{i}=\tilde{g}%
^{si}g_{ps}\left( \gamma _{kl}^{p}-\beta _{kl}^{p}\right)  \label{c5.18b}
\end{eqnarray}
where 
\begin{eqnarray}
\gamma _{kl}^{i}\left( x\right) &=&\frac{1}{2}g^{si}\left(
g_{ks,l}+g_{sl,k}-g_{lk,s}\right) ,\qquad g^{ik}g_{lk}=\delta _{l}^{i}
\label{c5.20} \\
\beta _{kl}^{i}\left( x\right) &=&g^{si}\left( -\frac{1}{2}\left(
a_{k,ls}+a_{l,ks}\right) +a_{kls}\right)  \label{c5.21}
\end{eqnarray}
Here $\gamma _{kl}^{i}\left( x\right) $ is the Christoffel symbol for the
symmetric case, when $A\left( x,x^{\prime }\right) \equiv 0$.

Note that the tensors $\tilde{g}^{ik},$ $\tilde{g}_{ik}$ are not symmetric
with respect to transposition indices, in general, whereas $\tilde{\gamma}%
_{\left( {\rm f}\right) }{}_{kl}^{i}=\left[ \tilde{\Gamma}_{kl}^{i}\right] $%
, $\;\tilde{\gamma}_{\left( {\rm p}\right) }{}_{kl}^{i}=\left[ \tilde{\Gamma}%
_{k^{\prime }l^{\prime }}^{i^{\prime }}\right] $, $\gamma _{kl}^{i}$ and $%
\beta _{kl}^{i}$ are symmetric with respect to transposition of indices $k$
and $l$. Besides it follows from (\ref{c5.18}), (\ref{c5.18b}), that 
\begin{equation}
\beta _{kl}^{i}\left( x\right) =g^{si}\left( -\frac{1}{2}\left(
a_{k,ls}+a_{l,ks}\right) +a_{kls}\right) =\frac{1}{2}g^{ip}\left( \tilde{g}%
_{sp}\tilde{\gamma}_{\left( {\rm f}\right) }{}_{kl}^{s}-\tilde{g}_{ps}\tilde{%
\gamma}_{\left( {\rm p}\right) }{}_{kl}^{s}\right)  \label{c5.22}
\end{equation}
In the case, when $a_{i}\equiv 0$ and tensor $\tilde{g}_{ik}$ is symmetric,
the quantity $\beta _{kl}^{i}$ is one-point tensor because difference of two
Christoffel symbols $\tilde{\gamma}_{\left( {\rm f}\right) }{}_{kl}^{s}-%
\tilde{\gamma}_{\left( {\rm p}\right) }{}_{kl}^{s}$ is a tensor.

\section{ Curvature tensors}

In the Riemannian geometry the Riemann-Christoffel curvature tensor $\tilde{r%
}_{\left( {\rm q}\right) }{}_{s.ik}^{.l}$ is defined as a commutator of
covariant derivatives $\tilde{D}_{\left( {\rm q}\right) i}$ with the
Christoffel symbol $\tilde{\gamma}_{\left( {\rm q}\right) }{}_{kl}^{i}$ 
\[
\left( \tilde{D}_{\left( {\rm q}\right) i}\tilde{D}_{\left( {\rm q}\right)
k}-\tilde{D}_{\left( {\rm q}\right) k}\tilde{D}_{\left( {\rm q}\right)
i}\right) t_{s}=\tilde{r}_{\left( {\rm q}\right) }{}_{s.ik}^{.l}t_{l} 
\]
where $\tilde{D}_{\left( {\rm q}\right) i}$ is the usual covariant
derivative in the Riemannian space with the Christoffel symbol $\tilde{\gamma%
}_{\left( {\rm q}\right) }{}_{si}^{l}$, 
\begin{equation}
\tilde{r}_{\left( {\rm q}\right) }{}_{s.ik}^{.l}=\tilde{\gamma}_{\left( {\rm %
q}\right) }{}_{si,k}^{l}-\tilde{\gamma}_{\left( {\rm q}\right)
}{}_{sk,i}^{l}+\tilde{\gamma}_{\left( {\rm q}\right) }{}_{si}^{p}\tilde{%
\gamma}_{\left( {\rm q}\right) }{}_{pk}^{l}-\tilde{\gamma}_{\left( {\rm q}%
\right) }{}_{sk}^{p}\tilde{\gamma}_{\left( {\rm q}\right) }{}_{pi}^{l}
\label{f6.2}
\end{equation}
and $t_{l}$ is an arbitrary vector at the point $x$. Index $q$ runs the
values $p$ and $f.$

In the $\Sigma $-space one can consider commutator of covariant derivatives $%
\tilde{\nabla}_{i}^{x^{\prime }}$ and $\tilde{\nabla}_{k^{\prime }}^{x}$
with respect to $x^{i}$ and $x^{\prime k}$ respectively. Calculation gives 
\begin{eqnarray}
&&\left( \tilde{\nabla}_{i}^{x^{\prime }}\tilde{\nabla}_{s^{\prime }}^{x}-%
\tilde{\nabla}_{s^{\prime }}^{x}\tilde{\nabla}_{i}^{x^{\prime }}\right)
T_{k\ldots l^{\prime }{\ldots }}^{j\ldots m^{\prime }{\ldots }}  \nonumber \\
&=&\tilde{F}_{ika^{\prime }s^{\prime }}\Sigma ^{ba^{\prime }}T_{b\ldots
l^{\prime }{\ldots }}^{j\ldots m^{\prime }{\ldots }}+\ldots -\Sigma
^{ja^{\prime }}\tilde{F}_{iba^{\prime }s^{\prime }}T_{k\ldots l^{\prime
}\ldots }^{b\ldots m^{\prime }{\ldots }}  \nonumber \\
&&-\ldots +\Sigma ^{am^{\prime }}\tilde{F}_{iab^{\prime }s^{\prime
}}T_{k\ldots l^{\prime }\ldots }^{j\ldots b^{\prime }\ldots }+\ldots -\tilde{%
F}_{ial^{\prime }s^{\prime }}\Sigma ^{ab^{\prime }}T_{k\ldots b^{\prime
}\ldots }^{j\ldots m^{\prime }\ldots }-\ldots  \label{f6.3}
\end{eqnarray}
where $T_{k\ldots l^{\prime }{\ldots }}^{j\ldots m^{\prime }{\ldots }}$ is
an arbitrary two-point tensor. $\tilde{F}$-tensor, defined by the relation 
\begin{equation}
\tilde{F}_{ilk^{\prime }j^{\prime }}\equiv \Sigma _{iq^{\prime }}\tilde{%
\Gamma}_{k^{\prime }j^{\prime }||l}^{q^{\prime }}=\Sigma _{pj^{\prime }}%
\tilde{\Gamma}_{il||k^{\prime }}^{p}=\Sigma _{,ilj^{\prime }\parallel
k^{\prime }}=\Sigma _{,ilj^{\prime }k^{\prime }}-\Sigma _{,sj^{\prime
}k^{\prime }}\Sigma ^{sm^{\prime }}\Sigma _{,ilm^{\prime }}  \label{f6.4}
\end{equation}
is a two-point analog of the one-point curvature tensor $%
r_{slik}=g_{lp}r_{s.ik}^{.p}$. To test that the quantity (\ref{f6.4}) is a
tensor, let us represent it in one of two forms 
\begin{equation}
\tilde{F}_{ilk^{\prime }j^{\prime }}\equiv \Sigma _{pj^{\prime }}\tilde{%
\Gamma}_{il||k^{\prime }}^{p}=\Sigma _{pj^{\prime }}\left( \tilde{\Gamma}%
_{il}^{p}-\tilde{\gamma}_{\left( {\rm f}\right) }{}_{il}^{p}\right)
_{||k^{\prime }}  \label{f6.5}
\end{equation}
\begin{equation}
\tilde{F}_{ilk^{\prime }j^{\prime }}\equiv \Sigma _{iq^{\prime }}\tilde{%
\Gamma}_{k^{\prime }j^{\prime }||l}^{q^{\prime }}=\Sigma _{iq^{\prime
}}\left( \tilde{\Gamma}_{k^{\prime }j^{\prime }}^{q^{\prime }}-\tilde{\gamma}%
_{\left( {\rm p}\right) }{}_{k^{\prime }j^{\prime }}^{q^{\prime }}\right)
_{||l}  \label{f6.5a}
\end{equation}

As far as the difference 
\begin{equation}
\tilde{Q}_{\left( {\rm f}\right) }{}_{il}^{p}=\tilde{\gamma}_{\left( {\rm f}%
\right) }{}_{il}^{p}-\tilde{\Gamma}_{il}^{p}\qquad \tilde{Q}_{\left( {\rm p}%
\right) }{}_{k^{\prime }j^{\prime }}^{q^{\prime }}=\tilde{\gamma}_{\left( 
{\rm p}\right) }{}_{k^{\prime }j^{\prime }}^{q^{\prime }}-\tilde{\Gamma}%
_{k^{\prime }j^{\prime }}^{q^{\prime }}  \label{f6.6}
\end{equation}
of two Christoffel symbols is a tensor, it follows from (\ref{f6.5}) and (%
\ref{f6.6}) that $\tilde{F}_{ilk^{\prime }j^{\prime }}$ is a tensor. $\tilde{%
F}$-tensor can be presented as a result of covariant differentiation of the $%
\Sigma $-function. Indeed 
\begin{eqnarray}
\tilde{Q}_{\left( {\rm f}\right) }{}_{il}^{s} &=&-\Sigma ^{sj^{\prime
}}(\Sigma _{,ilj^{\prime }}-\tilde{\gamma}_{\left( {\rm f}\right)
}{}_{il}^{m}\Sigma _{mj^{\prime }})=-\Sigma ^{sj^{\prime }}\tilde{D}_{\left( 
{\rm f}\right) l}\Sigma _{ij^{\prime }}=-\Sigma ^{sj^{\prime }}\tilde{D}%
_{\left( {\rm f}\right) l}\tilde{D}_{\left( {\rm p}\right) j^{\prime
}}\Sigma _{i}  \nonumber \\
&=&-\Sigma ^{sj^{\prime }}\tilde{D}_{\left( {\rm p}\right) j^{\prime }}%
\tilde{D}_{\left( {\rm f}\right) l}\Sigma _{i}=-\Sigma ^{sj^{\prime }}\left( 
\tilde{D}_{\left( {\rm f}\right) l}\Sigma _{i}\right) _{||j^{\prime
}}=-\left( \Sigma ^{sj^{\prime }}\tilde{D}_{\left( {\rm f}\right) l}\Sigma
_{i}\right) _{||j^{\prime }}  \label{f6.7}
\end{eqnarray}
\begin{eqnarray}
\tilde{Q}_{\left( {\rm p}\right) }{}_{k^{\prime }j^{\prime }}^{q^{\prime }}
&=&-\Sigma ^{sq^{\prime }}\left( \Sigma _{,sk^{\prime }j^{\prime }}-\tilde{%
\gamma}_{\left( {\rm p}\right) }{}_{k^{\prime }j^{\prime }}^{m^{\prime
}}\Sigma _{sm^{\prime }}\right) =-\Sigma ^{sq^{\prime }}\tilde{D}_{\left( 
{\rm p}\right) j^{\prime }}\Sigma _{sk^{\prime }}  \nonumber \\
&=&-\Sigma ^{sq^{\prime }}\tilde{D}_{\left( {\rm p}\right) j^{\prime }}%
\tilde{D}_{\left( {\rm f}\right) s}\Sigma _{k^{\prime }}=-\Sigma
^{sq^{\prime }}\tilde{D}_{\left( {\rm f}\right) s}\tilde{D}_{\left( {\rm p}%
\right) j^{\prime }}\Sigma _{k^{\prime }}=-\left( \Sigma ^{sq^{\prime }}%
\tilde{D}_{\left( {\rm p}\right) j^{\prime }}\Sigma _{k^{\prime }}\right)
_{||s}  \label{f6.7a}
\end{eqnarray}
Then according to (\ref{f6.5}) -- (\ref{f6.7a}), one obtains

\begin{equation}
\tilde{F}_{ilk^{\prime }j^{\prime }}=\left( \tilde{D}_{\left( {\rm f}\right)
l}\Sigma _{i}\right) _{||j^{\prime }||k^{\prime }}=\left( \tilde{D}_{\left( 
{\rm p}\right) j^{\prime }}\Sigma _{k^{\prime }}\right) _{||i||l}
\label{f6.8}
\end{equation}

The commutator of covariant derivatives $\nabla _{i}^{x^{\prime }}$ and $%
\nabla _{k^{\prime }}^{x}$, connected with the symmetric component $G$ of
the world function, has the property 
\[
T_{k\ldots l^{\prime }{\ldots |s}^{\prime }|i}^{j\ldots m^{\prime }{\ldots }%
}-T_{k\ldots l^{\prime }{\ldots }|i|s^{\prime }}^{j\ldots m^{\prime }{\ldots 
}}=F_{ika^{\prime }s^{\prime }}G^{ba^{\prime }}T_{b\ldots l^{\prime }{\ldots 
}}^{j\ldots m^{\prime }{\ldots }}+\ldots -G^{ja^{\prime }}F_{iba^{\prime
}s^{\prime }}T_{k\ldots l^{\prime }\ldots }^{b\ldots m^{\prime }{\ldots }} 
\]
\begin{equation}
-\ldots +G^{am^{\prime }}F_{iab^{\prime }s^{\prime }}T_{k\ldots l^{\prime
}\ldots }^{j\ldots b^{\prime }\ldots }+\ldots -F_{ial^{\prime }s^{\prime
}}G^{ab^{\prime }}T_{k\ldots b^{\prime }\ldots }^{j\ldots m^{\prime }\ldots
}-\ldots  \label{f6.16}
\end{equation}
where the curvature $F$-tensor has the form 
\begin{equation}
F_{ilk^{\prime }j^{\prime }}\equiv G_{pj^{\prime }}\Gamma _{il|k^{\prime
}}^{p}=G_{pj^{\prime }}\left( \Gamma _{il}^{p}-\gamma _{il}^{p}\right)
_{|k^{\prime }}=G_{i;l|k^{\prime }|j^{\prime }}  \label{f6.17}
\end{equation}
Here ($;$) denotes the usual covariant derivative with the Christoffel
symbol $\gamma _{kl}^{i}$, and the $Q$-tensor is written as follows

\begin{equation}
Q_{kl}^{i}=\gamma _{kl}^{i}-\Gamma _{kl}^{i}=-G^{sj^{\prime
}}G_{i;l|j^{\prime }}  \label{f6.18}
\end{equation}

Let us discover a connection between the $F$-tensor at the coincidence limit 
$\left[ F_{ilk^{\prime }j^{\prime }}\right] $ and the curvature tensor $%
r_{s.ik}^{.l}$, constructed of the Christoffel symbols $\gamma _{si}^{l}$ by
means of formula (\ref{f6.2}) 
\begin{equation}
r_{s.ik}^{.l}=\gamma _{si,k}^{l}-\gamma _{sk,i}^{l}+\gamma _{si}^{p}\gamma
_{pk}^{l}-\gamma _{sk}^{p}\gamma _{pi}^{l}.  \label{f6.18a}
\end{equation}
Let us take into account that 
\begin{equation}
\left[ \left( \nabla _{i}^{x^{\prime }}+\nabla _{i^{\prime }}^{x}\right)
T_{sp^{\prime }}\left( x,x^{\prime }\right) \right]
_{x}=D_{i}t_{sp}=t_{sp;i},\qquad t_{sp}\equiv \left[ T_{sp^{\prime }}\left(
x,x^{\prime }\right) \right] _{x}  \label{f6.9}
\end{equation}
Then, using (\ref{f6.16}), (\ref{f6.17}), one obtains 
\begin{eqnarray}
&&\left[ \left( \nabla _{i}^{x^{\prime }}+\nabla _{i^{\prime }}^{x}\right)
\left( \nabla _{k}^{x^{\prime }}+\nabla _{k^{\prime }}^{x}\right)
T_{s}\left( x,x^{\prime }\right) -\left( \nabla _{k}^{x^{\prime }}+\nabla
_{k^{\prime }}^{x}\right) \left( \nabla _{i}^{x^{\prime }}+\nabla
_{i^{\prime }}^{x}\right) T_{s}\left( x,x^{\prime }\right) \right] _{x} 
\nonumber \\
&=&\left( D_{i}D_{k}-D_{k}D_{i}\right) t_{s}=\left[ \Gamma _{is|k^{\prime
}}^{a}-\Gamma _{ks|i^{\prime }}^{a}\right] _{x}t_{a}  \label{f6.10}
\end{eqnarray}
where $D_{i}$ is the usual covariant derivative in the Riemannian space with
the Christoffel symbol $\gamma _{kl}^{i}=\left[ \Gamma _{kl}^{i}\right] _{x}$

On the other hand, the relation 
\begin{equation}
\left( D_{i}D_{k}-D_{k}D_{i}\right) t_{s}=r_{s.ik}^{.l}t_{l}  \label{f6.10a}
\end{equation}
takes place. Comparison of relations (\ref{f6.10}) and (\ref{f6.10a}) gives 
\begin{equation}
r_{s.ik}^{.l}=\left[ \Gamma _{is|k^{\prime }}^{l}-\Gamma _{ks|i^{\prime
}}^{l}\right] _{x}=\left[ \Gamma _{is,k^{\prime }}^{l}-\Gamma _{ks,i^{\prime
}}^{l}\right] _{x}=-g^{lp}f_{ispk}+g^{lp}f_{kspi}  \label{f6.13}
\end{equation}
where 
\begin{equation}
f_{ispk}\equiv \left[ F_{isp^{\prime }k^{\prime }}\right] _{x}=\left[
G_{lp^{\prime }}\Gamma _{is|k^{\prime }}^{l}\right] _{x}=-g_{lp}\left[
\Gamma _{is|k^{\prime }}^{l}\right] _{x}  \label{f6.14}
\end{equation}
According to (\ref{f6.17}) the one-point tensor $f_{ispk}$ is symmetric with
respect to transposition indices $i\leftrightarrow s$ and $p\leftrightarrow
k $ separately. 
\begin{equation}
f_{ispk}=f_{sipk},\qquad f_{ispk}=f_{iskp}  \label{f6.14a}
\end{equation}
Equation (\ref{f6.13}) can be written in the form 
\begin{equation}
g_{lp}r_{s.ik}^{.l}=-f_{ispk}+f_{kspi}  \label{f6.15}
\end{equation}
The metric tensor $g_{ik}$ is symmetric, and $f_{ispk}$ has the following
symmetry properties 
\begin{equation}
f_{ispk}=f_{sipk},\qquad f_{ispk}=f_{iskp},\qquad f_{ispk}=f_{pkis}
\label{f6.20}
\end{equation}
One can obtain connection of the type (\ref{f6.13}), (\ref{f6.14}) between
the $\tilde{F}$-tensor and the Riemannian -Christoffel curvature tensor in
the case of nonsymmetric T-geometry. Taking into account (\ref{f6.5}),
evident identity 
\begin{equation}
\frac{\partial }{\partial x^{k}}\left[ \tilde{\Gamma}_{il}^{p}\right]
_{x}\equiv \left[ \tilde{\Gamma}_{il,k}^{p}\right] _{x}+\left[ \tilde{\Gamma}%
_{il,k^{\prime }}^{p}\right] _{x}  \label{f6.21}
\end{equation}
and using relations (\ref{c5.15}), (\ref{c5.18}), one obtains 
\begin{equation}
\left[ \tilde{F}_{ilk^{\prime }j^{\prime }}\right] _{x}=-\tilde{g}_{pj}\left[
\tilde{\Gamma}_{il,k^{\prime }}^{p}\right] _{x}=-\tilde{g}_{pj}\left( \tilde{%
\gamma}_{\left( {\rm f}\right) }{}_{il,k}^{p}-\left[ \tilde{\Gamma}%
_{il,k}^{p}\right] _{x}\right)  \label{f6.22}
\end{equation}
Let us take into account identity 
\begin{equation}
\Sigma ^{pr^{\prime }}\Sigma _{,sr^{\prime }k}+\left( \Sigma ^{pr^{\prime
}}\right) _{,k}\Sigma _{sr^{\prime }}=0  \label{f6.23}
\end{equation}
obtained by differentiation of (\ref{c3.4}). Then using relations (\ref
{c5.15}), (\ref{c5.18}), one obtains from (\ref{f6.22}) 
\begin{eqnarray*}
\left[ \tilde{F}_{ilk^{\prime }j^{\prime }}\right] _{x} &=&-\tilde{g}_{pj}%
\tilde{\gamma}_{\left( {\rm f}\right) }{}_{il,k}^{p}+\tilde{g}_{pj}\left[
\Sigma ^{pq^{\prime }}\Sigma _{,ilkq^{\prime }}-\Sigma ^{pr^{\prime }}\Sigma
_{,sr^{\prime }k}\Sigma ^{sq^{\prime }}\Sigma _{,ilq^{\prime }}\right] _{x}
\\
&=&-\tilde{g}_{pj}\tilde{\gamma}_{\left( {\rm f}\right) }{}_{il,k}^{p}+%
\tilde{g}_{pj}\left[ -\tilde{g}^{pq}\Sigma _{,ilkq^{\prime }}-\tilde{\Gamma}%
_{sk}^{p}\tilde{\Gamma}_{il}^{s}\right] _{x}
\end{eqnarray*}
\begin{equation}
\tilde{f}_{ilkj}\equiv \left[ \tilde{F}_{ilk^{\prime }j^{\prime }}\right]
_{x}=-\left[ \Sigma _{,ilkj^{\prime }}\right] _{x}-\tilde{g}_{pj}\left( 
\tilde{\gamma}_{\left( {\rm f}\right) }{}_{il,k}^{p}+\tilde{\gamma}_{\left( 
{\rm f}\right) }{}_{il}^{s}\tilde{\gamma}_{\left( {\rm f}\right)
}{}_{sk}^{p}\right)  \label{f6.24}
\end{equation}
Alternating with respect to indices $k,l$, one obtains 
\begin{equation}
\tilde{f}_{ilkj}-\tilde{f}_{iklj}=\tilde{g}_{pj}\tilde{r}_{\left( {\rm f}%
\right) }{}_{i.kl}^{.p}  \label{f6.25}
\end{equation}
where $\tilde{r}_{\left( {\rm f}\right) }{}_{i.kl}^{.s}$ is the
Riemann-Christoffel curvature tensor, constructed on the base of the
Christoffel symbol $\tilde{\gamma}_{\left( {\rm f}\right) }{}_{ik}^{s}\ $ 
\begin{equation}
\tilde{r}_{\left( {\rm f}\right) }{}_{i.kl}^{.p}=\tilde{\gamma}_{\left( {\rm %
f}\right) }{}_{ik,l}^{p}-\tilde{\gamma}_{\left( {\rm f}\right)
}{}_{il,k}^{p}+\tilde{\gamma}_{\left( {\rm f}\right) }{}_{ik}^{s}\tilde{%
\gamma}_{\left( {\rm f}\right) }{}_{sl}^{p}-\tilde{\gamma}_{\left( {\rm f}%
\right) }{}_{il}^{p}\tilde{\gamma}_{\left( {\rm f}\right) }{}_{sk}^{k}
\label{f6.26}
\end{equation}

In the same way one can express $\tilde{f}_{ilkj}-\tilde{f}_{iklj}$ via the
Riemann-Christoffel curvature tensor $\tilde{r}_{\left( {\rm p}\right)
}{}_{i.kl}^{.s}$, constructed on the base of the Christoffel symbol $\tilde{%
\gamma}_{\left( {\rm p}\right) }{}_{ik}^{s}$ 
\begin{equation}
\tilde{f}_{ilkj}-\tilde{f}_{iklj}=\tilde{g}_{lp}\tilde{r}_{\left( {\rm p}%
\right) }{}_{k.ij}^{.p}  \label{f6.27}
\end{equation}
where 
\begin{equation}
\tilde{r}_{\left( {\rm p}\right) }{}_{i.kl}^{.p}=\tilde{\gamma}_{\left( {\rm %
p}\right) }{}_{ik,l}^{p}-\tilde{\gamma}_{\left( {\rm p}\right)
}{}_{il,k}^{p}+\tilde{\gamma}_{\left( {\rm p}\right) }{}_{ik}^{s}\tilde{%
\gamma}_{\left( {\rm p}\right) }{}_{sl}^{p}-\tilde{\gamma}_{\left( {\rm p}%
\right) }{}_{il}^{p}\tilde{\gamma}_{\left( {\rm p}\right) }{}_{sk}^{k}
\label{f6.28}
\end{equation}

To obtain representation (\ref{f6.27}), let us use another representation (%
\ref{f6.5a}) 
\[
\tilde{F}_{ilk^{\prime }j^{\prime }}=\Sigma _{iq^{\prime }}\tilde{\Gamma}%
_{k^{\prime }j^{\prime },l}^{q^{\prime }} 
\]
of the $\tilde{F}$-tensor, which differs from the representation (\ref{f6.5}%
) by a change $x\leftrightarrow x^{\prime }$. Producing the same operations (%
\ref{f6.22}) -- (\ref{f6.24}), one obtains (\ref{f6.27}) instead of (\ref
{f6.25}).

Note that relations (\ref{f6.25}) and (\ref{f6.27}) are different, because
the tensor $\tilde{g}_{ik}$ is not symmetric. In (\ref{f6.25}) summation is
produced over the first index, whereas in (\ref{f6.27}) it is produced over
the second index. In the symmetric T-geometry, when $\tilde{g}_{ik}$ is
symmetric, three expressions (\ref{f6.13}) (\ref{f6.25}) and (\ref{f6.27})
coincide.

There are two essentially different cases of asymmetric T-geometry:

1. Rough antisymmetry, when the field $a_{i}\left( x\right) \neq 0$. In this
case the field $a_{i}\left( x\right) $ dominates at small distances $%
x-x^{\prime }$, and the world function is determined by the linear form 
\[
\Sigma \left( x,x^{\prime }\right) =a_{i}\left( x\right) \left(
x^{i}-x^{\prime i}\right) +... 
\]
In this case the antisymmetry is the main phenomenon at small distances.

2. Fine antisymmetry, when the field $a_{i}\left( x\right) \equiv 0$. In
this case the antisymmetric effects are described by the field $a_{ikl}$. At
small distances $x-x^{\prime }$ the symmetric structure dominates, and the
world function is determined by the quadratic form 
\[
\Sigma \left( x,x^{\prime }\right) =\frac{1}{2}g_{ik}\left( x\right) \left(
x^{i}-x^{\prime i}\right) \left( x^{k}-x^{\prime k}\right) +... 
\]
as in the symmetric T-geometry. In this case the antisymmetric effects may
be considered as corrections to gravitational effects. This corrections may
be essential at large distances $\xi ^{i}=x^{i}-x^{\prime i}$, when the form 
$\frac{1}{6}a_{ikl}\xi ^{i}\xi ^{k}\xi ^{l}$ becomes of the same order as
the form $\frac{1}{2}g_{ik}\xi ^{i}\xi ^{k}$,

The asymmetric T-geometry with fine antisymmetry is simpler, because it is
rather close to the usual symmetric T-geometry.

\section{Gradient lines on the manifold in the case of fine antisymmetry $%
a_{i}\equiv 0$}

Let us consider a one-dimensional line ${\cal L}_{({\rm f})}$, passing
through points $x^{\prime }$ and $x^{\prime \prime }.$ This line is defined
by the relations 
\begin{equation}
{\cal L}_{({\rm f})}:\;\;\;\Sigma _{,i^{\prime }}\left( x,x^{\prime }\right)
=\tau \Sigma _{,i^{\prime }}\left( x^{\prime \prime },x^{\prime }\right)
=\tau b_{i^{\prime }},\qquad i=0,1,...n  \label{c6.1}
\end{equation}
Let us suppose that $\det ||\Sigma _{,i^{\prime }k}\left( x,x^{\prime
}\right) ||\neq 0$. Then $n+1$ equations (\ref{c6.1}) can be resolved with
respect to $x$ in the form 
\begin{equation}
{\cal L}_{({\rm f})}:\;\;x^{i}=x^{i}\left( \tau \right) ,\qquad i=0,1,...n
\label{c6.2}
\end{equation}
where $\tau $ is a parameter along the line ${\cal L}_{({\rm f})}$. As it
follows from (\ref{c6.1}), this line passes through the point $x^{\prime }$
at $\tau =0$ and through the point $x^{\prime \prime }$ at $\tau =1$. Such a
line will be referred to as gradient line (curve) from the future. Let us
derive differential equation for the gradient curve ${\cal L}_{({\rm f})}$.

Differentiating (\ref{c6.1}) with respect to $\tau $, one obtains 
\begin{equation}
\Sigma _{,ki^{\prime }}\left( x,x^{\prime }\right) \frac{dx^{k}}{d\tau }%
=\Sigma _{,i^{\prime }}\left( x^{\prime \prime },x^{\prime }\right)
=b_{i^{\prime }},\qquad i=0,1,...n  \label{c6.3}
\end{equation}
Differentiating once more, one obtains 
\begin{equation}
\Sigma _{,ki^{\prime }}\left( x,x^{\prime }\right) \frac{d^{2}x^{k}}{d\tau
^{2}}+\Sigma _{,kli^{\prime }}\left( x,x^{\prime }\right) \frac{dx^{k}}{%
d\tau }\frac{dx^{l}}{d\tau }=0,\qquad i=0,1,...n  \label{c6.4}
\end{equation}
Using relation (\ref{c3.6}), one can write equations (\ref{c6.4}) in the
form 
\begin{equation}
\frac{d^{2}x^{i}}{d\tau ^{2}}+\tilde{\Gamma}_{kl}^{i}\left( x,x^{\prime
}\right) \frac{dx^{k}}{d\tau }\frac{dx^{l}}{d\tau }=0,\qquad i=0,1,...n
\label{c6.5}
\end{equation}
The equation (\ref{c6.5}) may be interpreted as an equation for a geodesic
in some $(n+1)$-dimensional Euclidean space with the Christoffel symbol $%
\tilde{\Gamma}_{kl}^{i}\left( x,x^{\prime }\right) $. This geodesic passes
through the points $x^{\prime }$ and $x^{\prime \prime }$.

Let the points $x^{\prime }$ and $x^{\prime \prime }$ be infinitesimally
close. Then equation (\ref{c6.5}) can be written in the form 
\begin{equation}
{\cal L}_{({\rm f})}:\;\;\frac{d^{2}x^{i}}{d\tau ^{2}}+\tilde{\gamma}%
_{\left( {\rm f}\right) }{}_{kl}^{i}\frac{dx^{k}}{d\tau }\frac{dx^{l}}{d\tau 
}=0,\qquad i=0,1,...n  \label{c6.6}
\end{equation}
where $\tilde{\gamma}_{\left( {\rm f}\right) }{}_{kl}^{i}=\tilde{\gamma}%
_{\left( {\rm f}\right) }{}_{kl}^{i}\left( x\right) =\left[ \tilde{\Gamma}%
_{kl}^{i}\right] _{x}$. Dividing the gradient line ${\cal L}_{({\rm f})}$
into infinitesimal segments and writing equations (\ref{c6.5}) in the form (%
\ref{c6.6}) on each segment, one obtains that the gradient line ${\cal L}_{(%
{\rm f})}$ is described by the equations (\ref{c6.6}) everywhere.

The equation (\ref{c6.6}) does not contain a reference to the point $%
x^{\prime }$, and any gradient line (\ref{c6.1}), (\ref{c6.2}) is to satisfy
this equation.

In the case of fine antisymmetry, when $a_{i}\equiv 0$, the equation (\ref
{c6.6}) can be written in other form. Using relations (\ref{c5.18}), and
taking into account that $a_{i}\equiv 0$, one obtains instead of (\ref{c6.6}%
) 
\begin{equation}
{\cal L}_{({\rm f})}:\qquad \frac{d^{2}x^{i}}{d\tau ^{2}}+\left( \gamma
_{kl}^{i}+\beta _{kl}^{i}\right) \frac{dx^{k}}{d\tau }\frac{dx^{l}}{d\tau }%
=0,\qquad a_{i}\equiv 0  \label{c6.6b}
\end{equation}
where 
\begin{eqnarray}
\gamma _{kl}^{i} &=&\gamma _{kl}^{i}\left( x\right) =\frac{1}{2}g^{si}\left(
g_{ks,l}+g_{sl,k}-g_{lk,s}\right) ,  \label{c6.6c} \\
\beta _{kl}^{i} &=&\beta _{kl}^{i}\left( x\right) =g^{si}a_{kls}
\label{c6.6d}
\end{eqnarray}
If $a_{ikl}=0$, the equations (\ref{c6.6b}) may be considered to be the
equations for a geodesic in a Riemannian space with the metric tensor $%
g_{ik} $.

In the case of rough antisymmetry, when $a_{i}\neq 0$, equations (\ref{c6.1}%
), (\ref{c6.5}) also describe a gradient line, but equation (\ref{c6.6}) is
not equivalent to (\ref{c6.5}), because the point $x^{\prime }$ does not
belong to ${\cal L}_{({\rm f})}$, in general. In this case one cannot choose
the points $x^{\prime }$and $x^{\prime \prime }$ infinitesimally close and
pass from equation (\ref{c6.5}) to (\ref{c6.6}). Thus, in the case of rough
antisymmetry the equation (\ref{c6.6}) does not describe a gradient line, in
general.

Now let us consider another type of gradient line ${\cal L}_{({\rm p})}$,
passing through the points $x$ and $x^{\prime \prime }$. Let the gradient
line ${\cal L}_{({\rm p})}$ be described by the equations (It is supposed
again that $a_{i}\equiv 0$). 
\begin{equation}
{\cal L}_{({\rm p})}:\qquad \Sigma _{,i}\left( x,x^{\prime }\right) =\tau
\Sigma _{,i}\left( x,x^{\prime \prime }\right) =\tau b_{i},\qquad i=0,1,...n
\label{c6.7}
\end{equation}
which determine 
\begin{equation}
{\cal L}_{({\rm p})}:\qquad x^{\prime i}=x^{\prime i}\left( \tau \right)
,\qquad i=0,1,...n  \label{c6.8}
\end{equation}
Equation (\ref{c6.7}) distinguishes from the equation (\ref{c6.1}) only in
transposition of the first and second arguments of the world function $%
\Sigma \left( x,x^{\prime }\right) .$ The gradient line ${\cal L}_{({\rm p}%
)} $, determined by the relation (\ref{c6.7}), may be referred to as the
gradient line from the past. Manipulating with the equation (\ref{c6.7}) in
the same way as with (\ref{c6.1}), one obtains instead of (\ref{c6.6}) 
\begin{equation}
{\cal L}_{({\rm p})}:\qquad \frac{d^{2}x^{i}}{d\tau ^{2}}+\tilde{\gamma}%
_{\left( {\rm p}\right) }{}_{kl}^{i}\frac{dx^{k}}{d\tau }\frac{dx^{l}}{d\tau 
}=0,\qquad i=0,1,...n  \label{c6.8a}
\end{equation}

In the case of fine antisymmetry, when $a_{i}\equiv 0$, the equation (\ref
{c6.8a}) can be written in other form. Using relation (\ref{c5.18b}), and
taking into account that $a_{i}\equiv 0$, one obtains instead of (\ref{c6.8a}%
) 
\begin{equation}
{\cal L}_{({\rm p})}:\qquad \frac{d^{2}x^{i}}{d\tau ^{2}}+\left( \gamma
_{kl}^{i}-\beta _{kl}^{i}\right) \frac{dx^{k}}{d\tau }\frac{dx^{l}}{d\tau }%
=0,\qquad a_{i}\equiv 0  \label{c6.9a}
\end{equation}
where $\gamma _{kl}^{i}$ and $\beta _{kl}^{i}$ are defined by the relations (%
\ref{c6.6c}), (\ref{c6.6d}).

In the case of symmetric T-geometry, when $a_{ikl}=0$ and $\beta _{kl}^{i}=0$%
, differential equations (\ref{c6.6b}) and (\ref{c6.9a}) respectively for
gradient line ${\cal L}_{({\rm f})}$ and for gradient line ${\cal L}_{({\rm p%
})}$ coincide.

In the case of asymmetric T-geometry the quantities $\left[ \Gamma _{kl}^{i}%
\right] _{x}$ and $\left[ \Gamma _{k^{\prime }l^{\prime }}^{i^{\prime }}%
\right] _{x}$ do not coincide, in general. In this case the equations (\ref
{c6.6}) and (\ref{c6.9a}) determine, in general, different gradient curves,
passing through the same points $x^{\prime }$ and $x^{\prime \prime }$.
Differential equations (\ref{c6.6}) and (\ref{c6.9a}) for the gradient
curves ${\cal L}_{({\rm p})}$ and ${\cal L}_{({\rm f})}$ differ in the sign
of the ''antisymmetric force'' 
\begin{equation}
\beta _{kl}^{i}\frac{dx^{k}}{d\tau }\frac{dx^{l}}{d\tau }=g^{si}\left( -%
\frac{1}{2}a_{k,ls}-\frac{1}{2}a_{l,ks}+a_{kls}\right) \frac{dx^{k}}{d\tau }%
\frac{dx^{l}}{d\tau }  \label{c6.10}
\end{equation}

Finally, one can introduce the neutral gradient line ${\cal L}_{({\rm n})}$,
defining it by the relations 
\begin{equation}
{\cal L}_{({\rm n})}:\;\;G_{,i^{\prime }}\left( x,x^{\prime }\right) =\tau
G_{,i^{\prime }}\left( x^{\prime \prime },x^{\prime }\right) =\tau
b_{i^{\prime }},\qquad i=0,1,...n  \label{c6.10a}
\end{equation}
which determine 
\begin{equation}
{\cal L}_{({\rm n})}:\;\;x^{i}=x^{i}\left( \tau \right) ,\qquad i=0,1,...n
\label{c6.10b}
\end{equation}
Equation (\ref{c6.10a}) distinguishes from the equation (\ref{c6.1}) only in
replacement of the world function $\Sigma \left( x,x^{\prime }\right) $ by
its symmetric component $G\left( x,x^{\prime }\right) .$ Manipulating with
the equation (\ref{c6.10a}) in the same way as with (\ref{c6.1}), one
obtains instead of (\ref{c6.6b}) 
\begin{equation}
{\cal L}_{({\rm n})}:\qquad \frac{d^{2}x^{i}}{d\tau ^{2}}+\gamma _{kl}^{i}%
\frac{dx^{k}}{d\tau }\frac{dx^{l}}{d\tau }=0  \label{c6.10c}
\end{equation}
where the ''antisymmetric force'' is absent.

The gradient lines (\ref{c6.1}) and (\ref{c6.7}) are insensitive with
respect to transformation of the world function of the form 
\begin{equation}
\Sigma \rightarrow \tilde{\Sigma}=f\left( \Sigma \right) ,\qquad \left|
f^{\prime }\left( \Sigma \right) \right| >0  \label{c6.10d}
\end{equation}
where $f$ is an arbitrary function, because for determination of the
gradient line only direction of the gradient $\Sigma _{i}$ or $\Sigma
_{i^{\prime }}$ is important, but not its module. Indeed, after substitution
of $\tilde{\Sigma}$ from (\ref{c6.10d}) in (\ref{c6.1}) one obtains the
equation 
\begin{equation}
\Sigma _{,i^{\prime }}\left( x,x^{\prime }\right) =\tau ^{\prime }\Sigma
_{,i^{\prime }}\left( x^{\prime \prime },x^{\prime }\right) ,\qquad
i=0,1,...n,\qquad \tau ^{\prime }=\tau \frac{f^{\prime }\left( \Sigma \left(
x^{\prime \prime },x^{\prime }\right) \right) }{f^{\prime }\left( \Sigma
\left( x,x^{\prime }\right) \right) }  \label{c6.10e}
\end{equation}
which describes the same gradient line, but with another parametrization.

\section{Conditions of degeneracy of the neutral first order tube}

In general, the asymmetric T-geometry is nondegenerate geometry, even if it
is given on $n$-dimensional manifold ${\cal M}_{n}$. In general case the
first order tube ${\cal T}_{({\rm n}){\bf x}^{\prime }{\bf x}^{\prime \prime
}}$, passing through points $x^{\prime }$ and $x^{\prime \prime }$, does not
coincide with gradient line ${\cal L}_{\left( {\rm f}\right) }$ or ${\cal L}%
_{\left( {\rm p}\right) },$ passing through the points $x^{\prime }$ and $%
x^{\prime \prime }$ and defined by the relations (\ref{c6.1}) and (\ref{c6.7}%
) respectively.

Let us investigate, under what conditions the neutral first order tube $%
{\cal T}_{({\rm n}){\bf x}^{\prime }{\bf x}^{\prime \prime }}$ degenerates
into gradient line ${\cal L}_{\left( {\rm f}\right) }$ or ${\cal L}_{\left( 
{\rm p}\right) }$. The first order tube ${\cal T}_{({\rm n}){\bf P}_{0}{\bf P%
}_{1}}$, passing through the points $P_{0}=\left\{ x\right\} $, $%
\;P_{1}=\left\{ x^{\prime }\right\} $ is defined by the relation 
\begin{equation}
F_{2}\left( P_{0},P_{1},R\right) =\left| 
\begin{array}{cc}
\left( \overrightarrow{P_{0}P_{1}}.\overrightarrow{P_{0}P_{1}}\right) & 
\left( \overrightarrow{P_{0}P_{1}}.\overrightarrow{P_{0}R}\right) \\ 
\left( \overrightarrow{P_{0}R}.\overrightarrow{P_{0}P_{1}}\right) & \left( 
\overrightarrow{P_{0}R}.\overrightarrow{P_{0}R}\right)
\end{array}
\right| =0  \label{c7.1}
\end{equation}
where $R=\left\{ x^{\prime }+dx^{\prime }\right\} $ is a running point on
the tube ${\cal T}_{P_{0}P_{1}}$.

One has the following expansion for the scalar $\Sigma $-products $\left( 
\overrightarrow{P_{0}P_{1}}.\overrightarrow{P_{0}R}\right) $ and $\left( 
\overrightarrow{P_{0}R}.\overrightarrow{P_{0}P_{1}}\right) $%
\[
\left( \overrightarrow{P_{0}P_{1}}.\overrightarrow{P_{0}R}\right) =\Sigma
\left( P_{1},P_{0}\right) +\Sigma \left( P_{0},R\right) -\Sigma \left(
P_{1},R\right) 
\]
\begin{eqnarray}
&=&\Sigma \left( x^{\prime },x\right) +\Sigma \left( x,x^{\prime
}+dx^{\prime }\right) -\Sigma \left( x^{\prime },x^{\prime }+dx^{\prime
}\right)  \nonumber \\
&=&2G+\left( \Sigma _{,i^{\prime }}-\left[ \Sigma _{,i^{\prime }}\right]
_{x^{\prime }}\right) dx^{\prime i^{\prime }}+\frac{1}{2}\left( \Sigma
_{,i^{\prime }k^{\prime }}-\left[ \Sigma _{,i^{\prime }k^{\prime }}\right]
_{x^{\prime }}\right) dx^{\prime i^{\prime }}dx^{\prime k^{\prime }}
\label{c7.2}
\end{eqnarray}
\[
\left( \overrightarrow{P_{0}R}.\overrightarrow{P_{0}P_{1}}\right) =\Sigma
\left( P_{0},P_{1}\right) +\Sigma \left( R,P_{0}\right) -\Sigma \left(
R,P_{1}\right) 
\]
\begin{eqnarray}
&=&\Sigma \left( x,x^{\prime }\right) +\Sigma \left( x^{\prime }+dx^{\prime
},x\right) -\Sigma \left( x^{\prime }+dx^{\prime },x^{\prime }\right) 
\nonumber \\
&=&2G+\left( G_{,i^{\prime }}-A_{,i^{\prime }}-\left[ \Sigma _{,i}\right]
_{x^{\prime }}\right) dx^{\prime i^{\prime }}+\frac{1}{2}\left(
G_{,i^{\prime }k^{\prime }}-A_{,i^{\prime }k^{\prime }}-\left[ \Sigma _{,ik}%
\right] _{x^{\prime }}\right) dx^{\prime i}dx^{\prime k}  \label{c7.3}
\end{eqnarray}
where unprimed indices are associated with the first argument and the primed
ones with the second argument of the $\Sigma $-function. 
\begin{equation}
\left( \overrightarrow{P_{0}P_{1}}.\overrightarrow{P_{0}P_{1}}\right) =2G
\label{c7.4}
\end{equation}
\begin{equation}
\left( \overrightarrow{P_{0}R}.\overrightarrow{P_{0}R}\right) =2G\left(
x,x^{\prime }+dx^{\prime }\right) =2G+2G_{,i^{\prime }}dx^{\prime i^{\prime
}}+G_{,i^{\prime }k^{\prime }}dx^{\prime i^{\prime }}dx^{\prime k^{\prime }}
\label{c7.5}
\end{equation}
Using relations (\ref{c7.2}) -- (\ref{c7.5}) and (\ref{c5.14}), (\ref{c5.15}%
), one reduces equation (\ref{c7.1}) to the form 
\begin{equation}
\left( \left( G_{,i^{\prime }}-A_{,i^{\prime }}-a_{i}\left( x^{\prime
}\right) \right) dx^{\prime i^{\prime }}\right) \left( \left( G_{,l^{\prime
}}+A_{,l^{\prime }}+a_{l}\left( x^{\prime }\right) \right) dx^{\prime
^{\prime }}\right) =2Gg_{l^{\prime }m^{\prime }}\left( x^{\prime }\right)
dx^{\prime l^{\prime }}dx^{\prime m^{\prime }}  \label{c7.6}
\end{equation}

We suppose that the world function is such, that the tube ${\cal T}_{{\bf x}%
^{\prime }{\bf x}}$ degenerates to a line. Then the solution of (\ref{c7.6})
has either the form 
\begin{equation}
dx^{\prime k^{\prime }}=g^{k^{\prime }l^{\prime }}\left( x^{\prime }\right)
\left( G_{,i^{\prime }}-A_{,i^{\prime }}-a_{i^{\prime }}\left( x^{\prime
}\right) \right) d\tau  \label{c7.9}
\end{equation}
or the form 
\begin{equation}
dx^{\prime k^{\prime }}=g^{k^{\prime }l^{\prime }}\left( x^{\prime }\right)
\left( G_{,k^{\prime }}+A_{,k^{\prime }}+a_{k^{\prime }}\left( x^{\prime
}\right) \right) d\tau  \label{c7.10}
\end{equation}
where $d\tau $ is an infinitesimal parameter. The relations (\ref{c7.9}), (%
\ref{c7.10}) describe a one-dimensional lines in vicinity of the point $%
x^{\prime }$. Both expressions (\ref{c7.9}), (\ref{c7.10}) are solutions of
the equation (\ref{c7.6}), provided the relation 
\begin{equation}
\left( G_{,i^{\prime }}-A_{,i^{\prime }}-a_{i^{\prime }}\left( x^{\prime
}\right) \right) g^{i^{\prime }k^{\prime }}\left( x^{\prime }\right) \left(
G_{,k^{\prime }}+A_{,k^{\prime }}+a_{k^{\prime }}\left( x^{\prime }\right)
\right) =2G  \label{c7.11}
\end{equation}
is fulfilled.

There is only one solution, provided solutions (\ref{c7.9}) and (\ref{c7.10}%
) coincide. It means that 
\begin{equation}
A_{,k^{\prime }}+a_{k^{\prime }}\left( x^{\prime }\right) =0  \label{c7.11a}
\end{equation}
and the condition (\ref{c7.11}) transforms to the equation 
\begin{equation}
G_{,i^{\prime }}g^{i^{\prime }k^{\prime }}\left( x^{\prime }\right)
G_{,k^{\prime }}=2G  \label{c7.12}
\end{equation}
This is well known equation for the world function of a Riemannian space 
\cite{S60}. The Riemannian geometry is locally degenerate in the sense of
definition \ref{d3.14}, and the equation (\ref{c7.12}) describes this
property of Riemannian geometry. The world function (\ref{g5.5}), (\ref{g5.6}%
) of the distorted space-time geometry ${\cal G}_{{\rm D}}$ does satisfy the
relation (\ref{c7.12})

According to expansion (\ref{c5.5}) the condition (\ref{c7.11a}) may take
place, only if 
\begin{equation}
A\left( x,x^{\prime }\right) =a_{i}\left( x^{i}-x^{\prime i}\right) ,\qquad
a_{i}=\text{const}  \label{c7.13}
\end{equation}
In this case the quantities (\ref{f2.2ee}) vanish, i.e. 
\begin{equation}
\eta _{{\rm f}}=A\left( x,x^{\prime }\right) +A\left( x^{\prime },y\right)
+A\left( y,x\right) =0,\qquad \forall x,x^{\prime },y\in {\cal M}_{n}
\label{c7.14}
\end{equation}
and the first order tubes are similar in symmetric and nonsymmetric
geometries.

For the case of the past first order tube and the future one the conditions
of degeneracy are also rather rigid. In this case instead of (\ref{c7.6})
one obtains two conditions 
\begin{equation}
4G\left( A_{,i^{\prime }}+a_{i^{\prime }}\right) dx^{\prime i^{\prime }}=0
\label{c7.15}
\end{equation}
\begin{equation}
\left( 2G\left( A_{,i^{\prime }k^{\prime }}-\left[ \Sigma _{,i^{\prime
}k^{\prime }}\right] _{x^{\prime }}\right) -G_{,i^{\prime }}G_{,k^{\prime
}}\right) dx^{\prime i^{\prime }}dx^{\prime k^{\prime }}=0  \label{c7.16}
\end{equation}
In the case (\ref{c7.13}) the condition (\ref{c7.15}) is fulfilled, and (\ref
{c7.16}) reduces to (\ref{c7.12}).

If the first order tube is nondegenerate in symmetric T-geometry, it cannot
degenerate after addition of antisymmetric component, because the local
degeneracy condition (\ref{c7.12}) remains to be not fulfilled.

Thus, practically any antisymmetric component of the world function destroys
degeneracy of the neutral first order tube. If one connects quantum effects
with the first order tube degeneracy \cite{R91}, one concludes that the
possible asymmetry of the space-time geometry is connected with quantum
effects.

\section{Examples of the first order tubes in \newline
nonsymmetric T-geometry.}

To imagine the possible corollaries of asymmetry in T-geometry, let us
construct the first order tube ${\cal T}_{P_{0}P_{1}}$ in the $\Sigma $%
-space. Let us consider $\Sigma $-space on the $4$-dimensional manifold with
the world function 
\begin{eqnarray}
\Sigma \left( x,x^{\prime }\right)  &=&a_{i}\xi ^{i}+\frac{1}{2}g_{ki}\xi
^{i}\xi ^{k},\qquad a_{i}=b_{i}\left( 1+\alpha f\left( \xi ^{2}\right)
\right) ,  \label{cc3.1} \\
\xi ^{i} &=&x^{i}-x^{\prime i},\qquad \xi ^{2}\equiv g_{ki}\xi ^{i}\xi ^{k},%
\text{\qquad }\alpha ,b_{i},g_{ik}=\text{const,}  \nonumber
\end{eqnarray}
where $f$ is some function of $\xi ^{2}$ and summation is made over
repeating indices from $0$ to $3$. One can interpret the relation (\ref
{cc3.1}) as an Euclidean space with a linear structure $a_{i}\xi ^{i}$ given
on it. Such a $\Sigma $-space is  not isotropic, because there is a vector $%
a_{i}$, describing some preferable direction in the $\Sigma $-space.

Let us construct the first order neutral tube ${\cal T}_{P_{0}P_{1}}$.
Coordinates of points $P_{0}=\left\{ 0\right\} $, $P_{1}=\left\{ y\right\} $%
, $R=\left\{ x\right\} $, where $R$ is the running point. In the given case
the characteristic quantity (\ref{f2.2ee}) has the form 
\begin{equation}
\eta _{{\rm f}}=\eta _{{\rm f}}\left( P_{0},P_{1},R\right) =\alpha \left(
-b_{i}x^{i}f\left( x^{2}\right) +b_{i}y^{i}f\left( y^{2}\right) +b_{i}\left(
x^{i}-y^{i}\right) f\left( \left( x-y\right) ^{2}\right) \right) ,
\label{cc3.1a}
\end{equation}
The quantity $\eta _{{\rm f}}$ does not depend on the constant component of
the vector $a_{i}.$ Then according to (\ref{f2.3}) - (\ref{f2.1}) the shape
of the tube does not depend on the constant component of the vector $a_{i}$.
If $\alpha =0$ and $a_{i}=$const, shape of all first order tubes is the
same, as in the case of symmetric T-geometry, when $a_{i}=0$. In other
words, the shape of the first order tubes is insensitive to the space-time
anisotropy, described by the vector field $a_{i}=$const. We omit the
constant component of the field $a_{i}$ and consider the cases, when its
variable part has the form 
\begin{eqnarray}
1 &:&\quad f\left( \xi ^{2}\right) =\xi ^{2},\qquad 2:\quad f\left( \xi
^{2}\right) =\frac{1}{1+\beta \xi ^{2}},\;\;\;\;\beta =\text{const}
\label{cc3.1b} \\
\xi ^{2} &\equiv &g_{ik}\xi ^{i}\xi ^{k},\qquad \xi ^{i}\equiv
x^{i}-x^{\prime i}  \nonumber
\end{eqnarray}
In the first case the antisymmetric structure is essential at large
distances $\xi =x-x^{\prime }$. In the second case the antisymmetric
structure vanishes at large $\xi $.

The equation (\ref{f2.3}), determining the shape of the tube ${\cal T}_{0y}$
has the form 
\begin{equation}
\left| 
\begin{array}{cc}
2G\left( 0,y\right) & \left( {\bf 0y.0x}\right) \\ 
\left( {\bf 0x.0y}\right) & 2G\left( 0,x\right)
\end{array}
\right| =4G\left( 0,y\right) G\left( 0,x\right) -\left( {\bf 0y.0x}\right)
\left( {\bf 0x.0y}\right) =0  \label{cc3.2}
\end{equation}
In the first case, when $f\left( \xi ^{2}\right) =\xi ^{2}$, calculation
gives for (\ref{cc3.2}) 
\begin{equation}
\left( x_{i}y^{i}\right) ^{2}-x^{2}y^{2}=\eta _{{\rm f}}^{2}  \label{cc3.4}
\end{equation}
\begin{equation}
\eta _{{\rm f}}=\alpha \left[ \left( x_{i}y^{i}\right) \left( -2\left(
b_{k}y^{k}\right) +2\left( b_{k}x^{k}\right) \right) -\left(
b_{i}x^{i}\right) y^{2}+\left( b_{i}y^{i}\right) x^{2}\right]  \label{cc3.4a}
\end{equation}
where $x^{2}=x^{i}x_{i}$, $y^{2}=y_{i}y^{i}$. In the case, when the metric
tensor $g_{ik}$ is the metric tensor of the proper Euclidean space, $%
x^{2}y^{2}\geq \left( x_{i}y^{i}\right) ^{2}$, the equation (\ref{cc3.4})
has an interesting solution, only if $\alpha =0$. Then 
\begin{equation}
x^{2}y^{2}=\left( x_{i}y^{i}\right) ^{2},\;\;\;\;{\cal T}_{0y}=\left\{
x\left| \bigwedge_{i=0}^{i=3}x^{i}=y^{i}\tau \right. \right\} ,\qquad \alpha
=0  \label{cc3.5}
\end{equation}
In the case $\alpha \neq 0$, the first order tube ${\cal T}_{0y}$
degenerates to the set of basic points $\left\{ 0,y\right\} $, because
substitution of $x^{i}=y^{i}\tau $ in the square bracket in (\ref{cc3.4})
shows that the bracket vanishes only at $\tau =0$ or $\tau =1$. Thus, in the
case of proper Euclidean metric tensor $g_{ik}$ the first order tube shape
does not depend on $a_{i},$ provided $a_{i}=$const.

Let us consider a more interesting case, when the metric tensor $g_{ik}$ of $%
\Sigma $-space is the Minkowski one. Then $x^{2}y^{2}\leq \left(
x_{i}y^{i}\right) ^{2},$ provided the ${\bf 0y}$ is timelike ($\left| {\bf 0y%
}\right| ^{2}=2G\left( 0,y\right) >0$). In this case the equation (\ref
{cc3.4}) has the solution (\ref{cc3.5}), if $\alpha =0$.

If $\alpha \neq 0,$ let us consider the special case, when vector $b_{i}$ is
the unit timelike vector, and the basic vector ${\bf 0y}$ is chosen in such
a way, that 
\begin{equation}
b_{i}=\frac{y_{i}}{\left| y\right| },\qquad y=\left\{ \left| y\right| ,{\bf 0%
}\right\} ,\qquad \left| y\right| =\sqrt{y^{i}y_{i}}  \label{cc3.5a}
\end{equation}
Vector ${\bf 0x}$ is presented in the form 
\begin{equation}
x=\left\{ x^{0},{\bf x}\right\} =\left\{ t\left| y\right| ,{\bf r}\left|
y\right| \right\} ,\qquad r=\sqrt{{\bf r}^{2}}  \label{cc3.5b}
\end{equation}
Using relations (\ref{cc3.5a}), (\ref{cc3.5b}), one obtains from (\ref
{cc3.4a}) 
\begin{equation}
\eta _{{\rm f}}=\alpha \left| y\right| ^{3}\left( 3t\left( t-1\right)
-r^{2}\right)  \label{cc3.5c}
\end{equation}
and equation (\ref{cc3.4}) is reduced to the form 
\begin{equation}
-r^{2}+\alpha ^{2}\left| y\right| ^{2}\left( 3t\left( t-1\right)
-r^{2}\right) ^{2}=0,  \label{cc3.5d}
\end{equation}
Its solution has the form 
\begin{equation}
r=\pm \frac{1}{2\alpha \left| y\right| }\left( -1\pm \sqrt{\left( 1+12\alpha
^{2}\left| y\right| ^{2}t\left( t-1\right) \right) }\right)  \label{cc3.8}
\end{equation}
Any section $t=$const of the three-dimensional surface ${\cal T}_{0y}$ form
two (or zero) spheres, whose radii $r=r\left( t\right) $ are determined by
the relation (\ref{cc3.8}). Equation (\ref{cc3.8}) gives four values of $r$
for any value of $t$, but only two of them are essential, because radii $\ r$
and $-r$ describe the same surface.

It follows from (\ref{cc3.8}) that 
\[
\lim_{t\rightarrow \infty }\frac{r}{t}=\pm \sqrt{3}, 
\]
It means the tube ${\cal T}_{0y}$ is infinite only in spacelike directions.
In the timelike directions the tube size is bounded.

In the vicinity of the vector ${\bf 0y}$, generating the tube, the shape of
the tube depends on interrelation between the intensity of the antisymmetry,
described by the constant $\alpha $, and the length of the vector ${\bf 0y}$%
{\bf . }The quantity $\alpha $ appears in the equation (\ref{cc3.8}) only in
the combination $g=\alpha \left| y\right| $. In any case, when $\alpha \neq
0 $, the tube ${\cal T}_{0y}$ does not degenerate into a one-dimensional
curve.

If the antisymmetric structure is strong enough, and $\alpha \left| y\right|
>1/\sqrt{3}$, the tube ${\cal T}_{0y}$ is empty in its center in the sense
that intersection of ${\cal T}_{0y}$ with the plane $t=0.5$ is empty. If $%
\alpha \left| y\right| <1/\sqrt{3}$ intersection of ${\cal T}_{0y}$ with the
plane $t=0.5$ forms two concentric spheres of radii 
\begin{equation}
r_{1}=\frac{3\alpha \left| y\right| }{2\left( \sqrt{\left( 1-3\alpha
^{2}\left| y\right| ^{2}\right) }+1\right) },\qquad r_{2}=\frac{3\alpha
\left| y\right| }{2\left( 1-\sqrt{\left( 1-3\alpha ^{2}\left| y\right|
^{2}\right) }\right) }  \label{cc3.7}
\end{equation}
If $\alpha \left| y\right| \ll 1$, one of radii is small $r_{1}=0.75\alpha
\left| y\right| $ and another one is large $r_{2}=1/\left( \alpha \left|
y\right| \right) $.

The shape of the tube ${\cal T}_{0y}$ is symmetric with respect to the
reflection $t\rightarrow 1-t$. (See figures 1,2). In the same time a
separation of the tube into internal and external segments is not symmetric
with respect to the reflection $t\rightarrow 1-t$. This is shown
schematically in the figures 3,4, where internal segment is drawn by a thick
line, whereas external segments are drawn by thin line. The shape of
internal segment, as well as that of external ones looks rather unexpected.
The internal segment ${\cal T}_{[0y]}$ is strongly deformed with respect to
the case of symmetric geometry. A part of the internal segment ${\cal T}%
_{[0y]}$ spreads to spatial infinity. Both external segments ${\cal T}_{0]y}$
and ${\cal T}_{0[y}$ are restricted in time direction. The external segment $%
{\cal T}_{0[y}$ is placed in a finite region. The segment ${\cal T}_{0]y}$
spreads to the spatial infinity, but it is bounded in any timelike direction.

We have seen that in the symmetric T-geometry the thickness of the internal
segment is responsible for non-relativistic quantum effects \cite{R91}. At
the strong antisymmetric field $a_{i}$ the internal segment thickness
becomes to be infinite. It increases quantum effects and may lead to
unexpected phenomena.

Let us consider now the second case (\ref{cc3.1b}), when the antisymmetric
structure is essential only at small distances. In this case one obtains
instead of equation (\ref{cc3.5d}). 
\begin{equation}
r^{2}=g^{2}\left( -\frac{\left( t-1\right) }{1+g_{1}\left( \left( t-1\right)
^{2}-r^{2}\right) }+\frac{t}{1+g_{1}\left( t^{2}-r^{2}\right) }-\frac{1}{%
1+g_{1}}\right) ^{2}  \label{cc3.9}
\end{equation}
where the same designations (\ref{cc3.5a}) - (\ref{cc3.5b}) are used, and $%
g=\alpha \left| y\right| ,\;\;g_{1}=\beta \left| y\right| ^{2}$. At large $t$
the equation (\ref{cc3.9}) transforms to the equation 
\begin{equation}
r^{2}=\frac{g^{2}}{\left( 1+\beta \left| y\right| ^{2}\right) ^{2}},\qquad
t\rightarrow \infty  \label{cc3.10}
\end{equation}
It means that the tube is unbounded in the timelike direction ${\bf 0y}$ and
has a finite radius at $t\rightarrow \infty $. The tube is bounded in any
spatial direction. The tube shape is rather fancy, and the section ${\cal T}%
_{0y}\cap S_{t}$ forms several concentric circles ($S_{t}$ is the surface $%
t= $const).

Thus, the local antisymmetric structure produces only local perturbation of
the tube shape. At the timelike infinity this perturbation reduces to a
nonvanishing radius of the tube. As we have seen in the fifth section,
geometrical stochasticity depends on the thickness of tube internal segment.
Any asymmetry of the world function increases this thickness and increases
stochasticity. It generates additional nondegeneracy of T-geometry, which is
connected with the particle mass geometrization and with quantum effects 
\cite{R91}.

\section{Concluding remarks.}

The main goal of the nonsymmetric T-geometry development is its possible
application as a space-time geometry, especially as a space-time geometry of
microcosm. Approach and methods of T-geometry distinguish from those of the
Riemannian (pseudo-Riemannian) geometry, which is used now as a space-time
geometry. The Riemannian geometry imposes on the space-time geometry a
series of unfounded constraints. These restrictions are generated by methods
used at the description of the Riemannian geometry. Let us list some of them.

1. The continuity of space-time. This is a very fine property which cannot
be tested by a direct experiment. T-geometry is insensitive to continuity,
and it is free of this constraint. For application of T-geometry is
unessential, whether the space-time geometry is continuous or only
fine-grained.

2. The Riemannian geometry is a geometry with fixed dimension. It is very
difficult to imagine a geometry with variable dimension in the scope of the
Riemannian geometry. Such a problem is absent in T-geometry.

3. For the Riemannian geometry construction, one uses a coordinate system
and the concept of a curve, which are essentially methods of the Riemannian
geometry description. The curve is considered conventionally to be a
geometrical object (but not as a method of the geometry description), and
separation of properties of geometry from properties imported by a use of
the description in terms of curves is not considered usually. In particular,
in the Riemannian geometry the absolute parallelism is absent, in general.
Parallelism of two vectors at remote points is established by means of a
reference to a curve, along which the parallel transport of the vector is
produced. In other words, geometrical property of parallelism of two vectors
is formulated in terms of the method of description, and it is not known,
how to remove this dependence on the methods of description. T-geometry is
free of this defect. The concept of a curve is not used at the T-geometry
construction. There is an absolute parallelism in T-geometry.

4. T-geometry uses a special geometrical language, which contains only
concepts immanent to the geometry in itself ($\Sigma $-function and finite
subspaces). One does not need to eliminate the means of the geometry
description.

5. The geometrical language admits one to consider and to investigate
effectively such a situation, when the future and the past are not
geometrically equivalent.

6. The means of the Riemannian geometry description suppress such an
important property of geometry as nondegeneracy. As a corollary the particle
mass geometrization appears to be impossible in the framework of Riemannian
geometry. Geometrization of the particle mass is important, when the mass of
a particle is unknown and must be determined from some geometrical or
physical relations. It may appear to be important for determination of the
mass spectrum of elementary particles. T-geometry admits geometrization of
the particle mass.

7. Consideration of nondegeneracy and geometrization of the particle mass
have admitted one to make the important step in understanding of the
microcosm space-time geometry. One succeeded in explanation of
non-relativistic quantum effects as geometrical effects, generated by
nondegeneracy of the space-time geometry. There is a hope that asymmetry of
the space-time geometry will admit one to explain important characteristics
of elementary particles geometrically.

Capacities of T-geometry as a space-time geometry are far in excess of the
Riemannian geometry capacities.

\newpage {\LARGE Captions to figures}.

Figure 1. Timelike first order neutral tube for .

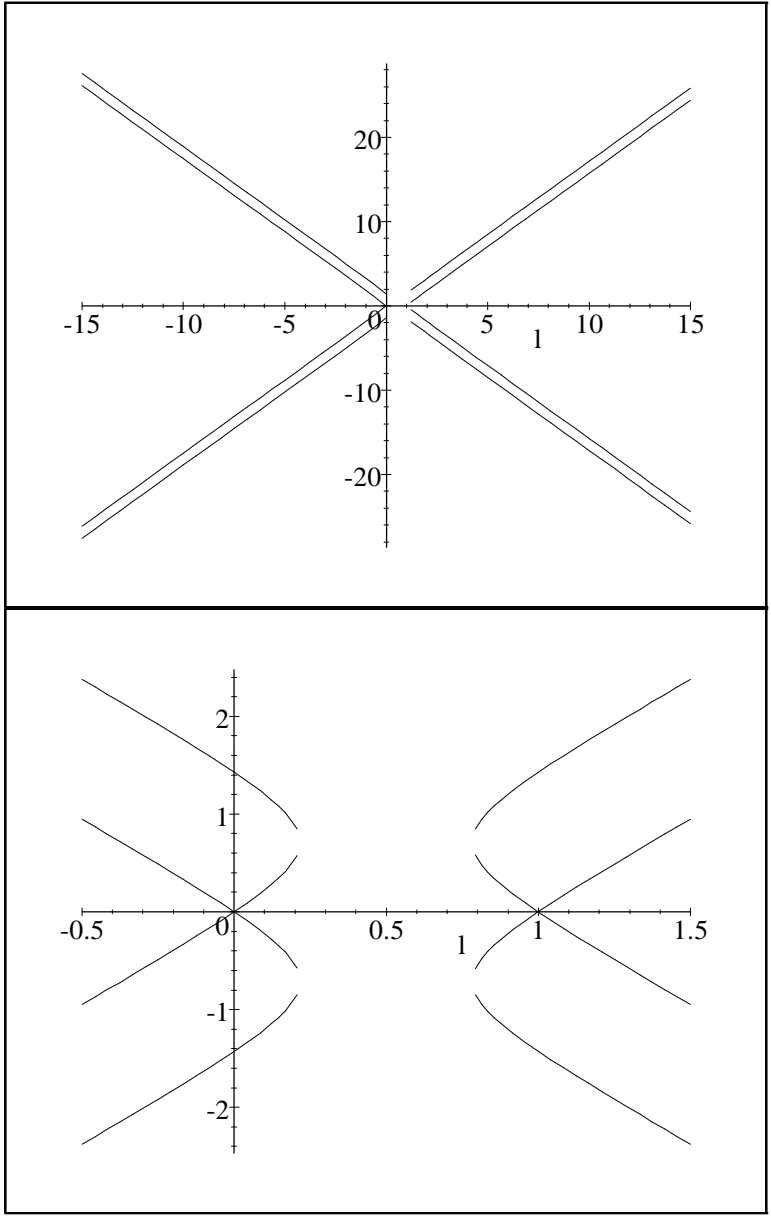
Figure 2. Timelike first order neutral tube for $\alpha \left| y\right| =0.7$

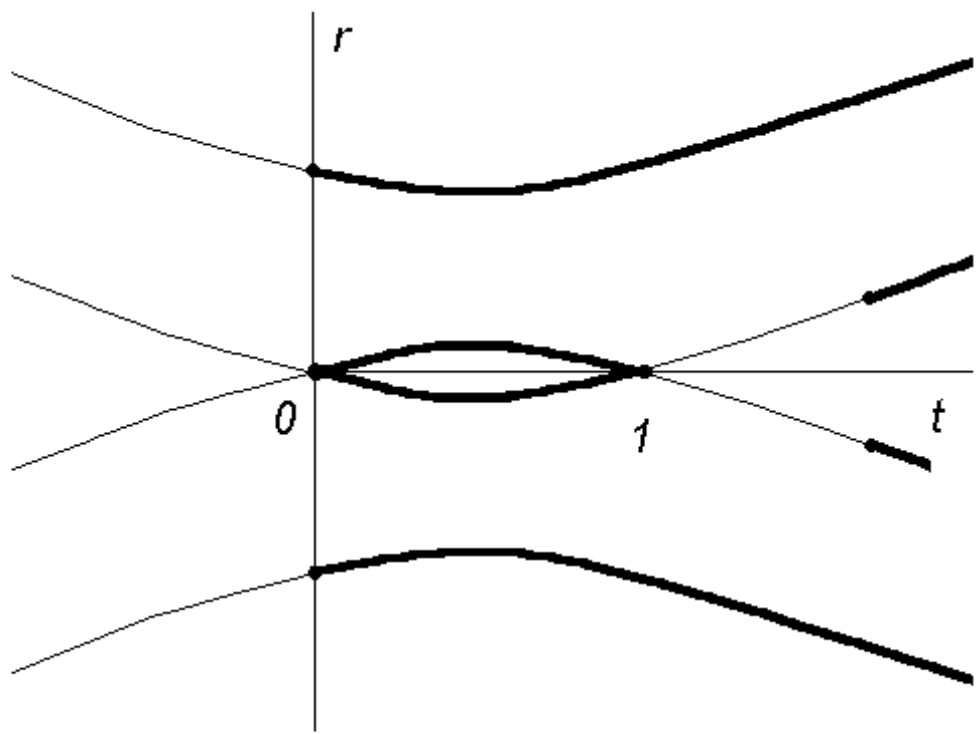
Figure 3. Schematic division of the timelike first order tube ($\alpha
\left| y\right| =0.4$) into internal and external segments. Internal segment
is drawn by thick line, external ones are drawn by thin line.

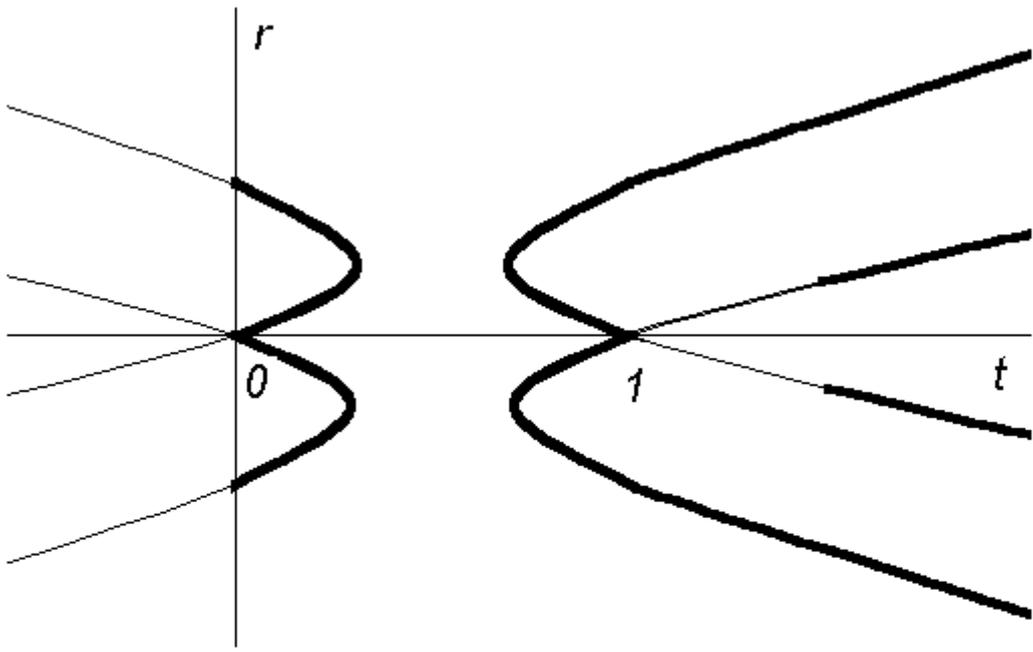
Figure 4. Schematic division of the timelike first order tube ($\alpha
\left| y\right| =0.7$) into internal and external segments. Internal segment
is drawn by thick line, external ones are drawn by thin line.

\end{document}